\documentclass[hidelinks, 11 pt]{article}
\usepackage{booktabs}
\usepackage{amsfonts}
\usepackage{amsmath}
\usepackage{amssymb}
\usepackage{amsthm}
\usepackage[normalem]{ulem}
\newsavebox{\redmathbox}

\newcommand{\redmathstrike}[2]{\text{\sbox{\redmathbox}{\ensuremath{#1#2}}\ifdim\wd\redmathbox>.88\linewidth\resizebox{.88\linewidth}{!}{\sout{\usebox{\redmathbox}}}\else\sout{\usebox{\redmathbox}}\fi}}
\usepackage{subcaption}
\usepackage[pdftex]{graphicx}
\usepackage{natbib}
\usepackage{setspace}
\usepackage[bottom]{footmisc}
\usepackage[hypertexnames=false]{hyperref}
\usepackage{pdfsync}
\usepackage{lscape}
\usepackage{placeins}
\usepackage[hmargin=1in,vmargin=1.25in]{geometry}
\usepackage{float}

\hypersetup{colorlinks, linkcolor = {teal}, citecolor = {blue}, urlcolor ={blue!80!black}}

\makeatletter
\renewcommand\paragraph{\@startsection{paragraph}{4}{\z@}%
            {-2.5ex\@plus -1ex \@minus -.25ex}%
            {1.25ex \@plus .25ex}%
            {\normalfont\normalsize\bfseries}}
\makeatother
\usepackage{mathrsfs,dsfont}
\usepackage[dvipsnames,svgnames, x11names]{xcolor}
\usepackage{pgfplots}
\pgfplotsset{compat=1.17} %suggested by sublime
\usetikzlibrary{patterns}
\usepackage{caption}

\DeclareGraphicsExtensions{.jpg}

\usepackage{tikz}

\newtheorem{theorem}{Theorem}
\newtheorem{assumption}{Assumption}

\newtheorem{lemma}{Lemma}

\newtheorem{remark}{Remark}
\newtheorem{definition}{Definition}
\newtheorem{algorithm}{Algorithm}

\renewcommand{\thesection}{\arabic{section}}
\renewcommand{\theequation}{\arabic{section}.\arabic{equation}}

\usepackage{xcolor}
\definecolor{ucla}{HTML}{005587}

\newcounter{bean}
\newcounter{beana}

\providecommand{\keywords}[1]{   
  \textbf{Keywords:} #1
}

\title{Continuous difference-in-differences with double/debiased machine learning}
\author{Lucas Z. Zhang
\thanks{The author would like to express his gratitude to Denis Chetverikov, Andres Santos, Rosa Matzkin, and referees for their generous time and extremely helpful discussions, which have led to substantial improvements to this paper. Additionally, he extends his thanks to Jinyong Hahn, Zhipeng Liao, Shuyang Sheng, and participants at the UCLA econometrics proseminars for their valuable comments and suggestions. Furthermore, the author is grateful to Kathleen McGarry, Daron Acemoglu, Amy Finkelstein, and the National Bureau of Economic Research for facilitating access to the data source in \cite{AF2008}. Finally, the author thanks the editor and the anonymous referees for their thorough review and constructive feedback.}\\
University of California, Los Angeles, Dept.\ of Economics;\\
Bates White Economic Consulting\\
\\ lucaszz@g.ucla.edu
}
\date{This Version: \today}

\begin{document}

\maketitle
\allowdisplaybreaks % can break the equations into multiple pages.

\begin{abstract}
This paper extends difference-in-differences to settings with continuous treatments. Specifically, the average treatment effect on the treated (ATT) at any level of treatment intensity is identified under a conditional parallel trends assumption. Estimating the ATT in this framework requires first estimating infinite-dimensional nuisance parameters, particularly the conditional density of the continuous treatment, which can introduce substantial bias. To address this challenge, we propose estimators for the causal parameters under the double/debiased machine learning framework and establish their asymptotic normality. Additionally, we provide consistent variance estimators and construct uniform confidence bands based on a multiplier bootstrap procedure. To demonstrate the effectiveness of our approach, we apply our estimators to the 1983 Medicare Prospective Payment System (PPS) reform studied by \cite{AF2008}, reframing it as a DiD with continuous treatment and nonparametrically estimating its effects.

  \keywords{Difference-in-differences, causal inference, continuous treatment, machine learning}

\end{abstract}

\newpage

\section{Introduction}
Difference-in-differences (DiD) is one of the most widely used research designs in empirical work. While conventional DiD settings typically focus on binary or discrete multi-valued treatments, there is growing interest in extending DiD to continuous treatments. The motivation for continuous DiD is clear: the treatment group rarely receives interventions at a constant level, and treatment effects can vary with the intensity or “dose” of the treatment. Thus, rather than comparing treated and control groups before and after an intervention at an aggregate level, one can further investigate how outcomes vary across different treatment intensities within the treated group.

Continuous treatments are prevalent in many empirical settings. For instance, individuals may experience varying levels of exposure to policy interventions, marketing campaigns, or environmental pollutants, all of which can be modeled as continuous treatments. Several recent studies have explored DiD with continuous treatments, including \cite{ZDS2022} on the impact of shutting down online advertising sites, \cite{CJLR23} on racial discrimination in public accommodations, and \cite{AGHP22} on the effects of the expanded child tax credit.

Despite its widespread use in empirical research, the theoretical foundation for continuous DiD remains relatively underdeveloped, particularly in comparison to the extensive body of literature on DiD with binary or discrete treatments (see \cite{RSBP23}, \cite{DD2023}, \cite{CALL2023} for recent overviews). A few recent studies have begun to bridge this gap, notably \cite{CDPV2022}, \cite{DHS2021}, and \cite{CGS2024}. For instance, \cite{DHS2021} extend the change-in-changes model of \cite{AI2006} to accommodate continuous treatments, while \cite{CDPV2022} examine the average slope of stayers in the continuous DiD setting. Our paper is closely related to \cite{CGS2024}, which studies continuous DiD in the commonly used two-way fixed effect (TWFE) regression framework. \cite{CGS2024} demonstrate that, under TWFE, the regression parameter of interest can be decomposed as weighted integrals of either the average treatment parameters across treatment intensities with potentially negative weights or average causal responses with selection bias but nonnegative weights. They also provide data-driven nonparametric estimators for these causal parameters that are rate optimal.

In this paper, we focus on the average treatment effect on the treated (ATT) for any given continuous treatment intensity. Although this parameter is one of several investigated in \cite{CGS2024}, our primary contribution is to incorporate covariates nonparametrically into both the identification and estimation procedures. Specifically, we modify the parallel trends assumption in \cite{CGS2024} by conditioning on covariates in a manner analogous to the ``conditional parallel trends" assumption used in DiD for binary or discrete treatments; see \cite{HIT97,HIST98}, \cite{Abadie2005}, \cite{Chang2020}, and \cite{SZ2020}, for example. As noted in \cite{Abadie2005}, an unconditional parallel trends assumption can be restrictive if covariates that influence outcome dynamics have different distributions across treatment and control groups. By conditioning on such covariates, we obtain a more robust framework for identifying and estimating the ATT in continuous treatment settings.

We first establish identification results analogous to those in \cite{Abadie2005}, adapted to the continuous treatment setting. Based on these identification results, a naive estimator for the ATT can be constructed in two steps. First, one estimates several nuisance parameters from the identification results, including the conditional density of the continuous treatment. In the second step, the nuisance estimates are substituted into a simple average to obtain the estimator of the causal parameter. However, for potentially high-dimensional controls, while one may employ machine learning methods to estimate the nuisance parameters, doing so can introduce substantial bias in the causal parameter estimation (see \cite{CCDDHNR} and the references therein). Moreover, reusing the same sample for both nuisance and causal parameter estimation can result in additional overfitting bias. To address these concerns, we adopt the double/debiased machine learning (DML) framework studied in \cite{CCDDHNR}, which uses orthogonalization and cross-fitting to reduce the influence of nuisance parameter estimation on causal estimates.

Previous studies have adopted similar strategies in related settings. For instance, \cite{Chang2020} considers the DML framework for DiD with binary or discrete treatments, and \cite{SZ2020} proposes efficient doubly robust estimators for DiD with binary treatment. We contribute to and extend this literature to the continuous treatment setting. In particular, in place of the usual propensity score for the treated group, our setting requires the conditional density of the continuous treatment, which poses additional difficulties for directly applying DML methods, often involving only conditional mean functions as the nuisance parameters. To circumvent this, we introduce an approximate causal parameter $ATT_h$ using a kernel function. As the kernel bandwidth shrinks, $ATT_h$ converges to the true $ATT$. Importantly, by focusing on $ATT_h$, we can replace the conditional density with a conditional mean, which allows us to apply the existing DML results. We then derive orthogonal scores for both panel and repeated cross-sectional cases and construct corresponding DML estimators. Building on \cite{CCK2014a, CCK2014b}, \cite{CCDDHNR}, and \cite{FHLZ22}, we establish the asymptotic normality of these estimators and show that the asymptotic bias becomes negligible under an appropriate undersmoothing kernel bandwidth. Additionally, we provide consistent variance estimators via cross-fitting and develop uniform confidence bands for the treatment curve using a multiplier bootstrap procedure. The results from our carefully designed simulation studies suggest that our estimators perform well.

To illustrate the usefulness of our method, we revisit \cite{AF2008}, which examines the impact of the 1983 Medicare payment system (PPS) reform on the healthcare industry. Since the PPS reform affected hospitals with varying proportions of Medicare inpatients differently, the share of Medicare inpatients can be interpreted as a continuous treatment variable. This makes \cite{AF2008} an exemplary case for applying our methods. Thus, we nonparametrically estimate the ATTs of the PPS reform in a continuous DiD context, providing a more detailed understanding of the effects of this policy reform. In particular, contrasting with the linear estimates from \cite{AF2008}, our results suggest significant heterogeneity in the impact of the PPS reform across hospitals with different shares of Medicare inpatients.

We note that the kernel smoothing has been previously considered in the causal inference literature with continuous treatment. For example, \cite{KMMS17} studies average potential outcomes under a continuous treatment, proposing a doubly robust signal and a two-step estimation procedure involving a pseudo-outcome and local kernel linear regression. Along similar lines, \cite{SC2021} employs series methods to establish uniform asymptotic results. \cite{HLM25} recently adopted a similar framework as \cite{KMMS17} to establish identification and estimation results on the average dose effect on treated. This causal parameter differs from ours in that it relies on different sets of parallel trends assumptions and it is an average dose-response on the entire treated group, akin to the average potential outcome. It is important to emphasize that while \cite{KMMS17} and \cite{HLM25} also employ the kernel techniques, their approach differs from ours in non-trivial ways. We use kernels primarily to approximate the original causal parameters, facilitating the construction of orthogonal scores, after which the final estimation proceeds as a simple average; see \cite{BL2017} for a more general discussion on this method. This contrasts with their approach, which uses kernel regressions to estimate the conditional mean of a pseudo-outcome. In this respect, our work is also related to \cite{KZ2018}, \cite{SUZ19}, and \cite{CYYL25}, all of which consider continuous treatments and employ kernel-based moment functions to study the average potential outcomes and partial effects. 

The remainder of this paper is organized as follows. Section 2 introduces continuous DiD and demonstrates the identification of the causal parameter. Section 3 provides the orthogonal scores. In Section 4, we present our estimators and establish their asymptotic properties. Section 5 showcases the simulation results, followed by a detailed empirical example in Section 6. Section 7 concludes.

\section{Setup and Identification}
\setcounter{equation}{0}
In this section, we formally set up the difference-in-differences with continuous treatment following \cite{Abadie2005} and \cite{CGS2024}. First, using the potential outcome notation (e.g. \cite{Rubin74}), let $Y_{i,t}(0)$ denote the potential outcome of individual $i$ in period $t$ when receiving no treatment, and similarly let $Y_{i,t}(d)$ denote the potential outcome of individual $i$ in period $t$ when receiving treatment with intensity $d$.

The treatment variable $D$ is modeled as a random variable with a mixture distribution: a probability mass at $0$ and a continuous distribution on an interval $[d_L,d_H]$ excluding $0$. Specifically, the control group consists of individuals who receive treatment $D=0$, and we need a relatively large number of individuals in the control group so that the comparison with the treated is meaningful. On the other hand, the treated individuals can receive varied treatments, each with a potentially different treatment dose/intensity $D=d\in[d_L,d_H]$. We restrict our attention to the two-period $(t-1, t)$ models and suppress the time notation in treatment $D_i$ in the panel setting. Let $X_i$ denote the set of individual-level covariates. We make the following assumptions:
\begin{assumption}[Panel]\label{didas1}
The observed data $\{Y_{i,t-1}, Y_{i,t}, D_i, X_i\}_{i=1}^N$ are independently and identically distributed.
\end{assumption}

\begin{assumption}[Repeated Cross-Sections]\label{didas2} (a) For each individual $i$ in the pooled sample, $T_i$ is a time indicator $=1$ if observation $i$ belongs to the post-treatment sample and $=0$ otherwise, and $Y_i = (1-T_i)Y_{i,t-1} + T_iY_{i,t}$; (b) $(D,X)\perp T$ and the following holds: (i) conditional on $T=0$, data are i.i.d. from the distribution of $(Y_{t-1},D,X)$; (ii) conditional on $T=1$, data are i.i.d. from the distribution of $(Y_t,D,X)$.
\end{assumption}

\begin{assumption}[Support]\label{didas3} (a) The support of $D$ is $\{0\}\sqcup[d_L, d_H]$ with $0<d_L<d_H<\infty$; (b) there exists a constant $0<\kappa <\frac{1}{2}$ such that, almost surely, $\kappa <P(D=0|X)<1-\kappa$ and $f_{D|X}(d|X) > \kappa$ for all $d\in [d_L, d_H]$.
\end{assumption}

\begin{assumption}[No Anticipation]\label{didas5} $Y_t = Y_t(D)$, $Y_{t-1} = Y_{t-1}(0)$.
\end{assumption}

\begin{assumption}[Conditional Parallel Trends]\label{didas4}
For all $d\in [d_L, d_H]$, the following holds
\begin{align}
E[Y_t(0)-Y_{t-1}(0)|X, D=d] = E[Y_{t}(0)-Y_{t-1}(0)|X, D=0].
\end{align}
\end{assumption}

Assumptions \ref{didas1} and \ref{didas2} are analogous to those in the DiD literature with a discrete treatment. While Assumption \ref{didas1} requires a balanced panel, Assumption \ref{didas2} allows for repeated cross-sections but imposes stationarity of $(D,X)$ and hence rules out compositional changes.\footnote{For DiD with compositional changes, see \cite{HONG2013}, \cite{ZIMMERT2020}, and \cite{SX2025} for detailed discussions in the discrete treatment setting, and \cite{HHZ2024} in the continuous treatment setting.} Assumption \ref{didas3} is the strong overlap assumption, ensuring sufficient support for both treated and untreated individuals, which is crucial for identification. Assumption \ref{didas5} formalizes the requirement that there is no anticipated treatment effect prior to the treatment. Assumption \ref{didas4}, a generalization of the discrete case in \cite{HIT97,HIST98}, is the key identifying condition for the causal parameter. This assumption essentially states that, conditional on covariates, the unobserved counterfactual trend of the treated \textit{at each given treatment intensity} is the same as the observed trend of the control group.

Next, we describe our target parameter. The causal parameter we are interested in is the average treatment effect on the treated (ATT for short) \textit{at any given treatment intensity} $d\in [d_L,d_H]$:
\begin{equation}\label{att}
ATT(d) := E[Y_t(d) - Y_t(0)|D=d].
\end{equation}
The interpretation of this parameter is analogous to the cases with discrete treatment variables: the expected effect of treatment with intensity $d$ for those who actually received treatment with intensity $d$. See also \cite{CGS2024} Section 3 for a comprehensive discussion on (\ref{att}) and an alternative parallel trends assumption under which the average treatment effect $ATE(d) := E[Y_t(d) - Y_t(0)]$ can be identified. The following theorem presents the main results of this section, in which we establish the identification of $ATT(d)$ for both panel and repeated cross-sectional settings.

\begin{theorem}[Identification of ATT]\label{thm:cdid1}
(a) (Panel) If Assumptions \ref{didas1}, \ref{didas3}, \ref{didas5}, and \ref{didas4} hold, then, for any $d\in[d_L, d_H]$,
\begin{align}
ATT(d) = E[Y_t-Y_{t-1}| D=d] - E\bigg[(Y_t-Y_{t-1})\mathbf{1}\{D=0\}\frac{f_{D|X}(d|X)}{f_{D}(d)P(D=0|X)}\bigg];
\end{align}
(b) (Repeated Cross-Sections) if Assumptions \ref{didas2}, \ref{didas3}, \ref{didas5}, and \ref{didas4} hold, then, for any $d\in[d_L, d_H]$,
\begin{align}
ATT(d) = E\bigg[\frac{T-\lambda}{\lambda(1-\lambda)}Y \bigg| D=d\bigg]- E\bigg[\frac{T-\lambda}{\lambda(1-\lambda)}Y\mathbf{1}\{D=0\}\frac{f_{D|X}(d|X)}{f_D(d)P(D=0|X)}\bigg]
\end{align}
where $\lambda := P(T=1)$.
\end{theorem}

With Theorem \ref{thm:cdid1}, one can build estimators for $ATT(d)$ using the estimated sample analogs. For potentially high-dimensional covariates, machine learning methods can be employed to estimate the nuisance parameters, including the conditional density $f_{D|X}(d|X)$ and the conditional probability $P(D=0|X)$. However, the use of machine learning methods can often result in non-trivial first-order biases in the estimation of the causal parameter, see e.g. \cite{CCDDHNR} and references therein for a detailed discussion. Therefore, we consider alternative estimating equations that reduce the influence of the nuisance parameters.

\section{Orthogonal Scores}
\setcounter{equation}{0}
In this section, we focus on the panel case for illustration as the repeated cross-sectional case only requires minor modifications. We begin by introducing \textit{Neyman orthogonality}. Let $\theta_0(d) \in \Theta\subset \mathbf{R}$ be the low-dimensional parameter of interest, e.g., $ATT(d)$, and let $\rho_0(d) \in\mathcal{H}(d)$ denote the true low-dimensional nuisance parameters, e.g., $\rho_0(d) = f_D(d)$. The true infinite-dimensional nuisance parameters $\eta_0(d)\in\mathcal{T}(d)$ include $f_{D|X}(d|X)$ and $P(D=0|X)$ with the estimated $\hat\eta(d)$ in the realization set $T_N(d)\subset \mathcal{T}(d)$ with high probability.\footnote{New infinite-dimensional nuisance parameters can arise when constructing the orthogonal scores. We also explicitly index the nuisance parameters and nuisance function spaces by treatment intensity $d$.} Let $Z$ be the observable random vector, e.g. $Z =(Y_{t-1},Y_t, D, X)$ in the panel setting, and let $\psi: (Z,\theta(d),\rho(d),\eta(d))\mapsto \mathbf{R}$ denote a score function.\footnote{We say $\psi$ is a score function if at the true nuisance parameters $(\rho_0(d),\eta_0(d))$ and the true $\theta_0(d)$, the moment condition $E[\psi(Z,\theta_0(d),\rho_0(d),\eta_0(d)] = 0$ holds.} With these notations, following \cite{CCDDHNR} and \cite{Chang2020}, we formally define the Neyman orthogonality with respect to the infinite-dimensional nuisance parameters.
\begin{definition}[Neyman Orthogonality]\label{neyman} A score $\psi$ satisfies the Neyman orthogonality condition at $(\theta_0(d),\rho_0(d),\eta_0(d))$ with respect to a nuisance realization set $T_N(d)\subset \mathcal{T}(d)$ if (a) $\theta_0(d)$ satisfies the moment condition 
\begin{align}
E_P[\psi(Z,\theta_0(d),\rho_0(d),\eta_0(d))] = 0;
\end{align}
(b) for $r\in[0,1)$ and $\eta(d)\in T_N(d)$, the Gateaux (directional) derivative satisfies
  \begin{align}
  \partial_r E_P[\psi(Z,\theta_0(d),\rho_0(d),\eta_0(d) + r(\eta(d)-\eta_0(d)))]|_{r=0} = 0.
  \end{align}
\end{definition}

In the above definition, (a) says that $\psi$ identifies the parameter of interests while (b) ensures the first-order bias from estimating the \textit{infinite-dimensional} nuisance parameters is zero. Recall that in the panel case,
\begin{align}
\theta_0(d) = ATT(d) = E[\Delta Y|D=d] - E\bigg[\Delta Y\mathbf{1}\{D=0\}\frac{f_{D|X}(d|X)}{f_D(d)P(D=0|X)}\bigg].
\end{align}
where $\Delta Y:= Y_t - Y_{t-1}$. First, given the continuous nature of the treatment intensity, $\theta_0(d)$ cannot be estimated nonparametrically at root-$N$ rate. This relates to a class of non-regular parameters involving continuous treatment variables; see \cite{GW2015}, \cite{KMMS17}, \cite{SUZ19}, \cite{SC2021}, \cite{FHLZ22}, and \cite{CYYL25} for example. Moreover, a score based on the above expression does not satisfy Neyman orthogonality, and an adjustment term has to be added.

To this end, we approximate the non-regular $ATT(d)$ with a family of smoothed parameters that are tractable. We note that this approach has been discussed extensively in \cite{BL2017} and \cite{CYYL25}, and specifically we rely on the following observation (e.g., \cite{FYT96}):
\begin{equation}\label{cd:kernel}
  f_{D|X}(d|x) = \lim_{h\to 0} E[K_h(D-d)|X=x],\quad K_h(u) := \frac{1}{h}K\Big(\frac{u}{h}\Big)
\end{equation}
where $K(\cdot)$ is a kernel function. Replacing $E[\Delta Y|D=d]$ and $f_{D|X}(d|x)$ by their kernel counterparts, we can define $ATT_h(d)$ as follows:
\begin{align}\label{atth}
  ATT_h(d) :=& E\bigg[\Delta Y\frac{K_h(D-d)}{f_D(d)}\bigg] - E\bigg[\Delta Y\mathbf{1}\{D=0\}\frac{E[K_h(D-d)|X]}{f_D(d)P(D=0|X)}\bigg] \notag \\
  =& E\bigg[\Delta Y\frac{K_h(D-d)P(D=0|X) - \mathbf{1}\{D=0\}E[K_h(D-d)|X]}{f_D(d)P(D=0|X)} \bigg],
\end{align}
which is an expression that consists of only conditional expectations, apart from the density normalization at a fixed $d$. Notably, it can be shown that 
\begin{align*}
  ATT(d) = \lim_{h\to 0} ATT_h(d),
\end{align*}
which suggests that we can work with $ATT_h(d)$ instead. In particular, define the bias $B_h(d):= ATT(d) - ATT_h(d)$, one can show that $B_h(d) = O(h^2)$, and we defer the formal result to the next section. For notation simplicity, we now formally define $ATT_h(d)$ in both settings.

\begin{definition}[Panel]
\begin{align}\label{eq:atth1}
ATT_h(d) = E\bigg[\Delta Y \frac{K_h(D-d)P(D=0|X) - \mathbf{1}\{D=0\}E[K_h(D-d)|X]}{f_D(d)P(D=0|X)} \bigg]
\end{align}
where $\Delta Y = Y_t - Y_{t-1}$.
\end{definition}

\begin{definition}[Repeated Cross-Sections]
\begin{align}\label{eq:atth2}
ATT_h(d) =& E\bigg[Y^\lambda\frac{K_h(D-d)P(D=0|X) - \mathbf{1}\{D=0\}E[K_h(D-d)|X]}{f_D(d)P(D=0|X)} \bigg]
\end{align}
where $Y^\lambda := \frac{T-\lambda}{\lambda(1-\lambda)}Y$ with $\lambda = P(T=1)$.
\end{definition}

Our goal is to construct scores that satisfy Neyman orthogonality for each $h$, and then take the limit as $h\to 0$. The next lemma presents such scores. To simplify the expressions, denote: $g(X) := P(D=0|X)$; $f_h(d|X):= E[K_h(D-d)|X] $; $\mathcal{E}_{\Delta Y}(X) := E[\Delta Y|X,D=0]$; $\mathcal{E}_{\lambda Y}(X) := E\big[\frac{T-\lambda}{\lambda(1-\lambda)}Y\big|X, D=0\big]$ with $\lambda = P(T=1)$.

\begin{lemma}\label{lm:scores}
Define (a) for the panel setting,
\begin{equation}\label{psi1}
\psi_h^{(1)} := \frac{K_h(D-d)g(X) - \mathbf{1}\{D=0\}f_h(d|X)}{f_D(d)g(X)}\bigg(\Delta Y -  \mathcal{E}_{\Delta Y}(X)\bigg) -ATT_h(d),
\end{equation}
and (b) for the repeated cross-sectional setting,
\begin{equation}\label{psi2}
\psi_h^{(2)} := \frac{K_h(D-d)g(X) - \mathbf{1}\{D=0\}f_h(d|X)}{f_D(d)g(X)} \bigg(\frac{T-\lambda}{\lambda(1-\lambda)}Y - \mathcal{E}_{\lambda Y}(X) \bigg) -ATT_h(d).
\end{equation}
Suppose there exist $M_h^{(1)}\in L^1(P_{Y_{t-1},Y_t,D,X})$ and $M_h^{(2)}\in L^1(P_{Y,T,D,X})$ such that $|\psi_h^{(1)}|\leq M_h^{(1)}$ and $|\psi_h^{(2)}|\leq M_h^{(2)}$ almost surely. Then the scores $\psi_h^{(1)}$ and $\psi_h^{(2)}$ satisfy Neyman orthogonality defined in (\ref{neyman}).
\end{lemma}
The proof is provided in the appendix, where we construct the adjustment term and verify the Neyman orthogonality conditions from Definition \ref{neyman}. We also provide an alternative derivation of these orthogonal scores using the method proposed in \cite{HDDV22}. The assumption on the existence of integrable functions $M_h^{(1)}$ and $M_h^{(2)}$ is mild and it justifies interchanging expectation and differentiation. For simplicity, we omit superscripts on $\psi^{(1)}_h$ and $\psi^{(2)}_h$ whenever the context is clear. The infinite-dimensional nuisance parameters in these new scores include $f_h(d|X)$, $g(X)$, $\mathcal{E}_{\Delta Y}(X)$, and $\mathcal{E}_{\lambda Y}(X)$, with the latter two introduced by the adjustment terms. Notably, the estimating moments for $ATT_h(d)$ based on these orthogonal scores remain robust to the first-order biases introduced by the nuisance estimates. In the next section, we construct DML estimators of $ATT(d)$ using these scores and establish their asymptotic properties.

\section{Estimation and Inference}
\setcounter{equation}{0}
As mentioned in the introduction, constructing DML estimators involves two main steps. In the previous section, we established scores that satisfy Neyman orthogonality (Lemma \ref{lm:scores}). These scores are then used alongside a cross-fitting procedure, further reducing estimation bias. With these key components in place, we construct DML estimators following the procedure proposed by \cite{CCDDHNR}. 

First, we partition the sample $I_N$ into $K\geq 2$ disjoint subsets $\{I_k\}_{k=1}^K$ of equal size $n=N/K$. For each $k\in\{1,\cdots,K\}$, we use the auxiliary sample $I_k^c:= I_N\setminus I_k$ to estimate the nuisance parameters. We then compute sample averages according to (\ref{psi1}) and (\ref{psi2}) using these estimates, evaluated at $I_k$, to obtain $\widehat{ATT}_k(d)$. Finally, we average across the $K$ estimates to obtain the final estimator $\widehat{ATT}(d)$. We note that at each $k=1,\cdots, K$, the nuisance parameters and $\widehat{ATT}_k(d)$ are estimated using disjoint subsamples, which reduces the overfitting bias and significantly simplifies the asymptotic analysis. Moreover, since $K$ is fixed, it does not affect the asymptotic properties of the estimator. In practice, we recommend using $K=5$ as a rule of thumb and leave the optimal choice of $K$ to future research. The detailed algorithms are deferred to Appendix A.

Next, we outline the regularity conditions required to establish the asymptotic properties of our DML estimators. We focus on the panel case and present the analogous results for the repeated cross-sections in Appendix B. For notation simplicity, let $\mathcal{D} := (d_L, d_H)$, and let $\mathcal{X}$ and $\Delta\mathcal{Y}$ denote the support of $X$ and $\Delta Y$ respectively.

\begin{assumption}[Kernel]\label{cdidas1} The kernel function $K(\cdot)$ satisfies: (a) $K(\cdot)$ is bounded and differentiable, with uniformly bounded derivative; (b) $\int K(u) du = 1$, $\int uK(u)du = 0$, $0<\int u^2 K(u) du <\infty$; (c) $\sup_{u\in\mathbb R}|u|^3|K(u)|<\infty$. Moreover, for notation simplicity, define $K_h(u):= h^{-1}K(u/h)$.
\end{assumption}

\begin{assumption}[Bounds and Smoothness, Panel]\label{cdidas2} (a) There exist constants $c>0$ and $0<C<\infty$ such that $\inf_{d\in\mathcal{D}} f_D^0(d)>c$, $|Y_{t-1}| < C$, $|Y_{t}| < C$, $c <f_h^0(d|X)<C$ $\forall d\in\mathcal{D}$, and $|\mathcal{E}_{\Delta Y}^0(X)|<C$ almost surely; (b) $f_D^0(d) \in C^2(\mathcal{D})$ and $\sup_{d\in\mathcal{D}}|\partial_d^2 f_D^{0}(d)| < \infty$; (c) $f_{D|X}^0(d|x) \in C^2(\mathcal{D})$ $\forall x\in \mathcal{X}$ and $\sup_{d,x \in \mathcal{D},\mathcal{X}} |\partial^2_d f_{D|X}^0(d|x)| < \infty$; (d) $f_{\Delta Y, D}^0(t, d) \in C^2(\mathcal{D})$ for every $\forall t\in \Delta\mathcal{Y}$ and $\sup_{t,d \in \Delta\mathcal{Y},\mathcal{D}} |\partial_d^2 f_{\Delta Y, D}^0(t, d)| < \infty$.
\end{assumption}

\begin{assumption}[Rates, Panel]\label{cdidas3} (a) The kernel bandwidth $h = h_N\to 0$ satisfies $Nh\to\infty$ and $\sqrt{Nh^5} = o(1)$; (b) there exists a sequence $\varepsilon_N\to 0$ such that $h^{-1}\varepsilon_N^2 = o(1)$; (c) with probability tending to $1$, $\|\hat{f}_h(d|X) - f_h^0(d|X)\|_{P,2}\leq h^{-1/2}\varepsilon_N$, $\|\hat{g}(X) - g_0(X)\|_{P,2}\leq \varepsilon_N$, $\|\hat{\mathcal{E}}_{\Delta Y}(X) - \mathcal{E}_{\Delta Y}^0(X)\|_{P,2}\leq \varepsilon_N$; (d) with probability tending to $1$, $\kappa<\hat{g}(X) < 1-\kappa$ and $c <\hat{f}_h(d|X)<C$ almost surely, and $\|\hat{\mathcal{E}}_{\Delta Y}(X)\|_{P,\infty}<C$.
\end{assumption}

The kernel function is central to our analysis. In addition to its well-established theoretical properties for estimating the density $f_D(d)$, we also use it to approximate the point mass at $D=d$ and the conditional density $f_{D|X}(d|X)$. Assumption \ref{cdidas1} imposes the standard regularity conditions on the kernel function, which are essential for establishing the asymptotic normality of our estimator. Assumption \ref{cdidas2} requires smoothness and boundedness of the outcome variable and relevant distributions, while Assumption \ref{cdidas3} specifies conditions on the kernel bandwidth and the quality of the nonparametric nuisance estimators.

\begin{remark}
\textnormal{Assumption \ref{cdidas3} (a) and (b) give $h = o(N^{-1/5})$ and $h = \omega(\varepsilon_N^2)$. However, consistency of our variance estimator additionally requires $h^{-2}\varepsilon_N^2 + h^{-3}N^{-1} = o(1)$, which imposes a more restrictive lower bound on $h$. Moreover, whereas the standard DML literature assumes the nuisance estimators to converge at rate $\varepsilon_N = o(N^{-1/4})$, we allow the conditional density $\hat{f}_h$ to converge at a slower rate $h^{-1/2}\varepsilon_N$. This relaxation does not contradict the existing DML results for regular parameters; our target parameter is non-regular and cannot be estimated nonparametrically at $\sqrt{N}$ rate because of the continuous treatment. Finally, although our results assume a deterministic kernel bandwidth, they should extend to data-driven choices (for example, the adaptive procedure in \cite{BL2017}), which we leave to future work.}
\end{remark}

The following lemma characterizes the bias of using kernels to approximate $ATT(d)$.

\begin{lemma}[Bias of $ATT_h(d)$, Panel]\label{lm:bias}
Suppose Assumptions \ref{cdidas1}, \ref{cdidas2}, \ref{cdidas3} hold. Then $B_h(d) := ATT(d) - ATT_h(d)$ satisfies $B_h(d) = O(h^2)$ for any $d\in \mathcal{D}$.
\end{lemma}

The proof is given in the companion supplement. This lemma suggests that, for an undersmoothing bandwidth, the bias does not affect the asymptotic distribution of our estimators. The next theorem is the main result of this section that establishes the asymptotic normality of our estimator for $ATT(d)$.

\begin{theorem}[Asymptotic Normality, Panel]\label{thm:asymp}\hspace{-2.9pt} Suppose assumptions \ref{didas1}, \ref{didas3}, \ref{didas5}, \ref{didas4}, \ref{cdidas1}, \ref{cdidas2}, and \ref{cdidas3} hold. Then, for $d\in\mathcal{D}$, if $\varepsilon_N = o(N^{-1/4})$ and $\liminf_{N\to\infty}h\sigma_N^2(d)>0$,
\[
\frac{\widehat{ATT}(d) - ATT(d)}{\sigma_{N}(d)/\sqrt{N}}\quad \to^d\quad N(0,1)
\]
where
\begin{align}\label{var1}
\sigma_{N}^2(d):= E\bigg[\bigg(\psi_h^{(1)}(Z,\theta_{0h}(d),f_D^0(d),\eta_0(d)) - \frac{\theta_{0h}(d)}{f_D^0(d)}\big(K_h(D-d)-E[K_h(D-d)]\big)\bigg)^2\bigg]
\end{align}
for $\theta_{0h}(d):= ATT_h(d)$ defined in (\ref{eq:atth1}) and $\psi_h^{(1)}$ defined in (\ref{psi1}).
\end{theorem}

The proof builds on the DML framework of \cite{CCDDHNR}, modified to accommodate kernel smoothing. The asymptotic variance has two components, both depending inversely on the kernel bandwidth $h$: one arising from the kernels in the orthogonal score $\psi_h$, and the other from the linear expansion of the estimator with respect to the kernel density estimator $\hat{f}_D(d)$. Since $h$ is a function of the sample size $N$ under our assumptions, we index the asymptotic variance by $N$ to reflect this dependence. Therefore, our estimator $\widehat{ATT}(d)$ attains a convergence rate of $\sqrt{Nh}$, which, though slower than the parametric rate $\sqrt{N}$, is comparable to the optimal rate for one-dimensional nonparametric regression estimation. 

Next, following \cite{CCDDHNR} and \cite{Chang2020}, we consider a cross-fitted variance estimator. For notation simplicity, denote $\hat{\theta}_h(d):= \widehat{ATT}(d)$ and $E_{n,k} f(Z_i):= n^{-1}\sum_{i\in I_k}f(Z_i)$ as the empirical average of a function $f$ evaluated at $Z_i$'s in the subsample $I_k$. For the panel case, define
\begin{equation}\label{asvar}
\hat{\sigma}_{N}^2(d) := \frac{1}{K}\sum_{k=1}^K E_{n,k}\bigg[
\bigg(
\psi_h^{(1)}(Z,\hat{\theta}_h(d),\hat{f}_k(d), \hat{\eta}_k(d)) - \frac{\hat{\theta}_h(d)}{\hat{f}_k(d)}\big(K_h(D-d)-\hat{f}_k(d)\big)
\bigg)^2
\bigg].
\end{equation}

Then, with this variance estimator, the $1-\alpha$ confidence interval can be constructed as $[\widehat{ATT}(d) - z_{1-\alpha/2}\hat{\sigma}_N(d)/\sqrt{N}, \widehat{ATT}(d) + z_{1-\alpha/2}\hat{\sigma}_N(d)/\sqrt{N}]$ where $z_{1-\alpha/2}$ denotes the $1-\alpha/2$-th quantile of the standard normal random variable. The following theorem establishes the consistency of the cross-fitted variance estimator.

\begin{theorem}[Consistency of Variance Estimator, Panel]\label{thm:asvar}
Suppose the conditions of Theorem \ref{thm:asymp} hold and assume that $h^{-2}\varepsilon_N^2 + h^{-3}N^{-1} = o(1)$. Then
\begin{align*}
 \hat{\sigma}_{N}^2(d) = \sigma_{N}^2(d) + o_p(1)
\end{align*}
where $\hat{\sigma}_{N}^2(d)$ is defined in (\ref{asvar}) and $\sigma_{N}^2(d)$ is defined in (\ref{var1}).
\end{theorem}

Alternatively, we can consider a multiplier bootstrap procedure to construct confidence intervals. Such procedure has been discussed extensively in recent studies, see, e.g., \cite{CCK2014b}, \cite{BCFH17}, \cite{SUZ19},  \cite{CJ21}, \cite{FHLZ22}, and \cite{CYYL25}. First, we make the following assumption on the multiplier.
\begin{assumption}[Sub-exponential Multiplier]\label{as:subexp}
The random variable $\xi$ satisfies: (a) $\xi$ has a sub-exponential distribution; (b) $E[\xi]=Var(\xi) = 1$; (c) $\xi$ is independent of $(Y_{t-1}, Y_t, D, X)$ for the panel case and independent of $(Y, T, D, X)$ for the repeated cross-sectional case.
\end{assumption}
In practice, let $\{\xi_i\}_{i=1}^N$ be an i.i.d. sequence of random variables that satisfies Assumption \ref{as:subexp}. Then for each $b=1,\cdots, B$, we independently draw such a sequence $\{\xi_i\}_{i=1}^{N}$ and construct estimates based on the following expression. For the panel case, define
\begin{align}\label{bootstrap}
\widehat{ATT}(d)_{b}^* := \frac{1}{N}\sum_{k=1}^K\sum_{i\in I_k} \xi_i& \frac{K_h(D_i-d)\hat{g}_k(X_i) - \mathbf{1}\{D_i=0\}\hat{f}_{h,k}(d|X_i)}{\hat{f}_k^*(d)\hat{g}_k(X_i)}\notag \\
\times& \big(\Delta Y_i - \hat{\mathcal{E}}_{\Delta Y,k}(X_i)\big),
\end{align}
where $\hat f_k^*(d)=(N-n)^{-1}\sum_{i\in I_k^c}\xi_iK_h(D_i-d)$, and the other nuisance estimates are as in the original estimation procedure. Let $\hat{c}_{\alpha}$ denote the $\alpha$-th quantile of $\{\widehat{ATT}(d)_{b}^*- \widehat{ATT}(d)\}_{b=1}^B$, a $1-\alpha$ confidence interval can be constructed as $[\widehat{ATT}(d) - \hat{c}_{1-\alpha/2}, \widehat{ATT}(d) -\hat{c}_{\alpha/2}]$.

Moreover, we establish a uniform linear expansion of the bootstrap estimator. For the uniform results, let $\mathcal D$ be a fixed compact subset of $(d_L,d_H)$. The following assumption strengthens Assumption \ref{cdidas3}. 

\begin{assumption}[Uniform Inference Rates, Panel]\label{cdidas4} \sloppy
(a) The kernel bandwidth $h = h_N\to 0$ satisfies $Nh/\log^4(N)\to\infty$ and $\sqrt{Nh^5} = o(1)$; (b) there exists a sequence $\varepsilon_N\to 0$ such that $h^{-1}\varepsilon_N^2  = o(1)$; (c) with probability tending to $1$, $\sup_{d\in\mathcal{D}}\|\hat{f}_h(d|X) - f_h^0(d|X)\|_{P,2}\leq h^{-1/2}\varepsilon_N$, $\|\hat{g}(X) - g_0(X)\|_{P,2}\leq \varepsilon_N$, $\|\hat{\mathcal{E}}_{\Delta Y}(X) - \mathcal{E}_{\Delta Y}^0(X)\|_{P,2}\leq \varepsilon_N$; (d) with probability tending to $1$, $\kappa<\hat{g}(X)<1-\kappa$ and $c <\hat{f}_h(d|X)<C$ almost surely $\forall d\in\mathcal{D}$, $\sup_{d\in \mathcal{D}}|\hat{f}_D^{(1)}(d)|<C$, $\sup_{d\in\mathcal{D}}\|\partial_d \hat{f}_h(d|X)\|_{P,\infty}< C$, and $\|\hat{\mathcal{E}}_{\Delta Y}(X)\|_{P,\infty}<C$.
\end{assumption}

This assumption differs from the pointwise case in two key ways. First, we require that, uniformly over the compact domain $\mathcal{D}$, the nuisance estimator $\hat{f}_h(d|X)$ remains bounded and has rate $h^{-1/2}\varepsilon_N$. Second, we assume that the estimated density and conditional density to have bounded derivatives with probability tending to $1$, ensuring that the score functions are Lipschitz continuous on $\mathcal{D}$. These additional assumptions are mild and can be enforced during estimation procedures. With these modified assumptions, the linear expansion of the bootstrap estimators holds uniformly over $d\in\mathcal{D}$.

\begin{theorem}[Uniform Linear Expansion, Panel]\label{thm:unifexp} Suppose assumptions \ref{didas1}, \ref{didas3}, \ref{didas5}, \ref{didas4}, \ref{cdidas1}, \ref{cdidas2}, \ref{as:subexp}, and \ref{cdidas4} hold. Then for $\varepsilon_N = o(N^{-1/4})$
\begin{align}
&\widehat{ATT}(d)^* - \widehat{ATT}(d) \notag \\ &= \frac{1}{N}\sum_{i=1}^N \dot{\xi}_i\Bigg[\psi_h^{(1)}(Z_i,\theta_{0h}(d),f_D^0(d),\eta_0(d)) - \frac{\theta_{0h}(d)}{f_D^0(d)}\big(K_h(D_i-d)-E[K_h(D-d)]\big)\Bigg]\notag \\ &+ R^{(1)}(d)
\end{align}
where $\dot{\xi}_i := \xi_i - 1$ and $\sup_{d\in\mathcal{D}} |R^{(1)}(d)| = o_p( (Nh)^{-1/2})$.
\end{theorem}
This theorem is the basis for establishing uniform inference theory using the multiplier bootstrap estimator. We consider the following procedure, see \cite{CCK2014b} and \cite{FHLZ22} for example, for constructing uniform confidence bands.
\begin{itemize}
    \item[1.] Construct $\widehat{ATT}(d)$ and $\hat{\sigma}_N(d)$ on a finite grid of values $d\in \bar{\mathcal{D}} \subset \mathcal{D}$.
    \item[2.] For each $b=1, \cdots, B$, draw an i.i.d. sequence of multipliers $\{\xi\}_{i=1}^N$ from a $N(1,1)$ distribution, and construct $\widehat{ATT}(d)_{b}^*$ for all $d\in \bar{\mathcal{D}}$.
    \item[3.] Compute $\hat{c}(1-\alpha)$, which we denote as the $(1-\alpha)$-th quantile of 
    \[
    \Bigg\{\max_{d\in \bar{\mathcal{D}}} \frac{\sqrt{N}|\widehat{ATT}(d) - \widehat{ATT}(d)_{b}^*|}{\hat{\sigma}_N(d)} \Bigg \}_{b=1}^B.
    \]
    \item[4.] For all $d\in\mathcal{D}$, construct the $1-\alpha$ uniform confidence band as
    \[
    [\widehat{ATT}(d) - \hat{c}(1-\alpha)\hat{\sigma}_N(d)/\sqrt{N}, \quad \widehat{ATT}(d) + \hat{c}(1-\alpha)\hat{\sigma}_N(d)/\sqrt{N}].
    \]
\end{itemize}

With Assumption \ref{cdidas4} and suitable additional uniform rate conditions, we can adapt our proof of Theorem \ref{thm:asvar} to establish the uniform consistency of our cross-fitted variance estimator (\ref{asvar}) over $\mathcal{D}$. Then, with Theorem \ref{thm:unifexp} and additional regularity conditions, we can show that the proposed uniform confidence band achieves
asymptotic coverage of $1-\alpha$, using results from
\cite{CCK2014a} (Proposition 3.2 and Remark 3.2) and \cite{CCK2014b} (Corollary 3.1). Since this general approach is well established in the literature, e.g., see the discussion of Theorem 4.2 in \cite{FHLZ22},
we do not establish a formal coverage result here. Instead, we focus on presenting the new results in Theorem \ref{thm:unifexp} and defer its proof to the appendix.

\begin{remark}
\textnormal{A natural extension of our framework is to develop a test for the conditional parallel trends assumption, akin to the approach in \cite{CS2018}, Section 4, which examines differences between the not-yet-treated and the never-treated in the pre-treatment period. Extending such a test to the continuous treatment setting requires a multi-period generalization of the methods considered in this paper, and we suspect that stronger parallel trends assumptions would be necessary for a valid test. While our companion study, \cite{HHZ2024}, proposes estimators that could aid in this analysis, a formal testing procedure remains an open question. Additionally, drawing on insights from \cite{SZ2020}, we recognize that more efficient estimators may exist in the repeated cross-sectional settings than those considered in this paper (Appendix B), and we defer a detailed investigation of such estimators to \cite{HHZ2024}.}
\end{remark}

\section{Simulation}
\setcounter{equation}{0}
\noindent \textbf{Data-generating process} (a) $p = 100$ dimensional covariates $X \sim N(0.2,\Sigma)$, where $\Sigma$ has variances $1$ on the diagonal and covariances $0.1$ off-diagonal; (b) the control group propensity score follows $P(D=0|X) = 1/(1+\exp(-X'\gamma))$, with $\gamma_j = 0.5j^{-2}$; (c) for $D>0$, the continuous treatment is generated as $D= (1+\exp(X'\alpha))^{-1} + V$, where $V \perp X$, $V \sim Beta(2,2)$, $\alpha_j = 0.3j^{-2}$; (d) the potential outcomes are given by $Y_{t-1}(0) = \epsilon_1$, $Y_t(0) = Y_{t-1}(0) + X'\beta + 1 + \epsilon_2$, $Y_t(D) = Y_t(0) - 0.5D^2 + \epsilon_3$, where $\beta_j = 0.5/j$ for $j = 1, \cdots, 6$ and $0$ otherwise, and $(\epsilon_1, \epsilon_2, \epsilon_3)\sim N(0, I_3)$. For the panel setting, the generated data are $(Y_{i,t-1}, Y_{i,t}, X_i, D_i)$, with $Y_{t-1} = Y_{t-1}(0)$ and $Y_t = 1\{D>0\}Y_t(D) + 1\{D=0\}Y_t(0)$. Additionally, for the repeated cross-sectional setting, the generated data are $(Y_i, T_i, X_i, D_i)$, with time indicator $T \sim \text{Bern}(0.5)$ and $Y = TY_t + (1-T)Y_{t-1}$, $Y_{t-1} = Y_{t-1}(0)$, $Y_t = 1\{D>0\}Y_t(D) + 1\{D=0\}Y_t(0)$.

In our simulations, the nuisance parameters $P(D=0|X)$, $f_h(d|X) = E[K_h(D-d)|X]$, $E[Y_t - Y_{t-1}|X,D=0]$, and $E\big[\frac{T-\lambda}{\lambda(1-\lambda)}Y\big|X,D=0\big]$ are estimated nonparametrically using random forests each with 200 trees of maximum depth 20.  Throughout our simulations, we also use an undersmoothing kernel bandwidth $h = 1.06\hat{\sigma}_{\tilde{D}}N^{-1/4}$, where $\hat{\sigma}_{\tilde{D}}$ is the estimated standard deviation of positive treatment intensities. We consider sample sizes $N = 2000$ and $10000$ for both panel and repeated cross-sectional settings, and we conduct $B = 500$ simulations in each setting. The DGP implies the true $ATT(d) = -0.5d^2$, and we focus on a specific treatment intensity $d = 0.9$. Notably, the continuous treatment variable is dependent on the correlated high-dimensional covariates in a nonlinear way. Additionally, the DGPs suggest that the effective sample size should be small at the target intensity, which adds another layer of difficulty for estimation. 

\begin{figure}[htbp!]
\centering
\includegraphics[width=\textwidth]{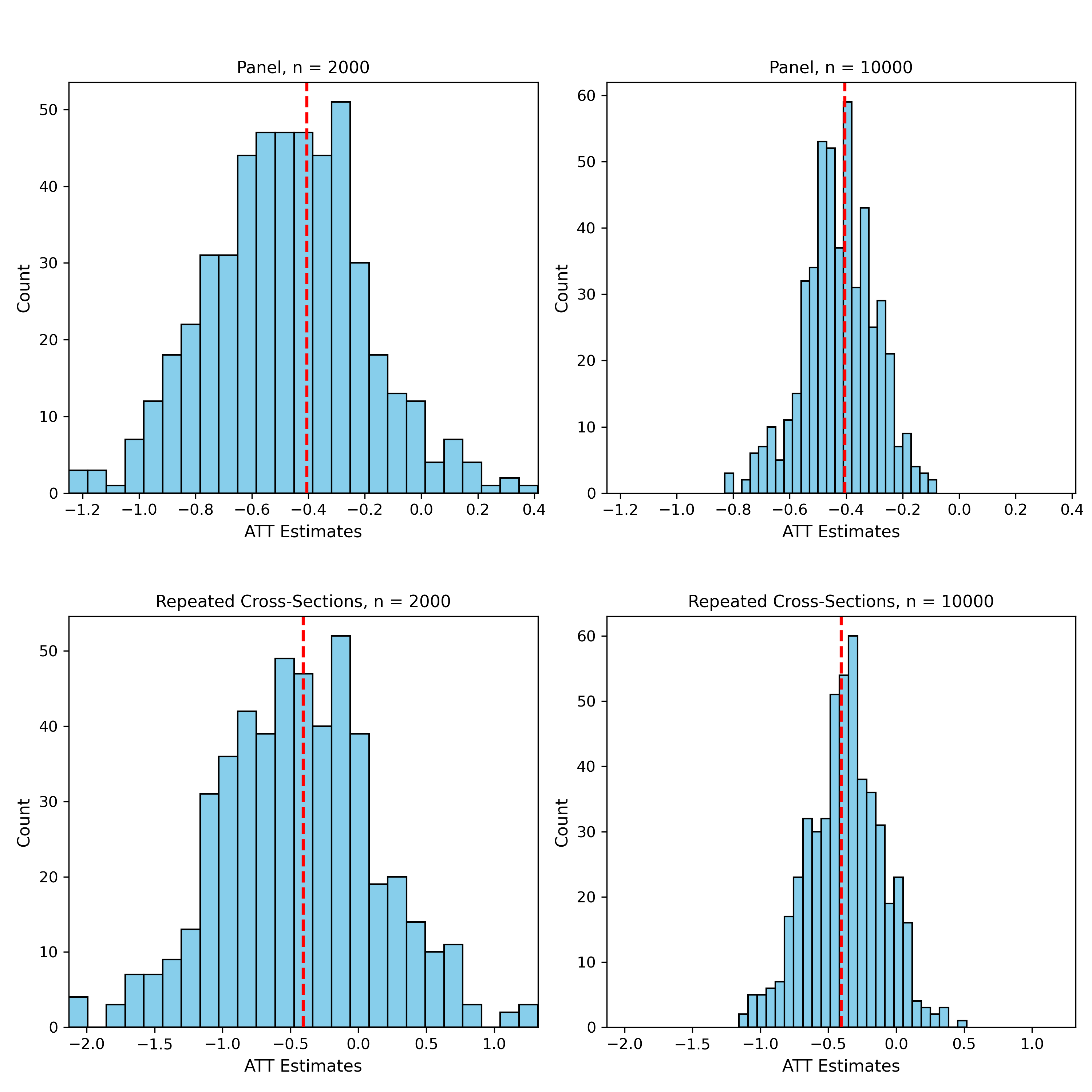}
\caption{The simulation results, true $ATT(d) = -0.405$.}
\label{figs1}
\end{figure}

Despite these challenges, the simulation results suggest that our estimators perform well. The histograms of these simulation estimates are shown in Figure \ref{figs1}, where the red lines indicate the true ATT. We see that as the sample size increases, both bias and variance decrease. The simulation estimates appear to follow a normal distribution in each case, which is consistent with our asymptotic theory. Moreover, in Table \ref{tab01}, we report the bias, the standard deviation of estimated ATTs (Std), the root-mean-squared error (RMSE), the average standard deviations (AVSE), and the coverage probability of 95 percent confidence intervals. In both settings, bias, standard deviation, and RMSE decrease as the sample size increases. The standard deviations of the simulation estimates are very close to the average estimated standard errors, suggesting that our variance estimators perform well. The coverage of the estimated confidence intervals is close to 95 percent, although there is a slight under-coverage in the panel setting.

\begin{table}[htbp!]
\caption{Monte Carlo simulation results.}\label{tab01}
\vspace{-10pt}
\begin{center}
\begin{tabular*}{\textwidth}{@{\extracolsep{\fill}}lccccc@{}}
\hline\hline
   Setting and Sample Size &      Bias &  Std &     RMSE &     AVSE &  Coverage \\
\hline
            panel, n=2000 &  -0.0720 &  0.2725 & 0.2819 & 0.2524 &  0.9180 \\
           panel, n=10000 &  -0.0198 &  0.1261 & 0.1277 & 0.1262 &  0.9300 \\
 cross-sections, n = 2000 &  -0.0340 &  0.5745 & 0.5755 & 0.5578 &  0.9440 \\
cross-sections, n = 10000 &   0.0290 &  0.2754 & 0.2769 & 0.2710 &  0.9500 \\
\hline\hline
\end{tabular*}
\end{center}
\end{table}

\section{Empirical Example}
\setcounter{equation}{0}
\subsection{Background}
The Medicare Prospective Payment System (PPS) reform, introduced in 1983, shifted Medicare hospital reimbursements from a full-cost model to a fixed payment per diagnosis. However, for the first three years, capital costs continued to be reimbursed based on actual expenses.\footnote{As noted in \cite{AF2008}, Medicare’s capital cost reimbursements remained unchanged until 1991 due to delays.} This created a relative increase in labor costs for hospitals treating Medicare inpatients. \cite{AF2008} highlights this feature, showing that the PPS reform significantly increased hospitals’ capital-labor ratios and encouraged technology adoption.

Theoretically, \cite{AF2008} predicts that PPS reform would lead to a higher capital-labor ratio and, if capital-labor substitution is sufficiently elastic, an increased demand for capital and technology. Since only hospitals with Medicare inpatients were affected, these effects likely varied with Medicare inpatient share. To test these predictions, \cite{AF2008} uses data from the 1980–1986 Annual American Hospital Association (AHA) survey, which provides hospital information including expenditures, employment, and technology adoption. Their baseline specification is a linear regression:
\begin{equation}\label{eq:af}
Y_{i,t} = \alpha_i + \gamma_t + X_{i,t}'\eta + \beta\cdot(D_i\cdot \text{Post}_t) + \varepsilon_{i,t},
\end{equation}
where $Y_{i,t}$ is the capital-labor ratio or total number of medical facilities for hospital $i$ in year $t$, $D_i$ is the pre-reform Medicare inpatient share (coded as zero for PPS-exempt hospitals), and $\text{Post}_t$ is a treatment-timing indicator. $X_{i,t}$ represents covariates, and $\alpha_i$ and $\gamma_t$ are hospital and year fixed effects, respectively. \cite{AF2008} argues that $\beta$ captures the causal effect of PPS reform on capital-labor ratios and technology adoption, relying on a parallel trends assumption: in the absence of the PPS reform, hospitals with different shares $D_i$ should have experienced similar changes in outcomes over time.

Recent work by \cite{CGS2024} examines the same empirical setting in detail and finds suggestive evidence that the parallel trends assumption may be too strong. This underscores the importance of incorporating covariates to improve the plausibility of the identifying assumption. By conditioning on covariates, our approach refines the parallel trends assumption, ensuring that hospitals are compared based on more similar characteristics. In this way, our analysis complements \cite{CGS2024}, offering an alternative perspective on the effects of the PPS reform.

\subsection{Setup as a continuous DiD}
Regression \eqref{eq:af} resembles a Two-Way Fixed Effects (TWFE) design but differs in that $D_i$ is continuous. As shown by \cite{CGS2024}, with continuous treatment, the coefficient $\beta$ in \eqref{eq:af} can be viewed as a weighted average of $ATT(d)$ with possible negative weights, which complicates interpretation.\footnote{See Proposition 10 in \cite{CGS2024}. They do not incorporate covariates, but the issue persists.} Our continuous DiD framework addresses this by reframing \cite{AF2008}’s design as follows:
\setcounter{bean}{0}
\begin{list}
{(\alph{beana})}{\usecounter{beana}}
  \item No Treatment Pre-PPS: Before the PPS reform, no hospital was treated.
  \item Control Group: Hospitals with $D_i = 0$ (no Medicare patients or PPS-exempt) are the control.
  \item Treatment Group: Hospitals with positive Medicare shares (treatment intensities) $D_i > 0$.
  \item Outcomes: $Y$ includes the capital-labor ratio or measures of technological adoption.
  \item Covariates: $X$ includes number of beds, metro status, private status, number of medical staff, and state dummies.\footnote{We exclude some additional characteristics in \cite{AF2008}—e.g., general, short-term, or federal status—to avoid conditioning on PPS exemption criteria.} For the capital-labor ratio, we also add binary indicators of specialized capital equipments (CT, MRI, etc.).
  \item \textit{Conditional} parallel trends:
  \[
  E[Y_{t}(0) - Y_{t-1}(0)|X,D=d] = E[Y_{t}(0) - Y_{t-1}(0)|X,D=0],
  \]
  i.e., absent the PPS reform, hospitals with share $D=d$ would have experienced similar changes over time as hospitals with no Medicare inpatients or PPS-exempt ($D = 0$), conditional on hospital-specific covariates $X$ determined before the PPS reform.
\end{list}
We identify the causal effect at intensity $d$ as:
\[
ATT(d) = E[Y_t(d) - Y_t(0)|D=d].
\]
Unlike the constant $\beta$ in (\ref{eq:af}), the causal effect curve $ATT(d)$ can be used to study the policy impact at a much more granular level. For example, if the PPS reform raised the capital-labor ratio, $ATT(d)$ should be positive for all $d>0$. Moreover, $ATT(d)$ should increase in $d$ if the impact of PPS reform is larger for hospitals with higher shares of Medicare inpatients. We apply our panel estimator and, for comparability with \cite{AF2008}, average pre-treatment outcomes ($Y_{t-1}$) over 1980–1983 and post-treatment outcomes ($Y_t$) over 1984–1986 (capital-labor ratio) or 1984–1985 (technology adoption). Our data source is the cleaned data file from \cite{AF2008}.

\subsection{Results}
First, we examine the results for capital-labor ratio. All estimated ATTs are positive, consistent with the findings in \cite{AF2008} and suggesting that the PPS reform led to an increase in the capital-labor ratio. For comparison, \cite{AF2008} reports an estimate $\hat{\beta} = 1.13$, corresponding to an treatment effect after multiplying by the treatment intensity $d$. Moreover, our estimates vary across treatment intensities and do not exhibit a strictly increasing trend, contradicting the theoretical prediction that hospitals with higher Medicare shares would see greater increases in the capital-labor ratio. At low and high treatment intensities, we note that the small effective sample sizes lead to noisier estimates, as reflected in the wider confidence intervals. For completeness, an effect curve estimated without covariates using a kernel method shows a similar pattern to our DML estimates.

Next, we present evidence of increased technological adoption following the PPS reform. Specifically, we consider the total number of specialized medical facilities per hospital as a proxy for technological adoption. All of our estimated ATTs for this outcome are positive, aligning with \cite{AF2008}’s prediction that the PPS reform would incentivize technological adoption. The estimated treatment curve initially rises with treatment intensity but then declines at higher intensities, again diverging from the theoretical prediction that hospitals with larger Medicare shares would invest more. For comparison, an effect curve estimated using a kernel method without covariates again shows a similar pattern to our DML estimates. As with the capital-labor ratio, estimates are especially noisy where data are sparse, an issue amplified by our undersmoothing bandwidth.

\begin{figure}[hbtp!]
    \centering
    \includegraphics[width=\textwidth, height=0.65\textheight]{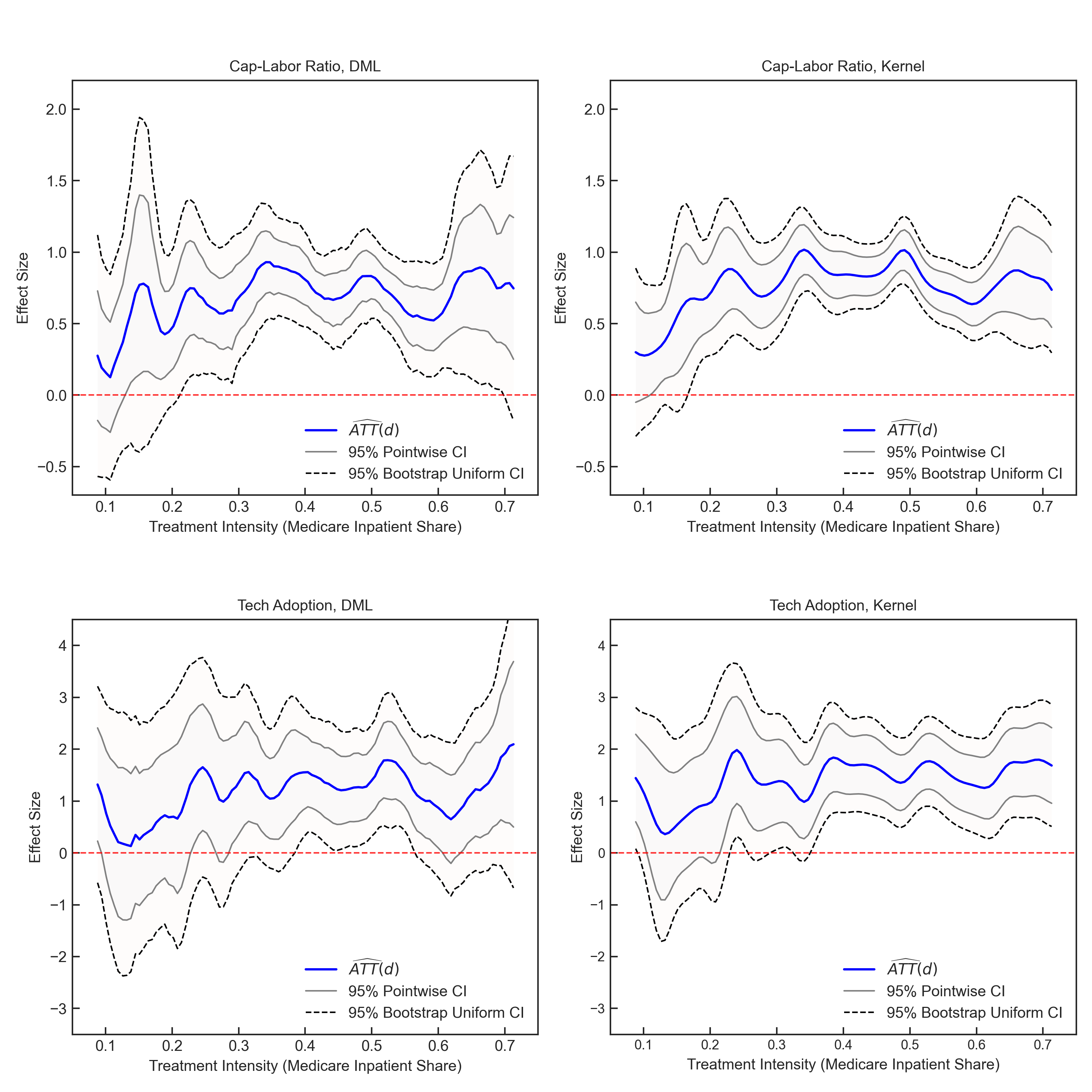}
    \vspace{-10pt}
    \caption{$\widehat{ATT}(d)$, panel data}
    \label{fig:cdid_avg_panel}
\end{figure}

\begin{remark}
\textnormal{
We apply 5-fold cross-fitting, shuffling the data before sample splitting to prevent over-representation in subsamples. A second-order Gaussian kernel with an undersmoothing bandwidth $h = 1.06\times\hat{\sigma}_{\tilde{D}}N^{-1/4}$ is used to estimate both the density $f_D(d)$ and the conditional mean $E[K_h(D-d)|X]$ (see \cite{S2018}). The infinite-dimensional nuisance parameters are estimated using the Random Forest (RF) from the Python scikit-learn package, with 200 trees of maximum depth 20 and fixed minimum leaf size 5. The RF is chosen for its flexibility to accommodate both continuous and discrete covariates, though other ML methods, such as deep neural networks, can be similarly employed. The standard errors are obtained from the cross-fitted estimator defined in (\ref{asvar}) and used to construct the 95-percent pointwise confidence intervals and the bootstrap uniform confidence bands. For the bootstrap CIs, we use Gaussian multipliers $\{\xi_i\}_{i=1}^N$ drawn from a normal distribution with $E[\xi_i]= Var[\xi_i] = 1$, with $B=1000$ repetitions.
}
\end{remark}

\section{Conclusion}
This paper studies difference-in-differences models with continuous treatments. Our identification results are based on a conditional parallel trends assumption, allowing researchers to account for covariates nonparametrically. Under the double/debiased machine learning framework, we develop nonparametric estimators for the average treatment effect on the treated at each continuous treatment intensity and establish their asymptotic properties. Monte Carlo simulations demonstrate that our estimators perform well despite the highly non-linear relationship between the continuous treatment and the high-dimensional covariates. To demonstrate the empirical relevance of our methodology, we re-examine the research questions posed in \cite{AF2008} by applying our estimators to their dataset and obtaining new empirical insights. The extension of difference-in-differences models to the continuous treatment setting has important implications for empirical research. Our methods provide researchers with new tools for examining the impacts of continuous treatment variables.

% \section*{Acknowledgements}
% The author would like to express his gratitude to Denis Chetverikov, Andres Santos, and Rosa Matzkin for their generous time and extremely helpful discussions, which have led to substantial improvements to this paper. Additionally, he extends his thanks to Jinyong Hahn, Zhipeng Liao, Shuyang Sheng, and participants at the UCLA econometrics proseminars for their valuable comments and suggestions. Furthermore, the author is grateful to Kathleen McGarry, Daron Acemoglu, Amy Finkelstein, and the National Bureau of Economic Research for facilitating access to the data source in \cite{AF2008}.

\bibliographystyle{chicago}

\newpage

\section*{Appendix A: Algorithms for Constructing DML Estimators}
\renewcommand{\theequation}{A.\arabic{equation}}
\renewcommand{\thesection}{A}
\setcounter{equation}{0}
\begin{algorithm}[CDID Estimator, Panel]\label{cdidalg}
Let $\{I_k\}_{k=1}^K$ denote a partition of the random sample $\{(Y_{i,t-1}, Y_{i,t},D_i, X_i\}_{i=1}^N$, each with equal size $n=N/K$, and for each $k\in\{1,\cdots,K\}$, let $I_k^c:= I_N\setminus I_k$ denote the complement.
 \setcounter{bean}{0}
       \begin{center}
        \begin{list}
         {\MakeUppercase{Step} \arabic{bean}.}{\usecounter{bean}}
  \item For each $k$, construct
  \begin{align}
  \widehat{ATT}_k(d) &:= \frac{1}{n}\sum_{i\in I_k} \frac{K_h(D_i-d)\hat{g}_k(X_i) - \mathbf{1}\{D_i=0\}\hat{f}_{h,k}(d|X_i)}{\hat{f}_k(d)\hat{g}_k(X_i)}\big(\Delta Y_i - \hat{\mathcal{E}}_{\Delta Y,k}(X_i)\big)
  \end{align}
  where $\hat{f}_k(d), \hat{f}_{h,k}, \hat{g}_k, \hat{\mathcal{E}}_{\Delta Y,k}$ are the estimators of $f_d, f_h(d|X), g(X)$ and $\mathcal{E}_{\Delta Y}(X)$ respectively using the rest of the sample $I_k^c$. In particular, $\hat{f}_k(d)$ is a kernel density estimator, and $\hat{f}_{h,k}$, $\hat{g}_k$ and $\hat{\mathcal{E}}_{\Delta Y,k}$ are estimated using ML methods (e.g. random forests or deep neural networks).
  \item[]
  \item Average through the $K$ estimators to obtain the final estimator
  \begin{align}
  \widehat{ATT}(d) := \frac{1}{K}\sum_{k=1}^K \widehat{ATT}_k(d).
  \end{align}
\end{list}
\end{center}
\end{algorithm}

\begin{algorithm}[CDID Estimator, Repeated Cross-Sections]\label{cdidalg2}
  Let $\{I_k\}_{k=1}^K$ denote a partition of the random sample $\{( Y_i,T_i,D_i, X_i)\}_{i=1}^N$, each with equal size $n=N/K$, and for each $k\in\{1,\cdots,K\}$, let $I_k^c:= I_N\setminus I_k$ denote the complement.
 \setcounter{bean}{0}
       \begin{center}
        \begin{list}
         {\MakeUppercase{Step} \arabic{bean}.}{\usecounter{bean}}
  \item For each $k$, construct
    \begin{align}
    \widehat{ATT}_k(d) := \frac{1}{n}\sum_{i\in I_k} &\frac{K_h(D_i-d)\hat{g}_k(X_i) - \mathbf{1}\{D_i=0\}\hat{f}_{h,k}(d|X_i)}{\hat{f}_k(d)\hat{g}_k(X_i)}\notag\\
    &\times\bigg(\frac{T_i-\hat{\lambda}_k}{\hat{\lambda}_k(1-\hat{\lambda}_k)}Y_i - \hat{\mathcal{E}}_{\lambda Y,k}(X_i)\bigg)
    \end{align}
    where $\hat{f}_k(d), \hat{f}_{h,k}, \hat{g}_k, \hat{\mathcal{E}}_{\lambda Y,k}$ are the estimators of $f_d, f_h(d|X), g(X)$ and $\mathcal{E}_{\lambda Y}(X)$ respectively using the rest of the sample $I_k^c$. In particular, $\hat{f}_k(d)$ is a kernel density estimator, and $\hat{f}_{h,k}$, $\hat{g}_k$ and $\hat{\mathcal{E}}_{\lambda Y,k}$ are estimated using ML methods (e.g. random forests or deep neural networks).
    \item Average through the $K$ estimators to obtain the final estimator
    \begin{align}
    \widehat{ATT}(d) := \frac{1}{K}\sum_{k=1}^K \widehat{ATT}_k(d).
    \end{align}
\end{list}
\end{center}
\end{algorithm}

\section*{Appendix B: Asymptotic Theories for Repeated Cross-Sections}
\renewcommand{\theequation}{B.\arabic{equation}}
\renewcommand{\thesection}{B}
\setcounter{equation}{0}
In this section, we present the asymptotic theory for the repeated cross-sectional setting, following a structure analogous to the panel case discussed in detail in the main text. For notation simplicity, let $\mathcal{D} = (d_L, d_H)$ denote the interior of the support of $D>0$. Let $\mathcal{X}$ and $\mathcal{Y}^\lambda$ denote the support of $X$ and $Y^\lambda$ respectively. We begin by introducing regularity conditions that allow us to establish the asymptotic normality of our estimator and the consistency of the corresponding variance estimator.

\begin{assumption}[Bounds and Smoothness, Repeated Cross-Sections]\label{csdidas1}\hspace{-4.6pt} (a) Th-ere exist constants $c>0$ and $0<C<\infty$ such that $\inf_{d\in\mathcal{D}} f_D^0(d)>c$, $|Y| < C$, $c<f_h^0(d|X)<C$ $\forall d\in\mathcal{D}$, and $|\mathcal{E}_{\lambda Y}^0(X)|<C$ almost surely; (b) $f_D^0(d) \in C^2(\mathcal{D})$ and $\sup_{d\in\mathcal{D}}|\partial_d^2 f_D^{0}(d)| < \infty$; (c) $f_{D|X}^0(d|x) \in C^2(\mathcal{D})$ $\forall x\in \mathcal{X}$ and $\sup_{d,x \in \mathcal{D},\mathcal{X}} |\partial^2_d f_{D|X}^0(d|x)| < \infty$; (d) $f_{Y^\lambda, D}^0(t,d) \in C^2(\mathcal D)$ $\forall t\in\mathcal Y^\lambda$ and $\sup_{t,d \in \mathcal{Y}^\lambda,\mathcal{D}} |\partial_d^2 f_{Y^\lambda, D}^0(t, d)| < \infty$.
\end{assumption}

\begin{assumption}[Rates, Repeated Cross-Sections]\label{csdidas2}\sloppy (a) The kernel bandwidth $h = h_N\to 0$ satisfies $Nh\to\infty$ and $\sqrt{Nh^5} = o(1)$; (b) there exists a sequence $\varepsilon_N\to 0$ such that $h^{-1}\varepsilon_N^2 = o(1)$; (c) with probability tending to $1$, $\|\hat{f}_h(d|X) - f_h^0(d|X)\|_{P,2}\leq h^{-1/2}\varepsilon_N$, $\|\hat{g}(X) - g_0(X)\|_{P,2}\leq \varepsilon_N$, $\|\hat{\mathcal{E}}_{\lambda Y}(X) - \mathcal{E}_{\lambda Y}^0(X)\|_{P,2}\leq \varepsilon_N$; (d) with probability tending to $1$, $\kappa<\hat{g}(X) <1-\kappa$, $c <\hat{f}_h(d|X)<C$ almost surely, and $\|\hat{\mathcal{E}}_{\lambda Y}(X)\|_{P,\infty}<C$.
\end{assumption}

Next, we state the lemma that characterizes the bias introduced by the kernel smoothing of $ATT(d)$. As in the panel case, this bias is asymptotically negligible with undersmoothing kernel bandwidth.
\begin{lemma}[Bias of $ATT_h(d)$, Repeated Cross-Sections]\label{lm:bias_rcs}\sloppy Suppose Assumptions \ref{cdidas1}, \ref{csdidas1}, and \ref{csdidas2} hold. Then $B_h(d) := ATT(d) - ATT_h(d)$ satisfies $B_h(d) = O(h^2)$ for any $d\in\mathcal{D}$.
\end{lemma}

We now state the asymptotic normality result for $\widehat{ATT}(d)$ in the repeated cross-sections.

\begin{theorem}[Asymptotic Normality, Repeated Cross-Sections]\label{thm:asymp_rcs}\hspace{-4.3pt}Suppose Assumptions \ref{didas2}, \ref{didas3}, \ref{didas5}, \ref{didas4}, \ref{cdidas1}, \ref{csdidas1}, and \ref{csdidas2} hold. Then, for $d\in\mathcal{D}$, if $\varepsilon_N = o(N^{-1/4})$ and $\liminf_{N\to\infty}h\sigma_N^2(d)>0$,
\[
\frac{\widehat{ATT}(d) - ATT(d)}{\sigma_{N}(d)/\sqrt{N}}\quad \to^d\quad N(0,1)
\]
where
\begin{align}\label{var2}
&\sigma_{N}^2(d)\notag \\&:= E\bigg[\bigg(\psi_h^{(2)}(Z,\theta_{0h}(d),\lambda_0, f_D^0(d),\eta_0(d)) - \frac{\theta_{0h}(d)}{f_D^0(d)}\big(K_h(D-d)-E[K_h(D-d)]\big)\bigg)^2\bigg]
\end{align}
for $\theta_{0h}(d):= ATT_h(d)$ defined as in (\ref{eq:atth2}) and $\psi_h^{(2)}$ defined as in (\ref{psi2}).
\end{theorem}

Next, we construct a cross-fitted variance estimator based on the theorem above, which is followed by a theorem that establishes its consistency.
\begin{equation}\label{asvar2}
  \hat{\sigma}_{N}^2(d) := \frac{1}{K}\sum_{k=1}^K E_{n,k}\bigg[
  \bigg(
  \psi_h^{(2)}(Z,\hat{\theta}_h(d), \hat{\lambda}_k, \hat{f}_k(d),\hat{\eta}_k(d)) - \frac{\hat{\theta}_h(d)}{\hat{f}_k(d)}\big(K_h(D-d)-\hat{f}_k(d)\big)
  \bigg)^2
  \bigg].
\end{equation}

\begin{theorem}[Variance Estimator Consistency, Repeated Cross-Sections]\label{thm:asvar_rcs}
If the conditions in Theorem \ref{thm:asymp_rcs} hold and assume $h^{-2}\varepsilon_N^2 + h^{-3}N^{-1}= o(1)$, then
\begin{align*}
 \hat{\sigma}_{N}^2(d) = \sigma_{N}^2(d) + o_p(1)
\end{align*}
where $\hat{\sigma}_{N}^2(d)$ is defined as in (\ref{asvar2}) and $\sigma_{N}^2(d)$ is defined as in (\ref{var2}).
\end{theorem}

Let $\{\xi_i\}_{i=1}^N$ be an i.i.d. sequence of random variables that satisfies Assumption \ref{as:subexp}. Then for each $b=1,\cdots, B$, we independently draw such a sequence $\{\xi_i\}_{i=1}^{N}$ and construct estimates based on the following expression. For the repeated cross-sections, define
\begin{align}\label{bootstrap2}
  \widehat{ATT}(d)_{b}^* := \frac{1}{N}\sum_{k=1}^K\sum_{i\in I_k} \xi_i& \frac{K_h(D_i-d)\hat{g}_k(X_i) - \mathbf{1}\{D_i=0\}\hat{f}_{h,k}(d|X_i)}{\hat{f}_k^*(d)\hat{g}_k(X_i)}\notag\\
  \times&\bigg(\frac{T_i-\hat{\lambda}_k}{\hat{\lambda}_k(1-\hat{\lambda}_k)}Y_i - \hat{\mathcal{E}}_{\lambda Y,k}(X_i)\bigg).
\end{align}
Let $\hat{c}_{\alpha}$ denote the $\alpha$-th quantile of $\{\widehat{ATT}(d)_{b}^*- \widehat{ATT}(d)\}_{b=1}^B$. Then a $1-\alpha$ confidence interval can be constructed as $[\widehat{ATT}(d) - \hat{c}_{1-\alpha/2}, \widehat{ATT}(d) -\hat{c}_{\alpha/2}]$.

For the uniform results, let $\mathcal D$ be a fixed compact subset of $(d_L,d_H)$. The following assumption strengthens Assumption \ref{csdidas2} and is used to establish the next theorem, which forms the foundation for constructing valid uniform confidence bands in the repeated cross-sectional setting using the proposed multiplier bootstrap. See Section 4 for the construction of the the uniform confidence bands.

\begin{assumption}[Uniform Inference Rates, Repeated Cross-Sections]\label{csdidas3}\sloppy 
(a) The kernel bandwidth $h = h_N\to 0$ satisfies $Nh/\log^4(N)\to\infty$ and $\sqrt{Nh^5} = o(1)$; (b) there exists a sequence $\varepsilon_N\to 0$ such that $h^{-1}\varepsilon_N^2 = o(1)$; (c) with probability tending to $1$, $\sup_{d\in\mathcal{D}}\|\hat{f}_h(d|X) - f_h^0(d|X)\|_{P,2}\leq h^{-1/2}\varepsilon_N$, $\|\hat{g}(X) - g_0(X)\|_{P,2}\leq \varepsilon_N$, $\|\hat{\mathcal{E}}_{\lambda Y}(X) - \mathcal{E}_{\lambda Y}^0(X)\|_{P,2}\leq \varepsilon_N$; (d) with probability tending to $1$, $\kappa<\hat{g}(X)<1-\kappa$ and $c <\hat{f}_h(d|X)<C$ almost surely $\forall d\in\mathcal{D}$, $\sup_{d\in \mathcal{D}}|\hat{f}_D^{(1)}(d)|<C$, $\sup_{d\in\mathcal{D}}\|\partial_d \hat{f}_h(d|X)\|_{P,\infty}< C$, and $\|\hat{\mathcal{E}}_{\lambda Y}(X)\|_{P,\infty}<C$.
\end{assumption}

\begin{theorem}[Uniform Linear Expansion, Repeated Cross-Sections]\label{thm:unifexp_rcs}\sloppy Suppose Assumptions \ref{didas2}, \ref{didas3}, \ref{didas5}, \ref{didas4}, \ref{cdidas1}, \ref{csdidas1}, \ref{as:subexp}, and \ref{csdidas3} hold. Then for $\varepsilon_N = o(N^{-1/4})$
\begin{align}
&\widehat{ATT}(d)^* - \widehat{ATT}(d) \notag\\ &= \frac{1}{N}\sum_{i=1}^N\dot{\xi}_i\Bigg[\psi_h^{(2)}(Z_i,\theta_{0h}(d),\lambda_0, f_D^0(d),\eta_0(d)) - \frac{\theta_{0h}(d)}{f_D^0(d)}\big(K_h(D_i-d)-E[K_h(D-d)]\big)\Bigg]\notag\\ &+ R^{(2)}(d)
\end{align}
where $\dot{\xi}_i := \xi_i - 1$ and $\sup_{d\in\mathcal{D}} |R^{(2)}(d)| = o_p( (Nh)^{-1/2})$.
\end{theorem}

\section*{Appendix C: Proofs}
\renewcommand{\theequation}{C.\arabic{equation}}
\renewcommand{\thesection}{C}
\setcounter{equation}{0}

\textbf{Proof of Theorem~\ref{thm:cdid1}:}
    By definition, $ATT(d) = E[Y_t(d)-Y_t(0)|D=d]$. First, 
\begin{align}
E[Y_t - Y_{t-1}|D = d] = E[Y_t(d) - Y_{t-1}(0)|D= d]
\end{align}
by the fact that $Y_t = Y_t(D)$ and $Y_{t-1} = Y_{t-1}(0)$.

Second, 
\begin{align}
&E\bigg[(Y_{t} - Y_{t-1})\mathbf{1}\{D=0\}\frac{f_{D|X}(d)}{f_D(d)P(D=0|X)}\bigg]\notag \\
&= E\bigg[(Y_{t} - Y_{t-1})\frac{f_{D|X}(d)}{f_D(d)P(D=0|X)}|D=0\bigg]P(D=0)\notag \\
&= \int E[(Y_{t}(0) - Y_{t-1}(0))|X=x,D=0]\frac{f_{D|X}(d|x)P(D=0)}{f_D(d)P(D=0|X=x)} f_{X|D=0}(x) dx\notag \\
&= \int E[(Y_{t}(0) - Y_{t-1}(0))|X=x, D=d]\notag \\&\times \frac{f_{D|X=x}(d)P(D=0)}{f_D(d)P(D=0|X=x)}\frac{P(D=0|X=x)f_X(x)}{P(D=0)}dx\notag \\
&= \int E[(Y_{t}(0) - Y_{t-1}(0))|X=x,D = d] f_{X|D=d}(x)dx\notag \\
&= E[(Y_{t}(0) - Y_{t-1}(0))|D = d]
\end{align}
where the first equality holds by the law of total probability, the second equality holds by the law of iterated expectation, the third equality holds by that $Y_t = Y_t(D)$ and $Y_{t-1} = Y_{t-1}(0)$, the fourth equality holds by Bayes' rule and conditional parallel trend, and the fifth equality holds by the law of iterated expectation.

Then combining the above results, we have
\begin{align}
&E[Y_t - Y_{t-1}|D = d] - E\bigg[(Y_{t} - Y_{t-1})\mathbf{1}\{D=0\}\frac{f_{D|X}(d)}{f_D(d)P(D=0|X)}\bigg]\notag \\
&= E[Y_{t}(d) - Y_{t-1}(0)|D = d] - E[Y_{t}(0) - Y_{t-1}(0)|D = d]\notag \\
&= E[Y_t(d)-Y_t(0)|D=d]\notag \\
&= ATT(d)
\end{align}

Next, for repeated cross-sections, we have
\begin{align}
&E\bigg[\frac{T-\lambda}{\lambda(1-\lambda)}Y|D=d\bigg]\notag \\
&= E\bigg[E\bigg[\frac{T-\lambda}{\lambda(1-\lambda)}Y|D=d,T\bigg]|D=d\bigg]\notag \\
&= E\bigg[\frac{T-\lambda}{\lambda(1-\lambda)}Y|D = d,T=1\bigg]P(T=1|D=d)\notag \\ &+ E\bigg[\frac{T-\lambda}{\lambda(1-\lambda)}Y|D = d,T=0\bigg]P(T=0|D= d)\notag \\
&= E\bigg[\frac{1-\lambda}{\lambda(1-\lambda)}Y|D=d,T=1\bigg]\lambda+ E\bigg[\frac{0-\lambda}{\lambda(1-\lambda)}Y|D = d,T=0\bigg](1-\lambda)\notag \\
&= E[Y_t|D=d] - E[Y_{t-1}|D=d]\notag \\
&= E[Y_t- Y_{t-1}|D=d]
\end{align}
where the first equality holds by law of iterated expectation, the second equality holds by definition, and the last two equalities hold by Assumption \ref{didas2}. 

Similarly, by law of iterated expectation and Assumption \ref{didas2}
\begin{align}
&E\bigg[\frac{T-\lambda}{\lambda(1-\lambda)}Y\mathbf{1}\{D=0\}\frac{f_{D|X}(d)}{f_D(d)P(D=0|X)}\bigg]\notag \\
& = E\bigg[\frac{1-\lambda}{\lambda(1-\lambda)}Y\mathbf{1}\{D=0\}\frac{f_{D|X}(d)}{f_D(d)P(D=0|X)}|T=1\bigg]P(T=1)\notag \\
& + E\bigg[\frac{0-\lambda}{\lambda(1-\lambda)}Y\mathbf{1}\{D=0\}\frac{f_{D|X}(d)}{f_D(d)P(D=0|X)}|T=0\bigg]P(T=0)\notag \\
& = E\bigg[\frac{1-\lambda}{\lambda(1-\lambda)}Y_t\mathbf{1}\{D=0\}\frac{f_{D|X}(d)}{f_D(d)P(D=0|X)}|T=1\bigg]\lambda\notag \\
& + E\bigg[\frac{0-\lambda}{\lambda(1-\lambda)}Y_{t-1}\mathbf{1}\{D=0\}\frac{f_{D|X}(d)}{f_D(d)P(D=0|X)}|T=0\bigg](1-\lambda)\notag \\
& = E\bigg[(Y_{t} - Y_{t-1})\mathbf{1}\{D=0\}\frac{f_{D|X}(d)}{f_D(d)P(D=0|X)}\bigg]
\end{align}
and the claim follows from the panel case.

\hfill$\square$\\

\noindent\textbf{Proof of Lemma~\ref{lm:scores}:}
First, consider the panel case. Define the unadjusted score $\varphi_h$ as
\begin{align}
  \varphi_h := \Delta Y\frac{K_h(D-d)g_0(X) - \mathbf{1}\{D=0\}f_h^0(d|X)}{f_D^0(d)g_0(X)} - ATT_h(d)
\end{align}
where we use the following notation: $\Delta Y = Y_t - Y_{t-1}$, $f_D^0(d):= f_{D}(d)$, $f_h^0(d|X):= E[K_h(D-d)|X]$, $g_0(X):= P(D=0|X)$. We will add an adjustment term to the original score so that the new score satisfies the Neyman orthogonality w.r.t. the infinite-dimensional parameters.  

The two infinite-dimensional nuisance parameters are $f_h^0(d|X)$ and $g_0(X)$, and in particular, they satisfy $f_h^0(d|X) =  E[K_h(D-d)|X]$ and $g_0(X) = E[\mathbf{1}\{D=0\}|X]$. Then the adjustment term $c_h$ takes the form
\begin{align}
c_h:= (K_h(D-d) - f_h^0(d|X))E[\partial_1\varphi_h|X] + (\mathbf{1}\{D=0\} - g_0(X))E[\partial_2\varphi_h|X]
\end{align}
where $\partial_1$ and $\partial_2$ denote the partial derivatives with respect to $f_h^0(d|X)$ and $g_0(X)$ respectively. Then, we have
\begin{align}
c_h &= (\mathbf{1}\{D=0\} -g_0(X))\frac{f_h^0(d|X)}{f_D^0(d)\cdot g_0^2(X)}E[\Delta Y\mathbf{1}\{D=0\}|X]\notag \\%_{:= \mathcal{E}_{\Delta Y}^0(X)}\\
&-(K_h(D-d) - f_h^0(d|X))\frac{1}{f_D^0(d)\cdot g_0(X)}E[\Delta Y\mathbf{1}\{D=0\}|X]\notag \\
&= \frac{[\mathbf{1}\{D=0\}-g_0(X)]f_h^0(d|X) - [K_h(D-d)-f_h^0(d|X)]g_0(X) }{f_D^0(d)\cdot g_0(X)}\underbrace{\frac{E[\Delta Y\mathbf{1}\{D=0\}|X]}{g_0(X)}}_{:= \mathcal{E}_{\Delta Y}^0(X)}\notag \\
&= \frac{\mathbf{1}\{D=0\}f_h^0(d|X)- K_h(D-d)g_0(X)}{f_D^0(d)\cdot g_0(X)}\mathcal{E}_{\Delta Y}^0(X)
\end{align}
where $\mathcal{E}_{\Delta Y}^0(X) = E[\Delta Y\mathbf{1}\{D=0\}|X]/g_0(X) = E[\Delta Y|D=0,X]$. In particular, note that $\psi_h$ in the lemma satisfies $\psi_h = \varphi_h + c_h$.

Now it remains to show the new score $\psi_h$ satisfies Neyman orthogonality w.r.t. the nuisance parameters, $f_h^0(d|X)$, $g_0(X)$, and $\mathcal{E}_{\Delta Y}^0(X)$. First, we need to check the moment condition $E[\psi_h] = 0$. Since $E[\varphi_h] = 0$, we only need to check $E[c_h] = 0$:
\begin{align}
E[c_h] &= E\bigg[\frac{\mathbf{1}\{D=0\}f_h^0(d|X)- K_h(D-d)g_0(X)}{f_D^0(d)\cdot g_0(X)}\mathcal{E}_{\Delta Y}^0(X)\bigg]\notag \\
&= E\bigg[\frac{E[\mathbf{1}\{D=0\}|X]f_h^0(d|X)- E[K_h(D-d)|X]g_0(X)}{f_D^0(d)\cdot g_0(X)}\mathcal{E}_{\Delta Y}^0(X)\bigg]\notag \\
&= E\bigg[\frac{g_0(X)f_h^0(d|X)- f_h^0(d|X)g_0(X)}{f_D^0(d)\cdot g_0(X)}\mathcal{E}_{\Delta Y}^0(X)\bigg]\notag \\
&= 0
\end{align}
where the second equality holds by the law of iterated expectation and the third equality holds by the fact that $E[K_h(D-d)|X] = f_h^0(d|X)$ and $E[\mathbf{1}\{D=0\}|X] = g_0(X)$.

Second, we need to show the Gateaux derivative of the score w.r.t. the nuisance parameters $\eta_0(d):= (f_h^0(d|X), g_0(X), \mathcal{E}_{\Delta Y}^0(X))$ vanishes at zero, that is, we need to show
\begin{align}
\partial_r E[\psi_h(\eta_0(d) + r(\eta(d) - \eta_0(d)))]|_{r=0} = 0.
\end{align}
We use the notation $\eta(d)$ without the subscript $0$ to denote generic nuisance parameters in the set $T_N(d)$. By the definition of Gateaux derivative, it suffices to show the partial derivative is zero w.r.t. each nuisance parameter separately.

\noindent \underline{w.r.t $f_h(d|X)$}:
\begin{align}
&\partial_rE[\psi_h(f_h^0(d|X) + r(f_h(d|X) - f_h^0(d|X)))]|_{r=0}\notag \\
&= -E\bigg[\frac{\mathbf{1}\{D=0\}\Delta f_h(d|X)}{f_D^0(d)\cdot g_0(X)}(\Delta Y - \mathcal{E}_{\Delta Y}^0(X))\bigg]\notag \\
&= -E\bigg[\frac{E[\mathbf{1}\{D=0\} \Delta Y |X]}{g_0(X)}\frac{\Delta f_h(d|X)}{f_D^0(d)} - \frac{E[\mathbf{1}\{D=0\}|X]}{g_0(X)}\frac{\Delta f_h(d|X)}{f_D^0(d)}\mathcal{E}_{\Delta Y}^0(X)\bigg]\notag \\
&= -E\bigg[\mathcal{E}_{\Delta Y}^0(X) \frac{\Delta f_h(d|X)}{f_D^0(d)} - \frac{g_0(X)}{g_0(X)}\frac{\Delta f_h(d|X)}{f_D^0(d)}\mathcal{E}_{\Delta Y}^0(X)\bigg]\notag \\
&= 0
\end{align}
where the first equality holds by definition with $\Delta f_h(d|X) := f_h(d|X) - f_h^0(d|X)$, the second equality holds by the law of iterated expectation, and the third equality holds by the fact that $E[\Delta Y\mathbf{1}\{D=0\}|X]/g_0(X) = \mathcal{E}_{\Delta Y}^0(X)$ and $E[\mathbf{1}\{D=0\}|X] = g_0(X)$.\\

\noindent \underline{w.r.t $g(X)$}:
\begin{align}
&\partial_rE[\psi_h(g_0(X) + r(g(X) - g_0(X)))]|_{r=0}\notag \\
&= E\bigg[\frac{\mathbf{1}\{D=0\}f_h^0(d|X)}{f_D^0(d)\cdot g_0^2(X)}(\Delta Y - \mathcal{E}_{\Delta Y}^0(X))\Delta g(X)\bigg]\notag \\
&= E\bigg[\Delta g(X)\frac{E[\mathbf{1}\{D=0\} \Delta Y |X]}{g_0(X)}\frac{f_h^0(d|X)}{f_D^0(d)g_0(X)}\notag\\ 
&- \Delta g(X) \frac{E[\mathbf{1}\{D=0\}|X]}{g_0(X)^2}\frac{f_h^0(d|X)}{f_D^0(d)}\mathcal{E}_{\Delta Y}^0(X)\bigg]\notag \\
&= E\bigg[\Delta g(X)\mathcal{E}_{\Delta Y}^0(X)\frac{f_h^0(d|X)}{f_D^0(d)g_0(X)} - \Delta g(X) \frac{g_0(X)}{g_0(X)^2}\frac{f_h^0(d|X)}{f_D^0(d)}\mathcal{E}_{\Delta Y}^0(X)\bigg]\notag \\
&= 0
\end{align}
where the first equality holds by chain rule and the definition $\Delta g(X) := g(X) - g_0(X)$, second equality holds by law of iterated expectation, and the third equality holds by that $E[\Delta Y\mathbf{1}\{D=0\}|X]/g_0(X) = \mathcal{E}_{\Delta Y}^0(X)$ and $E[\mathbf{1}\{D=0\}|X] = g_0(X)$.\\

\noindent \underline{w.r.t $\mathcal{E}_{\Delta Y}(X)$}:
\begin{align}
&\partial_rE[\psi_h(\mathcal{E}_{\Delta Y}^0(X) + r(\mathcal{E}_{\Delta Y}(X) - \mathcal{E}_{\Delta Y}^0(X)))]|_{r=0}\notag \\
&= -E[\frac{K_h(D-d)g_0(X) - \mathbf{1}\{D=0\}f_h^0(d|X)}{f_D^0(d)\cdot g_0(X)}\Delta\mathcal{E}(X)]\notag \\
&= -E[\frac{E[K_h(D-d)|X]g_0(X) - E[\mathbf{1}\{D=0\}|X]f_h^0(d|X)}{f_D^0(d)\cdot g_0(X)}\Delta\mathcal{E}(X)]\notag \\
&= 0
\end{align}
where the first line holds by definition with $\Delta\mathcal{E}(X) = \mathcal{E}_{\Delta Y}(X) - \mathcal{E}_{\Delta Y}^0(X)$, the second equality holds by law of iterated expectation, and the last equality holds by the definition that $E[K_h(D-d)|X] = f_h^0(d|X)$ and $E[\mathbf{1}\{D=0\}|X] = g_0(X)$.

This shows that the score $\psi_h$ is Neyman orthogonal w.r.t. the infinite-dimensional nuisance parameters. The proof for the repeated cross-sectional case follows the same argument by replacing $\Delta Y$ with $\frac{T-\lambda}{\lambda(1-\lambda)}Y$. \hfill$\square$\\

\noindent\textbf{Proof of Lemma~\ref{lm:scores} (Influence Function Approach):}
We introduce the method discussed in \cite{HDDV22} and adapt it to the panel case. Here we focus on the high-dimensional nuisance parameters and hold $f_D(d)$ fixed. For repeated cross-sections, the analogous score construction holds $f_D(d)$ and $\lambda$ fixed.

Let $\mathcal{P}$ denote the true distribution of the observed data and $\Psi(\mathcal{P})$ denote the target parameter. Let $\tilde{\mathcal{P}}$ denote another fixed distribution, and define a mixture
\begin{align}
    \mathcal{P}_t = t \tilde{\mathcal{P}} + (1-t)\mathcal{P}.
\end{align}
Note that for a distribution function $f$ at $\tilde{o}$, we have 
\begin{align}
    f_t(o) = t\delta_{\tilde{o}}(o) + (1-t)f(o) \implies \frac{df_t(o)}{dt}\bigg|_{t=0} = \delta_{\tilde{o}}(o) - f(o),
\end{align}
where $\delta_{\tilde{o}}$ denote the dirac delta at $\tilde{o}$.

The sensitivity of $\Psi$ to changes in $\mathcal{P}$ in the direction of $\tilde{\mathcal{P}}$ can be characterized by the Gateaux derivative (if exists)
\begin{align}
    \lim_{t \downarrow0} \frac{\Psi(\mathcal{P}_t) - \Psi(\mathcal{P})}{t} = \frac{d\Psi(\mathcal{P}_t)}{dt}\bigg|_{t=0} = \int \psi(o,\mathcal{P}) \Big(d\mathcal{\tilde{P}}(o) - d\mathcal{P}(o)\Big)
\end{align}
where the second equality holds by Riesz representation theorem. The canonical gradient $\psi$ is referred to as the efficient influence function. The efficient influence function has mean zero, which implies that
\begin{align*}
    \frac{d\Psi(\mathcal{P}_t)}{dt}\bigg|_{t=0} = \int \psi(o,\mathcal{P}) d\mathcal{\tilde{P}}(o) = E_{\mathcal{\tilde{P}}}[\psi(O,\mathcal{P})].
\end{align*}
\cite{HDDV22} considers perturbing $\Psi(\mathcal{P}_t)$ at a single point, and the above expression gives the efficient influence function directly as
\begin{align}
    \frac{d\Psi(\mathcal{P}_t)}{dt}\bigg|_{t=0} =  \psi(o,\mathcal{P}).
\end{align}

Fix $c=f_D^0(d)>0$ (remain fixed under perturbations of the law) and define $\Psi(\mathcal P)=\nu_h(\mathcal P)/c$, where $\nu_h(\mathcal P)$ denotes the numerator in the expression of $ATT_h(d)$, i.e., $\Psi(\mathcal P)=ATT_h(d)$. Then,
\begin{align}
    \Psi(\mathcal{P}) &= E\bigg[\Delta Y\frac{K_h(D-d)P(D=0|X) - \mathbf{1}\{D=0\}E[K_h(D-d)|X]}{f_D(d)P(D=0|X)}\bigg]\\
    & = \int \iota \frac{K_h(s-d)}{f_D(d)} f_{\Delta Y, D}(\iota, s) d\iota ds\notag\\
    &- \int \iota \frac{K_h(s-d)}{f_D(d)f_{D,X}(0,x)} f_{D,X}(s,x)f_{D,\Delta Y, X}(0,\iota,x)d\iota dsdx
\end{align}
Perturbing at a point $\tilde{o} = (\Delta Y =\tilde{\iota}, D=\tilde{s}, X = \tilde{x})$, we have (suppressing random variable subscripts), with $\delta_0=\mathbf1\{\tilde s=0\}$
\begin{align}
    &\frac{d\Psi(\mathcal{P}_t)}{dt} \notag\\
    =& \int \iota \frac{K_h(s-d)}{f_d} \frac{d}{dt} f_t(\iota, s)\bigg|_{t=0} d\iota ds\notag\\
    & + \int \iota \frac{K_h(s-d)}{f_d f^2(0,x)}f(s,x)f(0,\iota,x)\frac{d}{dt} f_t(0, x)\bigg|_{t=0}\notag\\
    & - \iota \frac{K_h(s-d)}{f_d f(0,x)}f(0,\iota,x)\frac{d}{dt} f_t(s, x)\bigg|_{t=0}\notag\\
    & - \iota \frac{K_h(s-d)}{f_d f(0,x)}f(s,x)\frac{d}{dt} f_t(0,\iota,x)\bigg|_{t=0} d\iota ds dx\\
    =& \tilde{\iota} \frac{K_h(\tilde{s}-d)}{f_d}+ \int \iota \frac{K_h(s-d)}{f_d f^2(0,\tilde{x})}f(s,\tilde{x})f(0,\iota,\tilde{x})\delta_{0} d\iota ds\notag\\
    & - \int \iota \frac{K_h(\tilde{s}-d)}{f_d f(0,\tilde{x})}f(0,\iota,\tilde{x})d\iota - \int \tilde{\iota} \frac{K_h(s-d)}{f_d f(0,\tilde{x})}f(s,\tilde{x})\delta_{0} ds\notag\\
    & - \underbrace{\bigg(\int \iota \frac{K_h(s-d)}{f_d} f(\iota, s) d\iota ds - \int \iota \frac{K_h(s-d)}{f_df(0,x)} f(s,x)f(0,\iota,x)d\iota dsdx\bigg)}_{=\Psi(\mathcal{P})}\\
    = & \tilde{\iota} \frac{K_h(\tilde{s}-d)}{f_d} + \frac{E[\Delta Y|D=0,X = \tilde{x}]}{f_dP(D=0|X = \tilde{x})} E[K_h(D-d)|X= \tilde{x}]\mathbf{1}\{D=0\}\notag\\
    & - \frac{K_h(\tilde{s}-d)}{f_d}E[\Delta Y| D=0, X = \tilde{x}]\notag\\
    & - \frac{\tilde{\iota}}{f_dP(D=0|X=\tilde{x})}E[K_h(D-d)|X = \tilde{x}]\mathbf{1}\{D=0\}\notag\\
    & - \Psi(\mathcal{P})\\
    = & (\tilde{\iota} - E[\Delta Y|D=0, X=\tilde{x}])\frac{K_h(\tilde{s}-d) P(D=0|X=\tilde{x}) - \mathbf{1}\{D=0\}E[K_h(D-d)|X= \tilde{x}]}{f_dP(D=0|X=\tilde{x})}\notag\\
    &- \Psi(\mathcal{P}).
\end{align}
Note that this is the same expression as the score we presented in Lemma~\ref{lm:scores}. \hfill$\square$\\

\noindent\textbf{Proof of~Lemma \ref{lm:bias}:}
We focus on the panel case. The bias $B_h(d)$ is defined as 
\begin{align}
  B_h(d) :=& ATT(d) - ATT_h(d)\notag \\
          =& E[\Delta Y|D=d] - E\bigg[\Delta Y\mathbf{1}\{D=0\}\frac{f_{D|X}(d|X)}{f_D(d)P(D=0|X)}\bigg]\notag \\
          -& E\bigg[\Delta Y \frac{K_h(D-d)P(D=0|X) - \mathbf{1}\{D=0\}E[K_h(D-d)|X]}{f_D(d)P(D=0|X)}\bigg]\notag \\
          =& \bigg(E[\Delta Y|D=d] -  E\bigg[\Delta Y \frac{K_h(D-d)}{f_D(d)}\bigg]\bigg)\notag \\
          -& E\bigg[\Delta Y\mathbf{1}\{D=0\} \frac{f_{D|X}(d|X) - E[K_h(D-d)|X]}{f_D(d)P(D=0|X)}\bigg].
\end{align}

First, note that 
\begin{align}
  &E[\Delta Y|D=d] -  E\bigg[\Delta Y \frac{K_h(D-d)}{f_D(d)}\bigg]\notag \\
  =&E[\Delta Y|D=d] -  E\bigg[\Delta Y \frac{K_h(D-d)}{f_D(d)}\mathbf{1}\{D>0\}\bigg] - E\bigg[\Delta Y \frac{K_h(D-d)}{f_D(d)}\mathbf{1}\{D=0\}\bigg]\notag \\
  =& \int t \frac{f_{\Delta Y, D}(t, d)}{f_D(d)}dt - \int t\frac{1}{f_D(d)} \int\frac{1}{h} K\Big(\frac{s-d}{h}\Big) f_{\Delta Y, D}(t, s) ds dt-\frac{K_h(-d)}{f_D(d)}E[\Delta Y\mathbf1\{D=0\}]\notag \\
  =& \int C_1\frac{t}{f_D(d)}h^2f_{\Delta Y, D}^{(2)}(t, d)dt + o(h^2)-\frac{K_h(-d)}{f_D(d)}E[\Delta Y\mathbf1\{D=0\}]\notag \\
  =& O(h^2)
\end{align}
where the two equalities hold by definition, the third equality holds by change of variables and Taylor expansion (see Lemma 5.1 in \cite{FY2003}), and the last equality holds by assumption. Note that since $f_D(d) > c$ and $\Delta Y$ is bounded, then by the assumption that $\sup_{u\in\mathbb{R}}|u|^3|K(u)| < \infty$, the term $K_h(-d)$ is also of order $O(h^2)$ (which holds uniformly on $(d_L, d_H)$).

Second, by the same argument using the change of variables and Taylor expansion, we have
\begin{align}
E[K_h(D-d)|X=x] &= g(x)K_h(-d)+\int \frac{1}{h}K\Big(\frac{s-d}{h}\Big)f_{D|X}(s|x)ds \notag \\
& = g(x)K_h(-d)+\int K(u)f_{D|X}(d+hu|x)du\notag \\
& = f_{D|X}(d|x) + C_2 h^2f_{D|X}^{(2)}(d|x) + o(h^2)+g(x)K_h(-d).
\end{align}
Then by the uniform boundedness of $f^{(2)}_{D|X}(d|x)$ and assumptions on $\Delta Y, f_D(d), P(D=0|X)$, applying the dominated convergence theorem, we have
\begin{align}
  &E\bigg[\Delta Y\mathbf{1}\{D=0\} \frac{f_{D|X}(d|X) - E[K_h(D-d)|X]}{f_D(d)P(D=0|X)}\bigg]\\
  =& C_3 h^2 E\bigg[\Delta Y\mathbf{1}\{D=0\} \frac{f_{D|X}^{(2)}(d|X)}{f_D(d)P(D=0|X)}\bigg] + o(h^2) -\frac{K_h(-d)}{f_D(d)}E[\Delta Y\mathbf1\{D=0\}]\\
  =& O(h^2).
\end{align}
Combining the two results, we have $B_h(d) = O(h^2)$, which completes the proof. The proof for the repeated cross-sectional case follows the same argument by replacing $\Delta Y$ with $\frac{T-\lambda}{\lambda(1-\lambda)}Y$. \hfill$\square$\\

\noindent\textbf{Proof of Theorem~\ref{thm:asymp}, Panel:}
{\sloppy\emergencystretch=2em Let $T_N(d)$ be the set of $\eta(d):= (f_h(d|X), g(X),\mathcal{E}_{\Delta Y}(X))$ such that $\|f_h(d|X) - f_h^0(d|X)\|_{P,2}\leq h^{-1/2}\varepsilon_N$, $\|g(X) - g_0(X)\|_{P,2}\leq \varepsilon_N$, $\|\mathcal{E}_{\Delta Y}(X) - \mathcal{E}_{\Delta Y}^0(X)\|_{P,2}\leq \varepsilon_N$, $\kappa<g(X)<1-\kappa\ \text{a.s.}$, $c < f_h(d|X)<C$, and $\|\mathcal{E}_{\Delta Y}(X)\|_{P,\infty}<C$. Let $F_N(d)$ be the set of functions $f>c$ such that $|f-f_D^0(d)|\leq M(Nh)^{-1/2}$. Then Assumption \ref{cdidas3} implies that, for every $\epsilon>0$, a sufficiently large fixed $M$ makes $\hat\eta_k(d)\in T_N(d)$ and $\hat f_k(d)\in F_N(d)$ hold jointly with probability at least $1-\epsilon$ for all sufficiently large $N$. The following bounds are on this event. Throughout the proof, we use $N$ to denote the sample size and $n:= N/K$ to denote the size of the subsamples. In particular, since $K$ is fixed, $n\asymp N$.

To simplify notation, let $\theta_0(d)$ denote the true $ATT(d)$, $\theta_{0h}(d)$ denote the true $ATT_h(d)$, and $\hat{\theta}_h(d)$ denote our cross-fitted estimator. In particular, recall that our estimator is
\begin{align}
\hat{\theta}_h(d) &:= \frac{1}{K}\sum_{k=1}^K \frac{1}{n}\sum_{i\in I_k} \frac{K_h(D_i-d)\hat{g}_k(X_i) - \mathbf{1}\{D_i=0\}\hat{f}_{h,k}(d|X_i)}{\hat{f}_k(d)\hat{g}_k(X_i)}\big(\Delta Y_i - \hat{\mathcal{E}}_{\Delta Y,k}(X_i)\big).
\end{align}
Then we have the following decomposition
\begin{align}
\hat{\theta}_h(d) - \theta_0(d) = \underbrace{\hat{\theta}_h(d) - \theta_{0h}(d)}_{(1)} + \underbrace{\theta_{0h}(d)- \theta_0(d)}_{(2)}
\end{align}
where $(1)$ will be our main focus while the bias term $(2)$ is shown in Lemma \ref{lm:bias} to be $O(h^2)$ and asymptotically negligible by the assumption of the under-smoothing bandwidth.

By definition, 
\begin{align}
\sqrt{N}(\hat{\theta}_h(d) - \theta_{0h}(d)) = \sqrt{N}\frac{1}{K}\sum_{k=1}^KE_{n,k}[\psi_h(Z_i,\theta_{0,h},\hat{f}_k(d),\hat{\eta}_k(d))]
\end{align}
where $\psi_h$ is defined as in (\ref{psi1}), and $E_{n,k}(f) = \frac{1}{n}\sum_{i\in I_k} f(Z_i)$ denotes the empirical average of a generic function $f$ over the set $I_k$. Then we have the following decomposition, using Taylor's theorem: %multivariate version of Taylor's theorem,
\begin{align}
\sqrt{N}(\hat{\theta}_h(d) - \theta_{0h}(d))
&= \sqrt{N}\frac{1}{K}\sum_{k=1}^K E_{n,k}[\psi_h(Z,\theta_{0h}(d),f_D^0(d),\hat{\eta}_k(d))] \label{eq:1}\\
&+ \sqrt{N}\frac{1}{K}\sum_{k=1}^K E_{n,k}[\partial_{f}\psi_h(Z,\theta_{0h}(d), f_D^0(d),\hat{\eta}_k(d))](\hat{f}_k(d) - f_D^0(d))\label{eq:3}\\
&+ \sqrt{N} \frac{1}{2K} \sum_{k=1}^K E_{n,k}[\partial_f^2\psi_h(Z,\theta_{0h}(d), \bar{f}_k,\hat{\eta}_k(d))](\hat{f}_k(d) - f_D^0(d))^2\label{eq:5}
\end{align}
where $\bar{f}_k \in (f_D^0(d), \hat{f}_k(d))$. This decomposition provides a roadmap for the remainder of the proof. There are roughly four steps. In the first step, we show the second-order term (\ref{eq:5}) vanishes rapidly and does not contribute to the asymptotic variance. In the second step, we bound the first-order term (\ref{eq:3}), which potentially contributes to the asymptotic variance. In step 3, we expand (\ref{eq:1}) around the nuisance parameter $\hat{\eta}_k(d)$, in which the first-order bias disappears by Neyman orthogonality, and we show the second-order terms have no impact on the asymptotics under our assumptions. In the final step, we verify the results used in the first two steps and conclude.

Before we start the main proof, we state two well-known results that will be used in the proof. For an i.i.d. sample $\{D_i\}_{i=1}^n$, the kernel estimator for the density $f_D(d):= f_D^0(d)$ in our setting is defined as
\begin{align}
  \hat{f}_d := \frac{1}{n}\sum_{i=1}^n K_h(D_i-d).
\end{align}
Then, 
\begin{align}
  \hat{f}_d - f_D^0(d) = \hat{f}_d - E[K_h(D-d)]  - (f_D^0(d) - E[K_h(D-d)]).
\end{align}
One can show that (see for example, \cite{Hardle90} and \cite{FY2003})
\begin{align}
  \hat{f}_d - E[K_h(D-d)] &= O_p((nh)^{-1/2})\\
  f_D^0(d) - E[K_h(D-d)] &= O(h^2).
\end{align}
Therefore, for an under-smoothing $h = o(n^{-1/5})$, we have $\hat{f}_d - f_D^0(d) = O_p((nh)^{-1/2})$ and $(\hat{f}_d - f_D^0(d))^2 = O_p((nh)^{-1})$.\\

\noindent \textit{Step 1: Second Order Terms}

First, we consider (\ref{eq:5}). By triangle inequality, we have
\begin{align}
&|E_{n,k}[\partial_f^2\psi_h(Z,\theta_{0h}(d), \bar{f}_k,\hat{\eta}_k(d))] - E[\partial_f^2\psi_h(Z,\theta_{0h}(d), f_D^0(d),\eta_0(d))]|\notag\\
&\leq \underbrace{|E_{n,k}[\partial_f^2\psi_h(Z,\theta_{0h}(d), \bar{f}_k,\hat{\eta}_k(d))] - E_{n,k}[\partial_f^2\psi_h(Z,\theta_{0h}(d), f_D^0(d),\eta_0(d))]|}_{J_{1k}} \\
&+ \underbrace{|E_{n,k}[\partial_f^2\psi_h(Z,\theta_{0h}(d), f_D^0(d),\eta_0(d))] - E[\partial_f^2\psi_h(Z,\theta_{0h}(d), f_D^0(d),\eta_0(d))]|}_{J_{2k}}.
\end{align}
To bound $J_{2k}$, note that since $f_D^0(d)$ is bounded away from zero, 
\begin{align}
\partial_f^2\psi_h(Z,\theta_{0h}(d), f_D^0(d),\eta_0(d)) = \frac{2}{(f_D^0(d))^2}(\psi_h(Z,\theta_{0h}(d),f_D^0(d),\eta_0(d)) + \theta_{0h}(d))
\end{align}
which implies that 
\begin{align}
E[J_{2k}^2] & = E\bigg[\bigg(\frac{1}{n}\sum_{i\in I_k} \partial_f^2\psi_h(Z_{i},\theta_{0h}(d), f_D^0(d),\eta_0(d)) - E[\partial_f^2\psi_h(Z,\theta_{0h}(d), f_D^0(d),\eta_0(d))] \bigg)^2\bigg] \notag \\
& = E\bigg[\bigg(\frac{1}{n}\sum_{i\in I_k} \partial_f^2\psi_h(Z_{i},\theta_{0h}(d), f_D^0(d),\eta_0(d))\bigg)^2\bigg]\notag\\
&- \big(E[\partial_f^2\psi_h(Z,\theta_{0h}(d), f_D^0(d),\eta_0(d))] \big)^2\notag \\
&\leq \frac{1}{n}E[(\partial_f^2\psi_h(Z,\theta_{0h}(d), f_D^0(d),\eta_0(d)))^2]\notag \\
&\lesssim E[K_h^2(D-d)]/N\notag \\
&\lesssim (hN)^{-1},
\end{align}
where the first inequality holds by Cauchy-Schwarz inequality, Jensen's inequality, and independence; the next bound holds by the boundedness assumption on the components of the score, and the last line holds by the assumption on the kernel function $K$. Then by the Markov's inequality, we have $J_{2k}\leq O_p\big((Nh)^{-1/2}\big)$.

Next, for $J_{1k}$, we have
\begin{align*}
&E[J_{1k}|I_k^c]\notag\\
&= E[|E_{n,k}[\partial_f^2\psi_h(Z,\theta_{0h}(d), \bar{f}_k,\hat{\eta}_k(d))] - E_{n,k}[\partial_f^2\psi_h(Z,\theta_{0h}(d), f_D^0(d),\eta_0(d))]||I_k^c]\\
&\leq \sup_{\eta(d)\in T_N(d)} E[\sup_{f\in F_N(d)}|\partial_f^2\psi_h(Z,\theta_{0h}(d),f,\eta(d))-\partial_f^2\psi_h(Z,\theta_{0h}(d), f_D^0(d),\eta_0(d))||I_k^c]\\
&\leq \sup_{\eta(d)\in T_N(d)} E[\sup_{f\in F_N(d)}|\partial_f^2\psi_h(Z,\theta_{0h}(d),f,\eta(d))-\partial_f^2\psi_h(Z,\theta_{0h}(d), f_D^0(d),\eta_0(d))|]\\
&\lesssim h^{-1/2}\varepsilon_N+(Nh)^{-1/2}, \quad\quad\quad \text{(a)}
\end{align*}
where the first inequality holds by the triangle inequality and the definition of supremum over the sets $F_N(d)$ and $T_N(d)$, and the second inequality holds since the supremum does not depend on the sample $I_k^c$ and the evaluation sample is independent of it. Then by conditional Markov's inequality, $J_{1k}\leq O_p(h^{-1/2}\varepsilon_N+(Nh)^{-1/2})=o_p(1)$. Also, $E[\partial_f^2\psi_h(Z,\theta_{0h}(d),f_D^0(d),\eta_0(d))]=2\theta_{0h}(d)/(f_D^0(d))^2=O(1)$, so the empirical second derivative is $O_p(1)$. Using the previous result that $(\hat{f}_k(d)-f_D^0(d))^2 = O_p((Nh)^{-1})$, we conclude that (\ref{eq:5}) $=o_p(h^{-1/2})$. We will verify (a) at the end of this section. \\

\noindent \textit{Step 2: First-Order Terms}

To bound (\ref{eq:3}), we first use the triangle inequality to obtain the decomposition
\begin{align}
&|E_{n,k}[\partial_f\psi_h(Z,\theta_{0h}(d), f_D^0(d),\hat{\eta}_k(d))] - E[\partial_f\psi_h(Z,\theta_{0h}(d), f_D^0(d),\eta_0(d))]|\notag \\
\leq& \underbrace{|E_{n,k}[\partial_f\psi_h(Z,\theta_{0h}(d), f_D^0(d),\hat{\eta}_k(d))] - E_{n,k}[\partial_f\psi_h(Z,\theta_{0h}(d), f_D^0(d),\eta_0(d))]|}_{J_{3k}} \\
+& \underbrace{|E_{n,k}[\partial_f\psi_h(Z,\theta_{0h}(d), f_D^0(d),\eta_0(d))] - E[\partial_f\psi_h(Z,\theta_{0h}(d), f_D^0(d),\eta_0(d))]|}_{J_{4k}}.
\end{align}
We first bound $J_{4k}$. By definition, we have
\begin{align}
\partial_f\psi_h(Z,\theta_{0h}(d), f_D^0(d),\eta_0(d)) = -\frac{1}{f_D^0(d)}(\psi_h(Z,\theta_{0h}(d), f_D^0(d),\eta_0(d)) + \theta_{0h}(d)).
\end{align}
By the boundedness assumption,
\begin{align}
E[J_{4k}^2] \leq \frac{1}{n} E[(\partial_f\psi_h(Z,\theta_{0h}(d), f_D^0(d),\eta_0(d)))^2]\lesssim (Nh)^{-1}.
\end{align}
Then by Markov's inequality, we have $J_{4k}\leq O_p((Nh)^{-1/2})$. With the assumption that $Nh\to\infty$, we have $J_{4k} = o_p(1)$.

Second, to bound $J_{3k}$, note that
\begin{align*}
&E[J_{3k}^2|I_k^c]\notag\\
&= E[|E_{n,k}[\partial_f\psi_h(Z,\theta_{0h}(d), f_D^0(d),\hat{\eta}_k(d))] - E_{n,k}[\partial_f\psi_h(Z,\theta_{0h}(d), f_D^0(d),\eta_0(d))]|^2|I_k^c]\\
&\leq \sup_{\eta(d)\in T_N(d)} E[|\partial_f\psi_h(Z,\theta_{0h}(d), f_D^0(d),\eta(d))-\partial_f\psi_h(Z,\theta_{0h}(d), f_D^0(d),\eta_0(d))|^2|I_k^c]\\
&\leq \sup_{\eta(d)\in T_N(d)} E[|\partial_f\psi_h(Z,\theta_{0h}(d),f_D^0(d),\eta(d))-\partial_f\psi_h(Z,\theta_{0h}(d), f_D^0(d),\eta_0(d))|^2]\\
&\lesssim h^{-1}\varepsilon_N^2 \quad\quad\quad \text{(b)}
\end{align*}
where the first equation holds by definition, the second line holds by Cauchy-Schwarz, and the third line holds by the construction that all the parameters are estimated using auxiliary sample $I_k^c$. Then we conclude with the conditional Markov's inequality that $J_{3k} = o_p(1)$ provided that $h^{-1}\varepsilon_N^2 = o(1)$, which holds by assumption. We will show (b) at the end of this section. Therefore,
\begin{align}
E_{n,k}[\partial_f\psi_h(Z,\theta_{0h}(d), f_D^0(d),\hat{\eta}_k(d))] = \underbrace{E[\partial_f\psi_h(Z,\theta_{0h}(d), f_D^0(d),\eta_0(d))]}_{:= S_f^0} + o_p(1). 
\end{align}

Note that the kernel density estimator satisfies $(\hat{f}_k(d) - f_D^0(d)) = O_p((Nh)^{-1/2})$, so we can rewrite (\ref{eq:3}) as
\begin{align}
(\ref{eq:3})&= \sqrt{N}\frac{1}{K}\sum_{k=1}^K E_{n,k}[\partial_f\psi_h(Z,\theta_{0h}(d), f_D^0(d),\hat{\eta}_k(d))](\hat{f}_k(d) - f_D^0(d))\notag \\
&= \sqrt{N}\frac{1}{K}\sum_{k=1}^K S_f^0(\hat{f}_k(d) - f_D^0(d)) + o_p(h^{-1/2})\notag \\
&= \sqrt{N}\frac{1}{N}\sum_{i=1}^N S_f^0(K_h(D_i-d) - E[K_h(D-d)]) + o_p(h^{-1/2})
\end{align}
where the last equality holds since
\begin{align*}
\hat{f}_k(d) - f_D^0(d) = (N-n)^{-1}\sum_{i\in I_k^c} K_h(D_i-d) - E[K_h(D-d)] + O(h^2)  
\end{align*}
with $N-n$ the sample size of each auxiliary subsample used to estimate the nuisance parameters, $h$ being an under-smoothing bandwidth, and the fact that $K^{-1}\sum_{k=1}^K (\hat{f}_k(d) - E[K_h(D-d)]) = N^{-1}\sum_{i=1}^N (K_h(D_i-d)-E[K_h(D-d)])$. In particular, the kernel expression in the last line is mean-zero and it will contribute to the asymptotic variance.\\

\noindent \textit{Step 3: ``Neyman Term''}

Now we consider (\ref{eq:1}), which we can rewrite as
\begin{align}
&\sqrt{N}\frac{1}{K}\sum_{k=1}^K E_{n,k}[\psi_h(Z,\theta_{0h}(d),f_D^0(d),\hat{\eta}_k(d))]\notag \\ &= \frac{1}{\sqrt{N}} \sum_{i=1}^N \psi_h(Z_i,\theta_{0h}(d),f_D^0(d),\eta_0(d))\\
&+ \sqrt{N}\frac{1}{K}\sum_{k=1}^K \underbrace{(E_{n,k}[\psi_h(Z,\theta_{0h}(d),f_D^0(d),\hat{\eta}_k(d))] - E_{n,k}[\psi_h(Z_i,\theta_{0h}(d),f_D^0(d),\eta_0(d))])}_{R_{nk}}.
\end{align}
Since $K$ is fixed, $n= O(N)$, it suffices to show that $R_{nk} = o_p(N^{-1/2}h^{-1/2})$, so it vanishes when scaled by the (square root of) asymptotic variance. Note that by triangle inequality, we have the following decomposition
\begin{align}
|R_{n,k}| \leq \frac{R_{1k} + R_{2k}}{\sqrt{n}}
\end{align}
where
\begin{align}
R_{1k}:= |G_{nk}[\psi_h(Z,\theta_{0h}(d),f_D^0(d),\hat{\eta}_k(d))] - G_{nk}[\psi_h(Z,\theta_{0h}(d),f_D^0(d),\eta_0(d))]|
\end{align}
with $G_{nk}(f) = \sqrt{n}\frac{1}{n}\sum_{i=1}^n (f(Z_i) - E[f(Z)])$ denoting the empirical process, and, with some abuse of notation, it will also be used to denote conditional version of the empirical process conditioning on the auxiliary sample $I_k^c$. Moreover,
\begin{align}
R_{2k}:= \sqrt{n}|E[\psi_h(Z,\theta_{0h}(d),f_D^0(d),\hat{\eta}_k(d))|I_k^c] - E[\psi_h(Z,\theta_{0h}(d),f_D^0(d),\eta_0(d))]|.
\end{align}

First, we consider $R_{1k}$. For simplicity, let's suppress other arguments in $\psi$ and denote $\psi_{\eta(d)}^i := \psi_h(Z_i,\theta_{0h}(d),f_D^0(d),\eta(d))$. Then, by the definition of the empirical process, we have
\begin{align}
G_{nk}\psi_{\hat{\eta}_k(d)} - G_{nk}\psi_{\eta_0(d)} = \sqrt{n} \frac{1}{n}\sum_{i=1}^n \underbrace{\psi_{\hat{\eta}_k(d)}^i - \psi_{\eta_0(d)}^i - E[\psi_{\hat{\eta}_k(d)}^i|I_k^c] + E[\psi_{\eta_0(d)}^i]}_{:= \Delta_{ik}}
\end{align}
In particular, it can be shown that $E[\Delta_{ik}\Delta_{jk}|I_k^c] =0$ for all $i\neq j$ using the law of iterated expectation, the i.i.d. assumption of the data, and the fact that the nuisance parameter $\hat{\eta}_k(d)$ is estimated using the auxiliary sample $I_k^c$. Then, we have
\begin{align*}
E[R_{1k}^2|I_k^c]&\leq E[\Delta_{ik}^2|I_k^c]\\
&\leq E[(\psi_{\hat{\eta}_k(d)}^i-\psi_{\eta_0(d)}^i)^2|I_k^c]\\
&\leq \sup_{\eta(d)\in T_N(d)} E[(\psi_{\eta(d)}^i-\psi_{\eta_0(d)}^i)^2|I_k^c]\\
&\leq \sup_{\eta(d)\in T_N(d)} E[(\psi_{\eta(d)}^i-\psi_{\eta_0(d)}^i)^2]\\
&\lesssim h^{-1}\varepsilon_N^2 \quad\quad\quad (c)
\end{align*}
and using the conditional Markov's inequality, we conclude that $R_{1k} = O_p(h^{-1/2}\varepsilon_N)$.

Now we bound $R_{2k}$. Note that by definition, $E[\psi_h(Z,\theta_{0h}(d), f_D^0(d),\eta_0(d))]= 0$, so it suffices to bound $E[\psi_h(Z,\theta_{0h}(d),f_D^0(d),\hat{\eta}_k(d))|I_k^c]$. Suppressing other arguments in the score, define
\begin{align}
h_k(r) := E[\psi_h(\eta_0(d) + r(\hat{\eta}_k(d)-\eta_0(d)))|I_k^c]
\end{align}
where by definition $h_k(0) = E[\psi_h(\eta_0(d))|I_k^c]= 0$ and $h_k(1) = E[\psi_h(\hat{\eta}_k(d))|I_k^c]$. Use Taylor's theorem, expand $h_k(1)$ around $0$, we have
\begin{align}
h_k(1) = h_k(0) + h_k'(0) + \frac{1}{2}h_k^{''}(\bar{r}),\quad \bar{r}\in(0,1).
\end{align}
Note that, by Neyman orthogonality,
\begin{align}
h_k'(0) = \partial_\eta(d) E[\psi_h(\eta_0(d))][\hat{\eta}_k(d) - \eta_0(d)] = 0
\end{align}
and use that fact that $h_k(0)=0$, we have
\begin{align*}
R_{2k} &= \sqrt{n}|h_k(1)| = \frac{\sqrt n}{2}|h_k^{''}(\bar{r})| \\
& \leq \sup_{r\in(0,1),\eta(d)\in T_N(d)} \sqrt{n}|\partial_r^2E[\psi_h(\eta_0(d) + r(\eta(d)-\eta_0(d)))]| \\
&\lesssim \sqrt{n}h^{-1/2}\varepsilon_N^2\quad\quad\quad (d)
\end{align*}

Combining the above results, we conclude that
\begin{align}
\sqrt{N} R_{n,k} \lesssim h^{-1/2}\varepsilon_N + \sqrt{N}h^{-1/2}\varepsilon_N^2,
\end{align}
and for $\varepsilon_N = o(N^{-1/4})$, we have $\sqrt{N} R_{n,k} = o_p(h^{-1/2})$.\\

\noindent \textit{Step 4: Auxiliary Results}

In this section, we show the auxiliary results (a)-(d) used in the previous steps. We first show (c) as it will also be used to bound other results. 

Recall that
\[
(c): \quad\quad \sup_{\eta(d)\in T_N(d)}E[(\psi_{\eta(d)} - \psi_{\eta_0(d)})^2]\lesssim h^{-1}\varepsilon_N^2.
\]
By definition,
\begin{align}
\psi_{\eta(d)} - \psi_{\eta_0(d)} =&\frac{K_h(D-d)g(X) - \mathbf{1}\{D=0\}f_h(d|X)}{f_D^0(d)g(X)}\big(\Delta Y -  \mathcal{E}_{\Delta Y}(X)\big)\notag \\
&- \frac{K_h(D-d)g_0(X) - \mathbf{1}\{D=0\}f_h^0(d|X)}{f_D^0(d)g_0(X)}\big(\Delta Y -  \mathcal{E}_{\Delta Y}^0(X)\big)\notag \\
=& \frac{K_h(D-d)}{f_D^0(d)}(\mathcal{E}_{\Delta Y}^0(X) - \mathcal{E}_{\Delta Y}(X))\notag \\
&- \frac{\mathbf{1}\{D=0\}}{f_D^0(d)}\bigg(\frac{f_h(d|X)}{g(X)}(\Delta Y - \mathcal{E}_{\Delta Y}(X)) - \frac{f_h^0(d|X)}{g_0(X)}(\Delta Y - \mathcal{E}_{\Delta Y}^0(X))\bigg)\notag \\
=& \frac{K_h(D-d)}{f_D^0(d)}(\mathcal{E}_{\Delta Y}^0(X) - \mathcal{E}_{\Delta Y}(X))\notag \\
&- \frac{\mathbf{1}\{D=0\}}{f_D^0(d)}\bigg(\frac{f_h(d|X)}{g(X)} - \frac{f_h^0(d|X)}{g_0(X)}\bigg)\Delta Y \notag \\
&+ \frac{\mathbf{1}\{D=0\}}{f_D^0(d)}\bigg(\frac{f_h(d|X)}{g(X)}\mathcal{E}_{\Delta Y}(X) - \frac{f_h^0(d|X)}{g_0(X)}\mathcal{E}_{\Delta Y}^0(X) \bigg)
\end{align}

Using overlap and boundedness,
\[
|\psi_{\eta(d)}-\psi_{\eta_0(d)}|
\lesssim |f_h-f_h^0|+|g-g_0|+(|K_h(D-d)|+1)|\mathcal E_{\Delta Y}-\mathcal E_{\Delta Y}^0|.
\]
Since $E[K_h^2(D-d)\mid X]\lesssim h^{-1}$, conditioning on $X$ gives
\[
E[(\psi_{\eta(d)}-\psi_{\eta_0(d)})^2]
\lesssim\|f_h-f_h^0\|_{P,2}^2+\|g-g_0\|_{P,2}^2+h^{-1}\|\mathcal E_{\Delta Y}-\mathcal E_{\Delta Y}^0\|_{P,2}^2
\lesssim h^{-1}\varepsilon_N^2.
\]
This shows (c).\\

Next, we consider (a). We want to show
\begin{align*}
(a):\quad \sup_{\eta(d)\in T_N(d)} E[\sup_{f\in F_N(d)}|\partial_f^2\psi_h(Z,\theta_{0h}(d),f,\eta(d))-\partial_f^2\psi_h(Z,\theta_{0h}(d), f_D^0(d),\eta_0(d))|]&\\\lesssim h^{-1/2}\varepsilon_N+(Nh)^{-1/2}
\end{align*}
By definition, 
\begin{align}
\partial_f^2\psi_h(Z,\theta_{0h}(d),f,\eta(d)) &= \frac{2}{f^2}(\psi_h(Z,\theta_{0h}(d),f,\eta(d))+\theta_{0h}(d))\notag\\
\partial_f^3\psi_h(Z,\theta_{0h}(d),f,\eta(d)) &= -\frac{6}{f^3}(\psi_h(Z,\theta_{0h}(d),f,\eta(d))+\theta_{0h}(d)).
\end{align}
Then using Taylor's theorem expand $\partial_f^2\psi_h(Z,\theta_{0h}(d),f,\eta(d))$ around $f_D^0(d)$, we have
\begin{align*}
&\partial_f^2\psi_h(Z,\theta_{0h}(d),f,\eta(d))-\partial_f^2\psi_h(Z,\theta_{0h}(d), f_D^0(d),\eta_0(d))\\
&= \partial_f^2\psi_h(Z,\theta_{0h}(d),f_D^0(d),\eta(d))-\partial_f^2\psi_h(Z,\theta_{0h}(d), f_D^0(d),\eta_0(d))\\
&+ \partial_f^3 \psi_h(Z,\theta_{0h}(d),\bar{f},\eta(d))(f-f_D^0(d))\\
&= \frac{2}{(f_D^0(d))^2}(\psi_h(Z,\theta_{0h}(d),f_D^0(d),\eta(d))-\psi_h(Z,\theta_{0h}(d), f_D^0(d),\eta_0(d)))\quad\text{(i)}\\
&-\frac{6}{\bar{f}^3}(\psi_h(Z,\theta_{0h}(d),\bar{f},\eta(d))+\theta_{0h}(d))(f-f_D^0(d))\quad\text{(ii)}
\end{align*}
By the assumption, on $F_N(d)$, $\bar{f}$ and $f_D^0(d)$ are bounded away from zero, so that (i) can be bounded with (c) and Cauchy--Schwarz as $E[|(i)|]\lesssim h^{-1/2}\varepsilon_N$. Moreover, by assumption, $E[\sup_{f\in F_N(d)}|(ii)|] = O((Nh)^{-1/2})$ since $\sup_{f\in F_N(d)}|f-f_D^0(d)|\lesssim(Nh)^{-1/2}$ and $\sup_{\bar f\in F_N(d),\eta(d)\in T_N(d)}|\psi_h(Z,\theta_{0h}(d),\bar f,\eta(d))+\theta_{0h}(d)|\lesssim |K_h(D-d)|+1$, whose expectation is $O(1)$. Therefore we conclude that
\begin{align}
&\sup_{\eta(d)\in T_N(d)} E[\sup_{f\in F_N(d)}|\partial_f^2\psi_h(Z,\theta_{0h}(d),f,\eta(d))-\partial_f^2\psi_h(Z,\theta_{0h}(d), f_D^0(d),\eta_0(d))|]\notag\\
&\lesssim h^{-1/2}\varepsilon_N+(Nh)^{-1/2}.
\end{align}
Similarly, by definition,
\begin{align}
&\partial_f \psi_h(Z,\theta_{0h}(d),f_D^0(d),\eta(d))-\partial_f\psi_h(Z,\theta_{0h}(d),  f_D^0(d),\eta_0(d))\notag \\ &= -\frac{1}{f_D^0(d)}(\psi_h(Z,\theta_{0h}(d),f_D^0(d),\eta(d))-\psi_h(Z,\theta_{0h}(d), f_D^0(d),\eta_0(d)))
\end{align}
and using the same arguments as before, (b) follows from (a) and (c). 

Last, we show (d). It suffices to show
\begin{align}
 \sup_{r\in(0,1),\eta(d)\in T_N(d)} |\partial_r^2E[\psi_h(\eta_0(d) + r(\eta(d)-\eta_0(d)))]|\lesssim h^{-1/2}\varepsilon_N^2.
\end{align}
By definition,
\begin{align}
&\psi_h(\eta_0(d) + r(\eta(d)-\eta_0(d)))\notag \\
&= \frac{K_h(D-d)}{f_D^0(d)}\big(\Delta Y -  (\mathcal{E}_{\Delta Y}^0(X) + r(\mathcal{E}_{\Delta Y}(X) -\mathcal{E}_{\Delta Y}^0(X)))\big)-\notag \\
&\frac{\mathbf{1}\{D=0\}(f_h^0(d|X) + r(f_h(d|X)-f_h^0(d|X)))}{f_D^0(d)(g_0(X) + r(g(X) -g_0(X)))}\big(\Delta Y -  \mathcal{E}_{\Delta Y}^0(X) - r(\mathcal{E}_{\Delta Y}(X) -\mathcal{E}_{\Delta Y}^0(X)) \big)-\theta_{0h}(d)
\end{align}
and we take the second-order partial derivatives w.r.t. $r$ term by term. For simplicity, we omit the derivations, and we have 
\begin{align}
|\partial_r^2 \psi_h(\eta_0(d) + r(\eta(d)-\eta_0(d)))|\lesssim |\Delta_f\Delta_g|+|\Delta_\mathcal E\Delta_g|+|\Delta_f\Delta_\mathcal E|+(\Delta_g)^2
\end{align}
where $\Delta_f := f_h- f_h^0$, $\Delta_g := g - g_0$, and $\Delta_\mathcal{E} := \mathcal{E}_{\Delta Y}- \mathcal{E}_{\Delta Y}^0$. Then by triangle inequality, Cauchy-Schwarz, and the assumption on the space of nuisance parameters $T_N(d)$, we have
\begin{align}
 &E[|\partial_r^2\psi_h(\eta_0(d) + r(\eta(d)-\eta_0(d)))|]\notag\\
 &\lesssim \|f_h - f_h^0\|_{P,2}\|g-g_0\|_{P,2} + \|f_h - f_h^0\|_{P,2}\|\mathcal{E}_{\Delta Y}- \mathcal{E}_{\Delta Y}^0\|_{P,2}\notag\\
&+ \|g-g_0\|_{P,2}\|\mathcal{E}_{\Delta Y}- \mathcal{E}_{\Delta Y}^0\|_{P,2} + \|g - g_0\|_{P,2}^2\notag\\
&\lesssim h^{-1/2}\varepsilon_N^2.
\end{align}
Then (d) follows by Jensen's inequality.

\noindent \textit{Step 5: Asymptotic Normality}

Combining previous results, we have
\begin{align}
&\widehat{ATT}(d) - ATT(d)\notag \\ 
&= \frac{1}{N}\sum_{i=1}^N\psi_h(Z_i,\theta_{0h}(d),f_D^0(d),\eta_0(d))\label{cdid:linear1} \\
&+ \frac{1}{N}\sum_{i=1}^N S_f^0(K_h(D_i -d) - E[K_h(D_i -d)]) \label{cdid:linear2}\\
&+ o_p((Nh)^{-1/2}) \label{cdid:linear3}\\
&+ \theta_{0h}(d)-\theta_0(d) \label{cdid:linear4}
\end{align}
where (\ref{cdid:linear1}) and (\ref{cdid:linear2}) are averages of i.i.d. zero-mean terms with the variance growing with kernel bandwidth $h$, and recall that $S_f^0 = E[\partial_f\psi_h(Z,\theta_{0h}(d), f_D^0(d),\eta_0(d))]$; (\ref{cdid:linear3}) are the terms that vanish when scaled by the (square root of) asymptotic variance; (\ref{cdid:linear4}) is the bias term which is shown to be of order $O(h^2)$ in Lemma \ref{lm:bias}.

Let
\[
\phi_{i,h}=\psi_h(Z_i,\theta_{0h}(d),f_D^0(d),\eta_0(d))
+S_f^0\{K_h(D_i-d)-E[K_h(D-d)]\}.
\]
Then $E[\phi_{i,h}]=0$, $E[|\phi_{i,h}|^3]=O(h^{-2})$, and the variance assumption gives $\sigma_N^2=E[\phi_{i,h}^2]\gtrsim h^{-1}$. Thus
\[
\frac{(N E[|\phi_{i,h}|^3])^{1/3}}{(N\sigma_N^2)^{1/2}}=O((Nh)^{-1/6})=o(1).
\]
Lyapunov's theorem and the remainder bounds then give the desired asymptotic normality result
\begin{align}
\frac{\widehat{ATT}(d) - ATT(d)}{\sigma_N/\sqrt{N}}\quad \to^d \quad N(0,1)
\end{align}
with $\sigma_N$ defined by
\begin{align}
\sigma_N^2:= E\bigg[\bigg(\psi_h - \frac{\theta_{0h}}{f_D^0(d)}(K_h(D-d)-E[K_h(D-d)])\bigg)^2\bigg]
\end{align}
where we have used the fact that $S_f^0 = -\theta_{0h}/f_D^0(d)$. \hfill$\square$\\

\noindent\textbf{Proof of Theorem~\ref{thm:asymp_rcs} (Repeated Cross-Sections)}
The proof for the repeated cross-sectional case follows very closely to that of the panel case, with only minor modifications due to the presence of a new parameter $\lambda = P(T=1)$, which can be estimated at the parametric rate. 

Let $T_N(d)$ be the set of functions $\eta(d):= (f_h(d|X), g(X),\mathcal{E}_{\lambda Y}(X))$ such that $\|f_h(d|X) - f_h^0(d|X)\|_{P,2}\leq h^{-1/2}\varepsilon_N$, $\|g(X) - g_0(X)\|_{P,2}\leq \varepsilon_N$, $\|\mathcal{E}_{\lambda Y}(X) - \mathcal{E}_{\lambda Y}^0(X)\|_{P,2}\leq \varepsilon_N$, $\kappa<g(X)<1-\kappa$ a.s., $c < f_h(d|X)<C$, and $\|\mathcal{E}_{\lambda Y}(X)\|_{P,\infty}<C$. Let $P_N$ be the set of $\lambda>0$ such that $|\lambda-\lambda_0|\leq MN^{-1/2}$. Let $F_N(d)$ be the set of $f>c$ such that $|f-f_D^0(d)|\leq M(Nh)^{-1/2}$. Then Assumption \ref{csdidas2} implies that, for every $\epsilon>0$, a sufficiently large fixed $M$ makes $\hat\eta_k(d)\in T_N(d)$, $\hat{\lambda}_k\in P_N$, and $\hat f_k(d)\in F_N(d)$ hold jointly with probability at least $1-\epsilon$ for all sufficiently large $N$. Throughout the proof, we use $N$ to denote the sample size and $n := N/K$ to denote the size of the subsamples. In particular, since $K$ is fixed, $n\asymp N$.

To simplify notation, let $\theta_0(d)$ denote the true $ATT(d)$, $\theta_{0h}(d)$ denote the true $ATT_h(d)$, and $\hat{\theta}_h(d)$ denote our cross-fitted estimator. In particular, recall that our estimator is
\begin{align}
\hat{\theta}_h(d) := \frac{1}{K}\sum_{k=1}^K \frac{1}{n}\sum_{i\in I_k} &\frac{K_h(D_i-d)\hat{g}_k(X_i) - \mathbf{1}\{D_i=0\}\hat{f}_{h,k}(d|X_i)}{\hat{f}_k(d)\hat{g}_k(X_i)}\\
&\times\bigg(\frac{T_i-\hat{\lambda}_k}{\hat{\lambda}_k(1-\hat{\lambda}_k)}Y_i - \hat{\mathcal E}_{\lambda Y,k}(X_i)\bigg)
\end{align}
Then we have the following
\begin{align}
\hat{\theta}_h(d) - \theta_0(d) = \underbrace{\hat{\theta}_h(d) - \theta_{0h}(d)}_{(1)} + \underbrace{\theta_{0h}(d)- \theta_0(d)}_{(2)}
\end{align}
where $(1)$ will be our main focus while the bias term $(2)$ is shown in Lemma \ref{lm:bias} to be $O(h^2)$ and asymptotically negligible by the assumption of the under-smoothing bandwidth $h$.

By definition, 
\begin{align}
\sqrt{N}(\hat{\theta}_h(d) - \theta_{0h}(d)) = \sqrt{N}\frac{1}{K}\sum_{k=1}^KE_{n,k}[\psi_h(Z_i,\theta_{0,h},\hat\lambda_k,\hat{f}_k(d),\hat{\eta}_k(d))]
\end{align}
where $\psi_h$ is defined as in (\ref{psi2}), and $E_{n,k}(f) = \frac{1}{n}\sum_{i\in I_k} f(Z_i)$ denotes the empirical average of a generic function $f$ over the set $I_k$. Then we have the following decomposition, using a multivariate Taylor's theorem,
\begin{align}
&\sqrt{N}(\hat{\theta}_h(d) - \theta_{0h}(d))\notag \\
&= \sqrt{N}\frac{1}{K}\sum_{k=1}^K E_{n,k}[\psi_h(Z,\theta_{0h}(d),\lambda_0,f_D^0(d),\hat{\eta}_k(d))] \label{cseq:1}\\
&+ \sqrt{N}\frac{1}{K}\sum_{k=1}^K E_{n,k}[\partial_\lambda\psi_h(Z,\theta_{0h}(d),\lambda_0, f_D^0(d),\hat{\eta}_k(d))](\hat{\lambda}_k - \lambda_0)\label{cseq:2}\\
&+ \sqrt{N}\frac{1}{K}\sum_{k=1}^K E_{n,k}[\partial_{f}\psi_h(Z,\theta_{0h}(d),\lambda_0, f_D^0(d),\hat{\eta}_k(d))](\hat{f}_k(d) - f_D^0(d))\label{cseq:3}\\
&+ \sqrt{N} \frac{1}{2K}\sum_{k=1}^K E_{n,k}[\partial_\lambda^2\psi_h(Z,\theta_{0h}(d),\bar{\lambda}_k, \bar{f}_k,\hat{\eta}_k(d))](\hat{\lambda}_k - \lambda_0)^2\label{cseq:4}\\
&+ \sqrt{N} \frac{1}{2K}\sum_{k=1}^K E_{n,k}[\partial_f^2\psi_h(Z,\theta_{0h}(d),\bar{\lambda}_k, \bar{f}_k,\hat{\eta}_k(d))](\hat{f}_k(d) - f_D^0(d))^2\label{cseq:5}\\
&+ \sqrt{N}\frac{1}{K}\sum_{k=1}^K E_{n,k}[\partial_\lambda\partial_f\psi_h(Z,\theta_{0h}(d),\bar{\lambda}_k, \bar{f}_k,\hat{\eta}_k(d))](\hat{f}_k(d) - f_D^0(d))(\hat{\lambda}_k - \lambda_0)\label{cseq:6}
\end{align}
where $\bar{\lambda}_k \in (\lambda_0,\hat{\lambda}_k)$ and $\bar{f}_k \in (f_D^0(d), \hat{f}_k(d))$. All the second-order terms (\ref{cseq:4})--(\ref{cseq:6}) are $o_p(h^{-1/2})$. The first-order term (\ref{cseq:3}) can be analyzed in the same way as the repeat outcomes case. Moreover, since $\hat\lambda_k=(N-n)^{-1}\sum_{i\in I_k^c}T_i$ converges at the parametric rate while the kernel estimator $\hat{f}_k(d)$ converges at a slower rate, the influence of (\ref{cseq:2}) on the asymptotic variance is negligible. The main term (\ref{cseq:1}) can be analyzed in the same way as in the panel case.\\ 

\noindent \textit{Step 1: Second Order Terms}

Following the same argument as the panel case, every second derivative in $(\lambda,f)$ is bounded in absolute value by $C(|K_h(D-d)|+1)$. Since $E_{n,k}[|K_h(D-d)|+1]=O_p(1)$, the terms (\ref{cseq:4}), (\ref{cseq:5}), and (\ref{cseq:6}) are respectively
\[
O_p(N^{-1/2}),\qquad O_p((\sqrt N h)^{-1}),\qquad O_p((Nh)^{-1/2}),
\]
and therefore are all $o_p(h^{-1/2})$.

\noindent \textit{Step 2: First-Order Terms}

We first consider (\ref{cseq:2}). By triangle inequality, we have
\begin{align}
&|E_{n,k}[\partial_\lambda\psi_h(Z,\theta_{0h}(d),\lambda_0, f_D^0(d),\hat{\eta}_k(d))] - E[\partial_\lambda\psi_h(Z,\theta_{0h}(d), \lambda_0, f_D^0(d),\eta_0(d))]|\notag \\
\leq& \underbrace{|E_{n,k}[\partial_\lambda\psi_h(Z,\theta_{0h}(d),\lambda_0, f_D^0(d),\hat{\eta}_k(d))] - E_{n,k}[\partial_\lambda\psi_h(Z,\theta_{0h}(d), \lambda_0, f_D^0(d),\eta_0(d))]|}_{J_{5k}}\\
+& \underbrace{|E_{n,k}[\partial_\lambda\psi_h(Z,\theta_{0h}(d), \lambda_0, f_D^0(d),\eta_0(d))] - E[\partial_\lambda\psi_h(Z,\theta_{0h}(d), \lambda_0, f_D^0(d),\eta_0(d))]|}_{J_{6k}}.
\end{align}
To bound $J_{6k}$, since $\lambda_0$ is bounded away from zero, the score $\psi$ satisfies, 
\begin{align}
|\partial_\lambda\psi_h(Z,\theta_{0h}(d), \lambda_0, f_D^0(d),\eta_0(d))| \lesssim |K_h(D-d)|+1.
\end{align}
This implies that 
\begin{align*}
E[J_{6k}^2] \leq \frac{1}{n} E[(\partial_\lambda\psi_h(Z,\theta_{0h}(d), \lambda_0, f_D^0(d),\eta_0(d)))^2]\lesssim E[K_h^2(D-d)]/N \lesssim (Nh)^{-1}.
\end{align*}
and by Markov's inequality, we have $J_{6k}\leq O_p((Nh)^{-1/2})$. With the assumption that $Nh\to\infty$, we have $J_{6k} = o_p(1)$. 

On the other hand, for $J_{5k}$, note that
\begin{align*}
&E[J_{5k}^2|I_k^c]\\ &= E[|E_{n,k}[\partial_\lambda\psi_h(Z,\theta_{0h}(d),\lambda_0, f_D^0(d),\hat{\eta}_k(d))]\notag\\
&\quad - E_{n,k}[\partial_\lambda\psi_h(Z,\theta_{0h}(d), \lambda_0, f_D^0(d),\eta_0(d))]|^2|I_k^c]\\
&\leq \sup_{\eta(d)\in T_N(d)} E[|\partial_\lambda\psi_h(Z,\theta_{0h}(d),\lambda_0, f_D^0(d),\eta(d))-\partial_\lambda\psi_h(Z,\theta_{0h}(d), \lambda_0, f_D^0(d),\eta_0(d))|^2|I_k^c]\\
&\leq \sup_{\eta(d)\in T_N(d)} E[|\partial_\lambda\psi_h(Z,\theta_{0h}(d),\lambda_0, f_D^0(d),\eta(d))-\partial_\lambda\psi_h(Z,\theta_{0h}(d), \lambda_0, f_D^0(d),\eta_0(d))|^2]\\
&\lesssim h^{-1}\varepsilon_N^2 \quad\quad\quad \text{(d)}
\end{align*}
where the first equation holds by definition, the second line holds by the Cauchy-Schwarz inequality, and the third line holds by the construction that all the parameters are estimated using auxiliary sample $I_k^c$ and hence can be treated as fixed. Then we conclude with conditional Markov's inequality that $J_{5k} = o_p(1)$. As before, we will show (d) at the end of this section.

Therefore,
\begin{align}
E_{n,k}[\partial_\lambda\psi_h(Z,\theta_{0h}(d),\lambda_0, f_D^0(d),\hat{\eta}_k(d))] \to^p E[\partial_\lambda\psi_h(Z,\theta_{0h}(d), \lambda_0, f_D^0(d),\eta_0(d))]:= S_\lambda^0
\end{align}
Note that $(\hat{\lambda}_k - \lambda_0) = O_p(N^{-1/2})$, we can rewrite (\ref{cseq:2}) as
\begin{align*}
(\ref{cseq:2})&= \sqrt{N}\frac{1}{K}\sum_{k=1}^K E_{n,k}[\partial_\lambda\psi_h(Z,\theta_{0h}(d),\lambda_0, f_D^0(d),\hat{\eta}_k(d))](\hat{\lambda}_k - \lambda_0)\\
&= \sqrt{N}\frac{1}{K}\sum_{k=1}^K S_\lambda^0(\hat{\lambda}_k - \lambda_0) + o_p(1)\\
&= \sqrt{N}\frac{1}{N}\sum_{i=1}^N S_\lambda^0(T_i - \lambda_0) + o_p(1)
\end{align*}
where the last equality holds by the definition that $\hat{\lambda}_k - \lambda_0 = (N-n)^{-1}\sum_{i\in I_k^c} T_i - \lambda_0$ and the fact that $K^{-1}\sum_{k=1}^K (\hat{\lambda}_k - \lambda_0) = \frac{1}{N}\sum_{i=1}^N (T_i - \lambda_0)$. We remark that, since $S_\lambda^0 =  E[\partial_\lambda\psi_h^0]$ is bounded by a constant and $\hat{\lambda}$ converges at parametric rate, (\ref{cseq:2}) vanishes when scaled by the square-root of the asymptotic variance that grows with sample size. 

Term (\ref{cseq:3}) will be bounded using the same argument as in the panel setting. First, by the triangle inequality
\begin{align}
&|E_{n,k}[\partial_f\psi_h(Z,\theta_{0h}(d), \lambda_0, f_D^0(d),\hat{\eta}_k(d))] - E[\partial_f\psi_h(Z,\theta_{0h}(d),\lambda_0, f_D^0(d),\eta_0(d))]|\notag \\
\leq& \underbrace{|E_{n,k}[\partial_f\psi_h(Z,\theta_{0h}(d),\lambda_0, f_D^0(d),\hat{\eta}_k(d))] - E_{n,k}[\partial_f\psi_h(Z,\theta_{0h}(d),\lambda_0, f_D^0(d),\eta_0(d))]|}_{J_{7k}}\\
+& \underbrace{|E_{n,k}[\partial_f\psi_h(Z,\theta_{0h}(d),\lambda_0, f_D^0(d),\eta_0(d))] - E[\partial_f\psi_h(Z,\theta_{0h}(d),\lambda_0, f_D^0(d),\eta_0(d))]|}_{J_{8k}}.
\end{align}
We first bound $J_{8k}$. Note that since $f_D^0(d)$ is bounded away from zero and the score $\psi$ satisfies
\begin{align}
&|\partial_f\psi_h(Z,\theta_{0h}(d),\lambda_0, f_D^0(d),\eta_0(d))|\\ &= |-\frac{1}{f_D^0(d)}(\psi_h(Z,\theta_{0h}(d),\lambda_0, f_D^0(d),\eta_0(d)) + \theta_{0h}(d))| \lesssim |K_h(D-d)|+1,
\end{align}
which implies that 
\begin{align}
E[J_{8k}^2] \leq \frac{1}{n}E[(\partial_f\psi_h(Z,\theta_{0h}(d),\lambda_0, f_D^0(d),\eta_0(d)))^2]\lesssim E[K_h^2(D-d)]/N \lesssim (Nh)^{-1}.
\end{align}
Then by Markov's inequality, we have $J_{8k}\leq O_p((Nh)^{-1/2})$. With the assumption that $Nh\to\infty$, we have $J_{8k} = o_p(1)$.

Second, to bound $J_{7k}$, note that
\begin{align*}
&E[J_{7k}^2|I_k^c]\\ &= E[|E_{n,k}[\partial_f\psi_h(Z,\theta_{0h}(d),\lambda_0, f_D^0(d),\hat{\eta}_k(d))]\notag\\
&\quad - E_{n,k}[\partial_f\psi_h(Z,\theta_{0h}(d),\lambda_0, f_D^0(d),\eta_0(d))]|^2|I_k^c]\\
&\leq \sup_{\eta(d)\in T_N(d)} E[|\partial_f\psi_h(Z,\theta_{0h}(d),\lambda_0, f_D^0(d),\eta(d))-\partial_f\psi_h(Z,\theta_{0h}(d),\lambda_0, f_D^0(d),\eta_0(d))|^2|I_k^c]\\
&\leq \sup_{\eta(d)\in T_N(d)} E[|\partial_f\psi_h(Z,\theta_{0h}(d),\lambda_0,f_D^0(d),\eta(d))-\partial_f\psi_h(Z,\theta_{0h}(d),\lambda_0, f_D^0(d),\eta_0(d))|^2]\\
&\lesssim h^{-1}\varepsilon_N^2 \quad\quad\quad \text{(e)}
\end{align*}
where the first equation holds by definition, the second line holds by Cauchy-Schwarz, and the third line holds by the construction that all the parameters are estimated using auxiliary sample $I_k^c$. Then we conclude with the conditional Markov's inequality that $J_{7k} = o_p(1)$. Therefore,
\begin{align}
E_{n,k}[\partial_f\psi_h(Z,\theta_{0h}(d),\lambda_0, f_D^0(d),\hat{\eta}_k(d))] \to^p E[\partial_f\psi_h(Z,\theta_{0h}(d), \lambda_0, f_D^0(d),\eta_0(d))]:= S_f^0
\end{align}
Since $(\hat{f}_k(d) - f_D^0(d)) = O_p((Nh)^{-1/2})$, we can rewrite (\ref{cseq:3}) as
\begin{align*}
(\ref{cseq:3})&= \sqrt{N}\frac{1}{K}\sum_{k=1}^K E_{n,k}[\partial_f\psi_h(Z,\theta_{0h}(d),\lambda_0, f_D^0(d),\hat{\eta}_k(d))](\hat{f}_k(d) - f_D^0(d))\\
&= \sqrt{N}\frac{1}{K}\sum_{k=1}^K S_f^0(\hat{f}_k(d) - f_D^0(d)) + o_p(h^{-1/2})\\
&= \sqrt{N}\frac{1}{N}\sum_{i=1}^N S_f^0(K_h(D_i-d) - E[K_h(D-d)]) + o_p(h^{-1/2})
\end{align*}
where the last equality holds by $\hat{f}_k(d) - f_D^0(d) = (N-n)^{-1}\sum_{i\in I_k^c} K_h(D_i-d) - E[K_h(D-d)] + O(h^2)$, the undersmoothing assumption $\sqrt{Nh}\,h^2=o(1)$, and the fact that $K^{-1}\sum_{k=1}^K (\hat{f}_k(d) - E[K_h(D-d)]) = \frac{1}{N}\sum_{i=1}^N (K_h(D_i-d)-E[K_h(D-d)])$. This term will contribute to the asymptotic variance.\\

\noindent \textit{Step 3: ``Neyman Term''}

Now we consider (\ref{cseq:1}), which can be shown using the same argument as the panel case.
\begin{align}
&\sqrt{N}\frac{1}{K}\sum_{k=1}^K E_{n,k}[\psi_h(Z,\theta_{0h}(d),\lambda_0,f_D^0(d),\hat{\eta}_k(d))]\notag \\ &= \frac{1}{\sqrt{N}} \sum_{i=1}^N \psi_h(Z_i,\theta_{0h}(d),\lambda_0,f_D^0(d),\eta_0(d))\notag\\
&+ \sqrt{N}\frac{1}{K}\sum_{k=1}^K \underbrace{(E_{n,k}[\psi_h(Z,\theta_{0h}(d),\lambda_0,f_D^0(d),\hat{\eta}_k(d))] - E_{n,k}[\psi_h(Z_i,\theta_{0h}(d),\lambda_0,f_D^0(d),\eta_0(d))])}_{R_{nk}}.
\end{align}
Since $K$ is fixed, $n= O(N)$, it suffices to show that $R_{nk} = o_p(N^{-1/2}h^{-1/2})$, so it vanishes when scaled by the (square root of) asymptotic variance. Note that by triangle inequality, we have the following decomposition
\begin{align}
|R_{n,k}| \leq \frac{R_{1k} + R_{2k}}{\sqrt{n}}
\end{align}
where
\begin{align}
R_{1k}:= |G_{nk}[\psi_h(Z,\theta_{0h}(d),\lambda_0,f_D^0(d),\hat{\eta}_k(d))] - G_{nk}[\psi_h(Z,\theta_{0h}(d),\lambda_0,f_D^0(d),\eta_0(d))]|
\end{align}
with $G_{nk}(f) = \sqrt{n}(P_n-P)(f)$ denote the empirical process, and with some abuse of notation, it will also be used to denote the conditional version of the empirical process conditioning on the auxiliary sample $I_k^c$. Moreover,
\begin{align}
R_{2k}:= \sqrt{n}|E[\psi_h(Z,\theta_{0h}(d),\lambda_0,f_D^0(d),\hat{\eta}_k(d))|I_k^c] - E[\psi_h(Z,\theta_{0h}(d),\lambda_0,f_D^0(d),\eta_0(d))]|.
\end{align}
For simplicity, denote $\psi_{\eta(d)}^i := \psi_h(Z_i,\theta_{0h}(d),\lambda_0,f_D^0(d),\eta(d))$.

First, we consider $R_{1k}$, in which 
\begin{align}
G_{nk}\psi_{\hat{\eta}_k(d)} - G_{nk}\psi_{\eta_0(d)} = \sqrt{n} \frac{1}{n}\sum_{i=1}^n \underbrace{\psi_{\hat{\eta}_k(d)}^i - \psi_{\eta_0(d)}^i - E[\psi_{\hat{\eta}_k(d)}^i|I_k^c] + E[\psi_{\eta_0(d)}^i]}_{:= \Delta_{ik}}.
\end{align}
In particular, it can be shown that $E[\Delta_{ik}\Delta_{jk}] =0$ for all $i\neq j$ using the i.i.d. assumption of the data and that the nuisance parameter $\hat{\eta}_k(d)$ is estimated using the auxiliary sample. Then, we have
\begin{align*}
E[R_{1k}^2|I_k^c]&\leq E[\Delta_{ik}^2|I_k^c]\\
&\leq E[(\psi_{\hat{\eta}_k(d)}^i-\psi_{\eta_0(d)}^i)^2|I_k^c]\\
&\leq \sup_{\eta(d)\in T_N(d)} E[(\psi_{\eta(d)}^i-\psi_{\eta_0(d)}^i)^2|I_k^c]\\
&\leq \sup_{\eta(d)\in T_N(d)} E[(\psi_{\eta(d)}^i-\psi_{\eta_0(d)}^i)^2]\\
&\lesssim h^{-1}\varepsilon_N^2 \quad\quad (f)
\end{align*}
and using the conditional Markov's inequality, we conclude that $R_{1k} = O_p(h^{-1/2}\varepsilon_N)$. 

Now we bound $R_{2k}$. Note that by definition, $E[\psi_h(Z,\theta_{0h}(d), \lambda_0,f_D^0(d),\eta_0(d))]= 0$, so it suffices to bound $E[\psi_h(Z,\theta_{0h}(d),\lambda_0,f_D^0(d),\hat{\eta}_k(d))|I_k^c]$. Suppressing other arguments in the score, define
\begin{align}
h_k(r) := E[\psi_h(\eta_0(d) + r(\hat{\eta}_k(d)-\eta_0(d)))|I_k^c]
\end{align}
where by definition $h_k(0) = E[\psi_h(\eta_0(d))|I_k^c]= 0$ and $h_k(1) = E[\psi_h(\hat{\eta}_k(d))|I_k^c]$. Use Taylor's theorem, expand $h_k(1)$ around $0$, we have
\begin{align}
h_k(1) = h_k(0) + h_k'(0) + \frac{1}{2}h_k^{''}(\bar{r}),\quad \bar{r}\in(0,1).
\end{align}
Note that, by Neyman orthogonality,
\begin{align}
h_k'(0) = \partial_\eta(d) E[\psi_h(\eta_0(d))][\hat{\eta}_k(d) - \eta_0(d)] = 0
\end{align}
and use that fact that $h_k(0)=0$, we have
\begin{align*}
R_{2k} &= \sqrt{n}|h_k(1)| = \frac{\sqrt n}{2}|h_k^{''}(\bar{r})|\\
& \leq \sup_{r\in(0,1),\eta(d)\in T_N(d)} \sqrt{n}|\partial_r^2E[\psi_h(\eta_0(d) + r(\hat{\eta}_k(d)-\eta_0(d)))]|\\
&\lesssim \sqrt n h^{-1/2}\varepsilon_N^2\quad\quad (g)
\end{align*}

Combining the above results, we conclude that
\begin{align}
\sqrt{N} R_{n,k} \lesssim h^{-1/2}\varepsilon_N + \sqrt{N}h^{-1/2}\varepsilon_N^2,
\end{align}
and for $\varepsilon_N = o(N^{-1/4})$, we have $\sqrt{N} R_{n,k} = o_p(h^{-1/2})$.\\

\noindent \textit{Step 4: Auxiliary Results}

Following the same argument as the panel case, writing $W_\eta=\{K_h(D-d)-\mathbf1\{D=0\}f_h(d\mid X)/g(X)\}/f_D^0(d)$, differentiation at fixed $\eta$ gives
\[
\partial_\lambda\psi_h
=W_\eta Y\left\{-\frac{T}{\lambda^2}-\frac{1-T}{(1-\lambda)^2}\right\}.
\]
Consequently,
\[
\begin{aligned}
&\partial_\lambda\psi_h(\lambda_0,f_D^0(d),\eta)-\partial_\lambda\psi_h(\lambda_0,f_D^0(d),\eta_0)\\
&=-\frac{\mathbf1\{D=0\}Y}{f_D^0(d)}
\left(\frac{f_h}{g}-\frac{f_h^0}{g_0}\right)
\left\{-\frac{T}{\lambda_0^2}-\frac{1-T}{(1-\lambda_0)^2}\right\}.
\end{aligned}
\]
Overlap and boundedness give the squared $L^2(P)$ bound $O(h^{-1}\varepsilon_N^2)$ in (d). The bounds (e), (f), and (g) follow from the panel arguments with $\Delta Y$ replaced by $\frac{T-\lambda_0}{\lambda_0(1-\lambda_0)}Y$, giving respectively $O(h^{-1}\varepsilon_N^2)$, $O(h^{-1}\varepsilon_N^2)$, and $O(\sqrt n h^{-1/2}\varepsilon_N^2)$.

\noindent \textit{Step 5: Asymptotic Normality}

Combining previous results, we have
\begin{align}
&\widehat{ATT}(d) - ATT(d)\notag \\ 
&= \frac{1}{N}\sum_{i=1}^N\psi_h(Z_i,\theta_{0h}(d),\lambda_0,f_D^0(d),\eta_0(d))\label{csdid:linear1} \\
&+ \frac{1}{N}\sum_{i=1}^N S_f^0(K_h(D_i -d) - E[K_h(D_i -d)]) \label{csdid:linear2}\\
&+ o_p((Nh)^{-1/2}) \label{csdid:linear3}\\
&+ \theta_{0h}(d)-\theta_0(d) \label{csdid:linear4}
\end{align}
where (\ref{csdid:linear1}) and (\ref{csdid:linear2}) are averages of i.i.d. zero-mean terms with the variance growing with kernel bandwidth $h$, and recall that $S_f^0 = E[\partial_f\psi_h(Z,\theta_{0h}(d),\lambda_0,f_D^0(d),\eta_0(d))]$; (\ref{csdid:linear3}) are the terms that vanish when scaled by the (square root of) asymptotic variance; (\ref{csdid:linear4}) is the bias term which is shown to be of order $O(h^2)$ in Lemma \ref{lm:bias}. 

Note that we have arrived at the identical decomposition as in the panel case, and by the same argument, we have
\[
\frac{\widehat{ATT}(d) - ATT(d)}{\sigma_N/\sqrt{N}}\quad \to^d \quad N(0,1)
\]
with $\sigma_N$ defined by
\begin{align}
\sigma_N^2:= E\bigg[\bigg(\psi_h - \frac{\theta_{0h}}{f_D^0(d)}\big(K_h(D-d)-E[K_h(D-d)]\big)\bigg)^2\bigg]
\end{align}
where we have used the fact that $S_f^0 = -\theta_{0h}/f_D^0(d)$. \hfill$\square$\\

\noindent\textbf{Proof of Theorem~\ref{thm:asvar} (Panel)}
The proof uses the same idea as in \cite{CCDDHNR} and \cite{Chang2020}. However, we need to adapt the proof to accommodate the presence of the kernel term. First, recall that the variance estimator is defined as
\begin{align*}
  \hat{\sigma}_N^2 :=& \frac{1}{K}\sum_{k=1}^K E_{n,k}\bigg[
  \bigg(
  \psi_h(Z,\hat{\theta}_h(d),\hat{f}_k(d), \hat{\eta}_k(d)) - \frac{\hat{\theta}_h(d)}{\hat{f}_k(d)}\big(K_h(D-d)-\hat{f}_k(d)\big)
  \bigg)^2
  \bigg]\\
  := & \frac{1}{K}\sum_{k=1}^K E_{n,k}\big[\big(\tilde{\psi}_h(Z,\hat{\theta}_h(d),\hat{f}_k(d), \hat{\eta}_k(d))\big)^2\big]
\end{align*}
where we define 
\begin{align}
  &\tilde{\psi}_h(Z,\theta,f_d,\eta(d))\notag\\ &:= \frac{K_h(D-d)g(X) - \mathbf{1}\{D=0\}f_h(d|X)}{f_dg(X)}\big(\Delta Y - \mathcal{E}_{\Delta Y}(X)\big) - \frac{\theta}{f_d}K_h(D-d).
\end{align}
In particular, note that 
\[
\sigma_N^2= E\big[\tilde{\psi}_h^2(Z,\theta_{0h}(d),f_D^0(d),\eta_0(d))\big] -\bigg(\frac{\theta_{0h}(d)}{f_D^0(d)}\big(E[K_h(D-d)]-f_D^0(d)\big)\bigg)^2.
\]
The subtracted term is $O(h^4)$. Therefore, we need to show that
\begin{align}
  J_k := \big| E_{n,k}\big[\tilde{\psi}_h^2(Z,\hat{\theta}_h(d),\hat{f}_k(d), \hat{\eta}_k(d))\big] -  E\big[\tilde{\psi}_h^2(Z,\theta_{0h}(d),f_D^0(d),\eta_0(d))\big]\big| = o_p(1).
\end{align}
By the triangle inequality, we have
\begin{align}
  J_k &\leq \underbrace{\big| E_{n,k}\big[\tilde{\psi}_h^2(Z,\hat{\theta}_h(d),\hat{f}_k(d), \hat{\eta}_k(d))\big] -  E_{n,k}\big[\tilde{\psi}_h^2(Z,\theta_{0h}(d),f_D^0(d),\eta_0(d))\big]\big|}_{:= J_{1k}}\\
  &+ \underbrace{\big| E_{n,k}\big[\tilde{\psi}_h^2(Z,\theta_{0h}(d),f_D^0(d),\eta_0(d))\big] -  E\big[\tilde{\psi}_h^2(Z,\theta_{0h}(d),f_D^0(d),\eta_0(d))\big]\big|}_{:=J_{2k}}.
\end{align}
We bound each term separately. 

First, we consider $J_{2k}$.
\begin{align}
  E[J_{2k}^2] & = E\big[\big( E_{n,k}\big[\tilde{\psi}_h^2(Z,\theta_{0h}(d),f_D^0(d),\eta_0(d))\big] -  E\big[\tilde{\psi}_h^2(Z,\theta_{0h}(d),f_D^0(d),\eta_0(d))\big]\big)^2\big]\notag \\  &=\operatorname{Var}\bigg[\tilde{\psi}_h^2(Z_i,\theta_{0h}(d),f_D^0(d),\eta_0(d))\bigg]\notag \\
  &\leq \frac{1}{n}E\big[\tilde{\psi}_h^4(Z,\theta_{0h}(d),f_D^0(d),\eta_0(d))\big]\notag \\
  &\lesssim \frac{1}{n}E\big[K_h^4(D-d)+1\big]\notag \\
  &\lesssim (Nh^3)^{-1},
\end{align}
where the third line holds by Cauchy-Schwarz inequality and independence, the fourth line holds by boundedness assumption, and the last line holds by change of variables using the assumptions on the kernel. Therefore, by Chebyshev's inequality, we have $J_{2k} = o_p(1)$ if $Nh^3\to\infty$.

Next, we consider $J_{1k}$. We first state the following convenient fact that will be used in the proof, see \cite{CCDDHNR} and \cite{Chang2020} for example. For any constants $a$ and $\delta$, 
\begin{align}
  |(a+\delta a)^2 - a^2| \leq 2|\delta a|(|a|+|\delta a|).
\end{align}
In our context, we define (for notation simplicity)
\begin{align}
a = \tilde{\psi}_h(Z_i,\theta_{0h}(d),f_D^0(d),\eta_0(d)):=& \psi_i  \\
a + \delta a = \tilde{\psi}_h(Z_i,\hat{\theta}_h(d),\hat{f}_k(d), \hat{\eta}_k(d)) :=& \hat{\psi}_i.
\end{align}
Then, we have
\begin{align}
  J_{1k} &= \bigg| \frac{1}{n}\sum_{i\in I_k} \hat{\psi}_i^2 - \psi_i^2\bigg| \leq \frac{1}{n}\sum_{i\in I_k} \big|\hat{\psi}_i^2 - \psi_i^2 \big|\notag \\
  &\leq \frac{2}{n}\sum_{i\in I_k} |\hat{\psi}_i - \psi_i|(|\psi_i| + |\hat{\psi}_i-\psi_i|)\notag \\
  &\leq 2\bigg(\frac{1}{n}\sum_{i\in I_k} |\hat{\psi}_i - \psi_i|^2\bigg)^{1/2}\bigg(\frac{1}{n}\sum_{i\in I_k} (|\psi_i| + |\hat{\psi}_i-\psi_i|)^2\bigg)^{1/2}\notag \\
  &\leq 2\bigg(\frac{1}{n}\sum_{i\in I_k} |\hat{\psi}_i - \psi_i|^2\bigg)^{1/2}\bigg[\bigg(\frac{1}{n}\sum_{i\in I_k} |\psi_i|^2\bigg)^{1/2} + \bigg(\frac{1}{n}\sum_{i\in I_k}|\hat{\psi}_i-\psi_i|^2\bigg)^{1/2}\bigg]
\end{align}
where the third line holds by Cauchy-Schwarz inequality, and the last line holds by the triangle inequality. Then, we have
\begin{align}
  J_{1k}^2 \lesssim S_N\bigg( S_N + \frac{1}{n}\sum_{i\in I_k} \psi_i^2 \bigg)
\end{align}
where $S_N := \frac{1}{n}\sum_{i\in I_k} |\hat{\psi}_i - \psi_i|^2$. 

We now bound $S_N$. By the definition of $\tilde{\psi}_h$, we have
\begin{align}
  S_N =& \frac{1}{n}\sum_{i\in I_k} \big(\tilde{\psi}_h(Z_i,\hat{\theta}_h(d),\hat{f}_k(d), \hat{\eta}_k(d)) - \tilde{\psi}_h(Z_i,\theta_{0h}(d),f_D^0(d),\eta_0(d))\big)^2\notag \\
  =& \frac{1}{n}\sum_{i\in I_k} \bigg( \tilde{\psi}_h(Z_i,\theta_{0h}(d),\hat{f}_k(d), \hat{\eta}_k(d)) - \tilde{\psi}_h(Z_i,\theta_{0h}(d),f_D^0(d),\eta_0(d))\notag \\
  &+ \frac{\partial}{\partial\theta} \tilde{\psi}_h(Z_i,\bar{\theta},\hat{f}_k(d), \hat{\eta}_k(d))(\hat{\theta}_h(d)-\theta_{0h}(d))\bigg)^2\notag \\
  \lesssim& \underbrace{\frac{1}{n}\sum_{i\in I_k} \big( \tilde{\psi}_h(Z_i,\theta_{0h}(d),\hat{f}_k(d), \hat{\eta}_k(d)) - \tilde{\psi}_h(Z_i,\theta_{0h}(d),f_D^0(d),\eta_0(d))\big)^2}_{:= S_{1N}}\\
  &+ \underbrace{\frac{1}{n}\sum_{i\in I_k} \bigg(\frac{K_h(D_i-d)}{\hat{f}_k(d)}(\hat{\theta}_h(d) - \theta_{0h}(d)) \bigg)^2}_{:= S_{2N}}
\end{align}
where the second line holds by Taylor's theorem with $\bar{\theta}$ between $\theta_{0h}(d)$ and $\hat{\theta}_h(d)$, and the last line holds by the fact that $\frac{\partial}{\partial_\theta }\tilde{\psi}_h(Z_i,\bar{\theta},\hat{f}_k(d), \hat{\eta}_k(d)) = -K_h(D_i-d)/\hat{f}_k(d)$. We bound $S_{1N}$ and $S_{2N}$ separately. 

To bound $S_{2N}$, note that 
\begin{align}
  S_{2N}  = \frac{(\hat{\theta}_h(d) - \theta_{0h}(d))^2}{\hat{f}_k(d)^2}\frac{1}{n}\sum_{i\in I_k} K_h^2(D_i-d).
\end{align}
Since $E[K_h^2(D-d)] = O(h^{-1})$, by Markov's inequality, we have $\frac{1}{n}\sum_{i\in I_k} K_h^2(D_i-d) = O_p(h^{-1})$. Moreover, by Theorem \ref{thm:asymp}, we have $(\hat{\theta}_h(d) - \theta_{0h}(d))^2 = O_p((Nh)^{-1})$. Therefore, we conclude
\begin{align*}
  S_{2N} \leq O_p((Nh^2)^{-1}).
\end{align*}

Next, we bound $S_{1N}$. By Taylor's theorem, for $\bar{f}$ between $f_D^0(d)$ and $\hat{f}_k(d)$, we have
\begin{align}
  &\tilde{\psi}_h(Z_i,\theta_{0h}(d),\hat{f}_k(d), \hat{\eta}_k(d)) - \tilde{\psi}_h(Z_i,\theta_{0h}(d),f_D^0(d),\eta_0(d))\notag\\
  &=\tilde{\psi}_h(Z_i,\theta_{0h}(d),f_D^0(d), \hat{\eta}_k(d)) - \tilde{\psi}_h(Z_i,\theta_{0h}(d),f_D^0(d),\eta_0(d))\\
  &+ \frac{\partial}{\partial f}\tilde{\psi}_h(Z_i,\theta_{0h}(d),\bar{f},\hat{\eta}_k(d))(\hat{f}_k(d) - f_D^0(d)).
\end{align}
Note that 
\begin{align}
  &\big|\frac{\partial}{\partial_f}\tilde{\psi}_h(Z,\theta,f,\eta(d))\big|\notag \\
  &= \bigg|\frac{\partial}{\partial f} \bigg(\frac{K_h(D-d)g(X) - \mathbf{1}\{D=0\}f_h(d|X)}{f g(X)}\big(\Delta Y - \mathcal{E}_{\Delta Y}(X)\big) - \frac{\theta_h}{f}K_h(D-d)\bigg)\bigg|\notag \\
  &\lesssim |K_h(D-d)|+1
\end{align}
where the last line holds by the boundedness assumption. Therefore, by the assumption on the kernel, we have
\begin{align*}
\|\partial_f\tilde{\psi}_h(Z,\theta_{0h}(d),\bar{f},\hat{\eta}_k(d))\|_{P,2} \lesssim \||K_h(D-d)\|+1\|_{P,2} = O(h^{-1/2}).
\end{align*}
Moreover, by definition
\begin{align}
  &\tilde{\psi}_h(Z_i,\theta_{0h}(d),f_D^0(d), \hat{\eta}_k(d)) - \tilde{\psi}_h(Z_i,\theta_{0h}(d),f_D^0(d),\eta_0(d))\notag \\
  &= \psi_h(Z_i,\theta_{0h}(d),f_D^0(d), \hat{\eta}_k(d)) - \psi_h(Z_i,\theta_{0h}(d),f_D^0(d),\eta_0(d)).
\end{align}
Then, by triangle inequality and Cauchy-Schwarz inequality, we have
\begin{align}
  &\|\tilde{\psi}_h(Z_i,\theta_{0h}(d),\hat{f}_k(d), \hat{\eta}_k(d)) - \tilde{\psi}_h(Z_i,\theta_{0h}(d),f_D^0(d),\eta_0(d))\|_{P,2}^2\notag \\
  \lesssim& \|\psi_h(Z_i,\theta_{0h}(d),f_D^0(d), \hat{\eta}_k(d)) - \psi_h(Z_i,\theta_{0h}(d),f_D^0(d),\eta_0(d))\|_{P,2}^2\notag\\
  &+ \|\partial_f\tilde{\psi}_h(Z_i,\theta_{0h}(d),\bar{f},\hat{\eta}_k(d))\|_{P,2}^2\|\hat{f}_k(d) - f_D^0(d)\|_{P,2}^2\notag \\
  \lesssim& h^{-1}\varepsilon_N^2 + h^{-2}N^{-1}.
\end{align}
where the last line holds by the assumptions on the rate of convergence of the $\hat{f}_k(d)$ and \begin{align}
\|\psi_h(Z_i,\theta_{0h}(d),f_D^0(d), \hat{\eta}_k(d)) - \psi_h(Z_i,\theta_{0h}(d),f_D^0(d),\eta_0(d))\|_{P,2}^2\lesssim h^{-1}\varepsilon_N^2
\end{align}
by the same arguments as in the proof of Theorem \ref{thm:asymp}. Then by Markov's inequality, we have
\begin{align}
  S_{1N} = O_p \big( h^{-1}\varepsilon_N^2 + h^{-2}N^{-1}\big).
\end{align}

Combining the results, we have
\begin{align*}
S_N = O_p\big(h^{-1}\varepsilon_N^2 + h^{-2}N^{-1}\big).
\end{align*}

Note that since $|\psi_i|\lesssim |K_h(D-d)|+1$, $\frac{1}{n}\sum_{i\in I_k} \psi_i^2 = O_p(h^{-1})$ by Markov's inequality. This implies that
\begin{align}
  J_{1k}^2 \lesssim S_N\bigg( S_N + \frac{1}{n}\sum_{i\in I_k} \psi_i^2 \bigg) = O_p\big(h^{-2}\varepsilon_N^2 + h^{-3}N^{-1}\big).
\end{align}
Then $J_{1k} = o_p(1)$ if $h^{-2}\varepsilon_N^2 + h^{-3}N^{-1}\to 0$.

Therefore, we conclude that $\hat{\sigma}_N^2 = \sigma_N^2 + o_p(1)$.\hfill$\square$\\

\noindent\textbf{Proof of Theorem~\ref{thm:asvar_rcs} (Repeated Cross-Sections)}
The proof is nearly identical to the panel case, and we only highlight the key differences. Again, the main idea follows \cite{CCDDHNR} and \cite{Chang2020}, and our proof requires modifications to account for the kernel function present in the score function.

First, recall that the variance estimator is defined as
\begin{align}
  \hat{\sigma}_N^2 :=& \frac{1}{K}\sum_{k=1}^K E_{n,k}\bigg[
  \bigg(
\psi_h(Z,\hat{\theta}_h(d),\hat{\lambda}_k,\hat{f}_k(d),\hat{\eta}_k(d)) - \frac{\hat{\theta}_h(d)}{\hat{f}_k(d)}\big(K_h(D-d)-\hat{f}_k(d)\big)
  \bigg)^2
  \bigg]\notag \\
  := & \frac{1}{K}\sum_{k=1}^K E_{n,k}\big[\big(\tilde{\psi}_h(Z,\hat{\theta}_h(d),\hat{\lambda}_k, \hat{f}_k(d), \hat{\eta}_k(d))\big)^2\big]
\end{align}
where we define 
\begin{align}
  \tilde{\psi}_h(Z,\theta,\lambda, f_d,\eta(d)) :=& \frac{K_h(D-d)g(X) - \mathbf{1}\{D=0\}f_h(d|X)}{f_dg(X)}\bigg(\frac{T-\lambda}{\lambda(1-\lambda)} Y - \mathcal{E}_{\lambda Y}(X)\bigg)\\
   &- \frac{\theta}{f_d}K_h(D-d).
\end{align}
In particular, note that 
\[
\sigma_N^2= E\big[\tilde{\psi}_h^2(Z,\theta_{0h}(d),\lambda_0, f_D^0(d),\eta_0(d))\bigg]
-\bigg(\frac{\theta_{0h}(d)}{f_D^0(d)}\big(E[K_h(D-d)]-f_D^0(d)\big)\bigg)^2.
\]
The subtracted term is $O(h^4)$. Therefore, we need to show that
\begin{align}
  J_k := \big| E_{n,k}\big[\tilde{\psi}_h^2(Z,\hat{\theta}_h(d),\hat{\lambda}_k, \hat{f}_k(d), \hat{\eta}_k(d))\big] -  E\big[\tilde{\psi}_h^2(Z,\theta_{0h}(d),\lambda_0, f_D^0(d),\eta_0(d))\big]\big| = o_p(1).
\end{align}
By the triangle inequality, we have
\begin{align}
  J_k &\leq \underbrace{\big| E_{n,k}\big[\tilde{\psi}_h^2(Z,\hat{\theta}_h(d),\hat{\lambda}_k, \hat{f}_k(d), \hat{\eta}_k(d))\big] -  E_{n,k}\big[\tilde{\psi}_h^2(Z,\theta_{0h}(d),\lambda_0, f_D^0(d),\eta_0(d))\big]\big|}_{:= J_{1k}}\\
  &+ \underbrace{\big| E_{n,k}\big[\tilde{\psi}_h^2(Z,\theta_{0h}(d),\lambda_0, f_D^0(d),\eta_0(d))\big] -  E\big[\tilde{\psi}_h^2(Z,\theta_{0h}(d),\lambda_0, f_D^0(d),\eta_0(d))\big]\big|}_{:=J_{2k}}.
\end{align}
We bound each term separately. 

Similar to the panel case, by boundedness and assumptions on the kernel function, $J_{2k}$ satisfies 
\begin{align}
  E[J_{2k}^2] \lesssim \frac{1}{n}E\big[K_h^4(D-d)+1\big] \lesssim (Nh^3)^{-1}.
\end{align}
Therefore, by Markov's inequality, we have $J_{2k} = o_p(1)$ if $nh^3\to\infty$.

Next, we bond $J_{1k}$. For notation simplicity, we define
\begin{align}
\psi_i :=& \tilde{\psi}_h(Z_i,\theta_{0h}(d),\lambda_0,f_D^0(d),\eta_0(d))  \\
\hat{\psi}_i:=& \tilde{\psi}_h(Z_i,\hat{\theta}_h(d),\hat{\lambda}_k,\hat{f}_k(d), \hat{\eta}_k(d)).
\end{align}
Then, using the same argument as in the panel case, we have
\begin{align}
  J_{1k}^2 \lesssim S_N\bigg( S_N + \frac{1}{n}\sum_{i\in I_k} \psi_i^2 \bigg)
\end{align}
where $S_N := \frac{1}{n}\sum_{i\in I_k} |\hat{\psi}_i - \psi_i|^2$. 

By triangle inequality, we have
\begin{align}
  S_N =& \frac{1}{n}\sum_{i\in I_k} \big(\tilde{\psi}_h(Z_i,\hat{\theta}_h(d),\hat{\lambda}_k,\hat{f}_k(d), \hat{\eta}_k(d)) - \tilde{\psi}_h(Z_i,\theta_{0h}(d),\lambda_0,f_D^0(d),\eta_0(d))\big)^2\notag \\
  =& \frac{1}{n}\sum_{i\in I_k} \bigg( \tilde{\psi}_h(Z_i,\theta_{0h}(d),\hat{\lambda}_k,\hat{f}_k(d), \hat{\eta}_k(d)) - \tilde{\psi}_h(Z_i,\theta_{0h}(d),\lambda_0, f_D^0(d),\eta_0(d))\notag \\
  &+ \frac{\partial}{\partial\theta} \tilde{\psi}_h(Z_i,\bar{\theta},\hat{\lambda}_k,\hat{f}_k(d), \hat{\eta}_k(d))(\hat{\theta}_h(d)-\theta_{0h}(d))\bigg)^2\notag \\
  \lesssim& \underbrace{\frac{1}{n}\sum_{i\in I_k} \big( \tilde{\psi}_h(Z_i,\theta_{0h}(d),\hat{\lambda}_k,\hat{f}_k(d), \hat{\eta}_k(d)) - \tilde{\psi}_h(Z_i,\theta_{0h}(d),\lambda_0,f_D^0(d),\eta_0(d))\big)^2}_{:= S_{1N}}\\
  & + \underbrace{\frac{1}{n}\sum_{i\in I_k} \bigg(\frac{K_h(D_i-d)}{\hat{f}_k(d)}(\hat{\theta}_h(d) - \theta_{0h}(d)) \bigg)^2}_{:= S_{2N}}
\end{align}
where the second line holds by Taylor's theorem with $\bar{\theta}$ between $\theta_{0h}(d)$ and $\hat{\theta}_h(d)$, and the last line holds by the fact that $\frac{\partial}{\partial_\theta }\tilde{\psi}_h(Z_i,\bar{\theta},\hat{\lambda}_k, \hat{f}_k(d), \hat{\eta}_k(d)) = -K_h(D_i-d)/\hat{f}_k(d)$. 

Note that, using the identical argument as in the panel case,
\begin{align}
S_{2N} = O_p(h^{-1})\times O_p((Nh)^{-1}).
\end{align}
Moreover, by Taylor's theorem, for $\bar{f}$ between $f_D^0(d)$ and $\hat{f}_k(d)$, and for $\bar{\lambda}$ between $\lambda_0$ and $\hat{\lambda}_k$, we have
\begin{align}
  &\|\tilde{\psi}_h(Z_i,\theta_{0h}(d),\hat{\lambda}_k,\hat{f}_k(d), \hat{\eta}_k(d)) - \tilde{\psi}_h(Z_i,\theta_{0h}(d),\lambda_0, f_D^0(d),\eta_0(d))\|_{P,2}^2\notag \\
  &\lesssim \|\psi_h(Z_i,\theta_{0h}(d),\lambda_0, f_D^0(d),\hat{\eta}_k(d)) - \psi_h(Z_i,\theta_{0h}(d),\lambda_0,f_D^0(d),\eta_0(d))\|_{P,2}^2 \\
  &+ \|\partial_\lambda\tilde{\psi}_h(Z_i,\theta_{0h}(d),\bar{\lambda},\bar{f},\hat{\eta}_k(d))\|_{P,2}^2\|\hat{\lambda}_k - \lambda_0\|_{P,2}^2 \\
  &+ \|\partial_f\tilde{\psi}_h(Z_i,\theta_{0h}(d),\bar{\lambda},\bar{f},\hat{\eta}_k(d))\|_{P,2}^2\|\hat{f}_k(d) - f_D^0(d)\|_{P,2}^2
\end{align}
By boundedness assumption, we have 
\begin{align}
\big|\frac{\partial}{\partial_f}\tilde{\psi}_h(Z,\theta,\lambda,f,\eta(d))\big| &\lesssim |K_h(D-d)|+1 \\
\big|\frac{\partial}{\partial_\lambda}\tilde{\psi}_h(Z,\theta,\lambda, f,\eta(d))\big| &\lesssim |K_h(D-d)|+1
\end{align}
and by the same argument as in the panel case, we have
\begin{align}
  \|\partial_\lambda\tilde{\psi}_h(Z_i,\theta_{0h}(d),\bar{\lambda},\bar{f},\hat{\eta}_k(d))\|_{P,2}^2 & = O\big(h^{-1}\big)\notag \\
  \|\partial_f\tilde{\psi}_h(Z_i,\theta_{0h}(d),\bar{\lambda},\bar{f},\hat{\eta}_k(d))\|_{P,2}^2 & = O\big(h^{-1}\big).
\end{align}
By assumptions, $\|\hat{\lambda}_k - \lambda_0\|_{P,2}^2 = O(N^{-1})$ and $\|\hat{f}_k(d) - f_D^0(d)\|_{P,2}^2 = O((Nh)^{-1})$. Moreover, $\|\psi_h(Z_i,\theta_{0h}(d),\lambda_0,f_D^0(d),\hat{\eta}_k(d)) - \psi_h(Z_i,\theta_{0h}(d),\lambda_0,f_D^0(d),\eta_0(d))\|_{P,2}^2\lesssim h^{-1}\varepsilon_N^2$ by the same arguments as in the proof of Theorem \ref{thm:asymp}. Therefore, by Markov's inequality, we have
\begin{align}
S_{1N} = O_p \big( h^{-1}\varepsilon_N^2 + h^{-2}N^{-1}\big).
\end{align}
Combining the results, we have
\begin{align}
S_N = O_p\big(h^{-1}\varepsilon_N^2 + h^{-2}N^{-1}\big).
\end{align}

Since $|\psi_i|\lesssim |K_h(D-d)|+1$, $\frac{1}{n}\sum_{i\in I_k} \psi_i^2 = O_p(h^{-1})$ by Markov's inequality. This implies that
\begin{align}
  J_{1k}^2 \lesssim S_N\bigg( S_N + \frac{1}{n}\sum_{i\in I_k} \psi_i^2 \bigg) = O_p\big(h^{-2}\varepsilon_N^2 + h^{-3}N^{-1}\big).
\end{align}
Then $J_{1k} = o_p(1)$ if $h^{-2}\varepsilon_N^2 + h^{-3}N^{-1}\to 0$.

Therefore, we conclude that $\hat{\sigma}_N^2 = \sigma_N^2 + o_p(1)$. \hfill$\square$\\

\noindent\textbf{Proof of Theorem~\ref{thm:unifexp}:}
We focus on the panel case as the repeated cross-sectional case only requires minor modifications. To simplify notation, let $\theta_0(d)$ denote the true $ATT(d)$, $\theta_{0h}(d)$ denote the true $ATT_h(d)$, and $\hat{\theta}_h(d)$ and $\hat{\theta}_h^*(d)$ denote our cross-fitted estimator and bootstrap estimators respectively. We only need to show the uniform asymptotic linear expansion of the bootstrap estimator since the cross-fitted estimator can be treated as a special case with $\xi_i = 1$.

Let $T_N$ be the set of $\eta(d) = (g(X), f_h(d|X), \mathcal{E}_{\Delta Y}(X))$ for $d\in\mathcal{D}$ s.t. $\sup_{d\in\mathcal{D}}\|f_h(d|X) - f_h^0(d|X)\|_{P,2}\leq h^{-1/2}\varepsilon_N$, $\|g(X) - g_0(X)\|_{P,2}\leq \varepsilon_N$, $\|\mathcal{E}_{\Delta Y}(X) - \mathcal{E}_{\Delta Y}^0(X)\|_{P,2}\leq \varepsilon_N$, $\kappa<g(X)<1-\kappa$, $c <f_h(d|X)<C$ a.s. $\forall d\in\mathcal{D}$, $\sup_{d\in\mathcal{D}}\|\partial_d f_h(d|X)\|_{P,\infty}< C$, and $\|\mathcal{E}_{\Delta Y}(X)\|_{P,\infty}<C$. Set $a_N=\sqrt{\log(N)/(Nh)}$. The uniform kernel bound gives $\sup_{d\in\mathcal D}|\hat f_k^*(d)-f_D^0(d)|=O_p(a_N)$ for all $k$, so $\inf_{d}\hat f_k^*(d)>c/2$ with probability tending to one. Expectations involving fitted functions below are conditional on the auxiliary sample and its multipliers.

Recall that the bootstrap estimator has the following form
\begin{align}
\hat{\theta}_h^*(d) &:= \frac{1}{K}\sum_{k=1}^K \frac{1}{n}\sum_{i\in I_k} \xi_i \frac{K_h(D_i-d)\hat{g}_k(X_i) - \mathbf{1}\{D_i=0\}\hat{f}_{h,k}(d|X_i)}{\hat{f}_k^*(d)\hat{g}_k(X_i)}\big(\Delta Y_i - \hat{\mathcal{E}}_{\Delta Y,k}(X_i)\big).
\end{align}
Then we have
\begin{align}
\hat{\theta}_h^*(d) - \theta_0(d) & = \hat{\theta}_h^*(d) - \theta_{0h}(d) + \theta_{0h}(d)- \theta_0(d) \end{align}
and we can focus on the stochastic part $\hat{\theta}_h^*(d) - \theta_{0h}(d)$.

As before, denote the full random sample by $I_N$, and each equal size subsample by $I_k$ for $k=1,\cdots, K$. Let the score $\psi_h$ be defined as in (\ref{psi1}), let $E_{n,k}(f) = \frac{1}{n}\sum_{i\in I_k} f(Z_i)$ denote the empirical average of a generic function $f$ over the subsample $I_k$, and similarly let $E_{n,k}^*(f) = \frac{1}{n}\sum_{i\in I_k} \xi_i f(Z_i)$ denote the multiplier version of the empirical average. Write $\mu_h(d)=E[K_h(D-d)]$, $b_h(d)=\mu_h(d)-f_D^0(d)$, and $\bar\xi_N=N^{-1}\sum_i\xi_i$. Since multiplier averaging multiplies a constant by $\bar\xi_N$,
\[
\hat\theta_h^*(d)-\theta_{0h}(d)
=\frac1K\sum_k E_{n,k}^*[\psi_h(Z,\theta_{0h}(d),\hat f_k^*(d),\hat\eta_k(d))]
+\theta_{0h}(d)(\bar\xi_N-1).
\]
Each observation belongs to $K-1$ auxiliary samples. Moreover, since
\[
\frac1K\sum_k(\hat f_k^*(d)-f_D^0(d))
=\frac1N\sum_i\xi_i\{K_h(D_i-d)-\mu_h(d)\}+b_h(d)+\mu_h(d)(\bar\xi_N-1),
\]
combining the $\theta_{0h}(d)(\bar\xi_N-1)$ with the non-centered terms from this density expansion gives
\begin{align*}
&\theta_{0h}(d)(\bar\xi_N-1) -\frac{\theta_{0h}(d)}{f_D^0(d)}
(b_h(d)+\mu_h(d)(\bar\xi_N-1))\\
&=-\frac{\theta_{0h}(d)b_h(d)\bar\xi_N}{f_D^0(d)}\\
&= E[\partial_{f}\psi_h(Z,\theta_{0h}(d), f_D^0(d),\eta_0(d))](E[K_h(D-d)] - f_D^0(d))\bar\xi_N.
\end{align*}
The last line holds by the fact that $E[\partial_f \psi_h(Z,\theta_{0h}(d),f_D^0(d),\eta_0(d))] = - \theta_{0h}(d)/f_D^0(d)$. Taylor's theorem therefore gives us the following expansion: 
\begin{align}
&\hat{\theta}_h^*(d) - \theta_{0h}(d)\notag \\
&= \frac{1}{K}\sum_{k=1}^K E_{n,k}^*[\psi_h(Z,\theta_{0h}(d),f_D^0(d),\hat\eta_k(d))]
+\theta_{0h}(d)(\bar\xi_N-1)\notag \\
&\quad+\frac{1}{K}\sum_{k=1}^K E_{n,k}^*[\partial_f\psi_h(Z,\theta_{0h}(d),f_D^0(d),\hat\eta_k(d))]
(\hat f_k^*(d)-f_D^0(d))\notag \\
&\quad+\frac{1}{2K}\sum_{k=1}^K E_{n,k}^*[\partial_f^2\psi_h(Z,\theta_{0h}(d),\bar f_k,\hat\eta_k(d))]
(\hat f_k^*(d)-f_D^0(d))^2\notag \\
&= \frac{1}{K}\sum_{k=1}^K E_{n,k}^*\Big[\psi_h(Z,\theta_{0h}(d),f_D^0(d),\eta_0(d)) - \frac{\theta_{0h}(d)}{f_D^0(d)}\Big(K_h(D - d) - E[K_h(D-d)]\Big)\Big]  \tag{M} \label{unif eq:1}\\
&+ \frac{1}{K}\sum_{k=1}^K \Big(E_{n,k}^*[\psi_h(Z,\theta_{0h}(d),f_D^0(d),\hat{\eta}_k(d))] - E_{n,k}^*[\psi_h(Z,\theta_{0h}(d),f_D^0(d),\eta_0(d))]\Big) \tag{R1}\label{unif eq:2}\\
&+ \frac{1}{K}\sum_{k=1}^K \Big(E_{n,k}^*[\partial_{f}\psi_h(Z,\theta_{0h}(d), f_D^0(d),\hat{\eta}_k(d))] - E[\partial_{f}\psi_h(Z,\theta_{0h}(d), f_D^0(d),\eta_0(d))]\Big)\notag\\
&\quad\quad\quad\times (\hat{f}_k^*(d) - f_D^0(d))\notag \\
& \quad + E[\partial_{f}\psi_h(Z,\theta_{0h}(d), f_D^0(d),\eta_0(d))](E[K_h(D-d)] - f_D^0(d))\bar\xi_N \tag{R2} \label{unif eq:3}\\
& + \frac{1}{2K}\sum_{k=1}^K E_{n,k}^*[\partial_f^2\psi_h(Z,\theta_{0h}(d), \bar{f}_k,\hat{\eta}_k(d))](\hat{f}_k^*(d) - f_D^0(d))^2 \tag{R3} \label{unif eq:4}
\end{align}
where $\bar{f}_k$ lies between $f_D^0(d)$ and $\hat f_k^*(d)$. Note that the identity $E[\partial_f \psi_h(Z,\theta_{0h}(d),f_D^0(d),\eta_0(d))] = - \theta_{0h}(d)/f_D^0(d)$ is also used in (\ref{unif eq:1}). Therefore, based on this decomposition, our goal is to show that the remainder terms satisfy
\begin{align}
    \sup_{d\in\mathcal{D}} |Rj(d)| = o_p((Nh)^{-1/2}).
\end{align}
for each $j = 1, 2, 3$.\\ 

\noindent \textit{Step 1: Second Order Term R3}

Let $A_k(Z,d):=f_D^0(d)\{\psi_h(Z,\theta_{0h}(d),f_D^0(d),\hat\eta_k(d))+\theta_{0h}(d)\}$.
% Taylor's theorem applied to $v\mapsto 1/v$ gives
% \[
% \frac1{\hat f_k^*(d)}
% =\frac1{f_D^0(d)}
% -\frac{\hat f_k^*(d)-f_D^0(d)}{[f_D^0(d)]^2}
% +\frac{(\hat f_k^*(d)-f_D^0(d))^2}{\bar f_k^3},
% \]
% for an intermediate value $\bar f_k$ common to all observations in $I_k$.
Since $\psi_h(Z,\theta_{0h}(d),v,\hat\eta_k(d))=A_k(Z,d)/v-\theta_{0h}(d)$,
\[
E_{n,k}^*[\partial_f^2\psi_h(Z,\theta_{0h}(d),\bar f_k,\hat\eta_k(d))]
=\frac{2}{\bar f_k^3}E_{n,k}^*[A_k(Z,d)].
\]
The bound $|A_k(Z,d)|\lesssim |K_h(D-d)|+1$ gives a uniformly bounded conditional mean. The bounds on the score and the maximal inequality used in Step 4 imply $\max_k\sup_d|E_{n,k}^*[A_k(Z,d)]|=O_p(1)$. Uniform consistency of $\hat f_k^*$ ensures $\bar f_k\ge c/2$ uniformly with probability tending to one, so the empirical second derivative is uniformly $O_p(1)$. Hence
\[
\sup_d|R3(d)|=O_p(a_N^2)=o_p((Nh)^{-1/2}).
\]
Here $\sqrt{Nh}\,a_N^2=\log N/\sqrt{Nh}\to0$ by the assumption on the bandwidth.

\noindent \textit{Step 2: First Order Term R2}

To bound the first order term R2, we follow the similar steps as above. By triangle inequality, we have
\begin{align}
&|E_{n,k}^*[\partial_f\psi_h(Z,\theta_{0h}(d), f_D^0(d), \hat{\eta}_k(d))] - E[\partial_f \psi_h(Z,\theta_{0h}(d), f_D^0(d), \eta_0(d))]| \notag \\
&\leq |E_{n,k}^*[\partial_f\psi_h(Z,\theta_{0h}(d), f_D^0(d), \hat{\eta}_k(d))] - E[\partial_f\psi_h(Z,\theta_{0h}(d), f_D^0(d), \hat{\eta}_k(d))]| \tag{R2.1}\\
& + |E[\partial_f\psi_h(Z,\theta_{0h}(d), f_D^0(d), \hat{\eta}_k(d))] - E[\partial_f\psi_h(Z,\theta_{0h}(d), f_D^0(d), \eta_0(d))]|\tag{R2.2}
\end{align}

Follow the same reasoning as before, we can show that, for the R2.1 term,
\begin{align}
    &\sup_{d\in\mathcal{D}} |E_{n,k}^*[\partial_f\psi_h(Z,\theta_{0h}(d), f_D^0(d),\hat{\eta}_k(d))] - E[\partial_f \psi_h(Z,\theta_{0h}(d), f_D^0(d),\hat{\eta}_k(d))]|\notag \\
    &\lesssim \sup_{d\in\mathcal{D}} |E_{n,k}^*[\psi_h(Z,\theta_{0h}(d), f_D^0(d),\hat{\eta}_k(d))] - E[\psi_h(Z,\theta_{0h}(d), f_D^0(d),\hat{\eta}_k(d))]| +|E_{n,k}^*1-1| \notag\\
    & = O_p\Big(\sqrt{\log(N)/(Nh)}\Big)\tag{c} \label{ep:2}
\end{align}
where we have used the fact that 
\begin{align}
\partial_f\psi_h(Z,\theta_{0h}(d), f_D^0(d),\hat{\eta}_k(d)) = -\frac{1}{f_D^0(d)}(\psi_h(Z,\theta_{0h}(d), f_D^0(d), \hat{\eta}_k(d)) + \theta_{0h}(d)). 
\end{align}
and that $f_D^0(d)$ is uniformly bounded below from zero. We argue that (\ref{ep:2}) holds at the end of this proof.

For (R2.2), by the definition of the derivative and the orthogonality bound (\ref{rate:3}) in the later step, we have
\[
\begin{aligned}
&\sup_d\big|E[\partial_f\psi_h(Z,\theta_{0h}(d),f_D^0(d),\hat\eta_k(d))\mid I_k^c]
-E[\partial_f\psi_h(Z,\theta_{0h}(d),f_D^0(d),\eta_0(d))]\big|\\
&=\sup_d\frac{|E[\psi_h(Z,\theta_{0h}(d),f_D^0(d),\hat\eta_k(d))\mid I_k^c]|}{f_D^0(d)}
\lesssim h^{-1/2}\varepsilon_N^2.
\end{aligned}
\]
Therefore, the product term in $R2$ is $O_p(a_N^2+a_Nh^{-1/2}\varepsilon_N^2)=o_p((Nh)^{-1/2})$ since $\sqrt N\varepsilon_N^2=o(1)$. Specifically, the two terms multiplied by $\sqrt{Nh}$ satisfy
\[
\sqrt{Nh}\,a_N^2=\frac{\log N}{\sqrt{Nh}}\to0,
\qquad
\sqrt{Nh}\,a_Nh^{-1/2}\varepsilon_N^2
=a_N\sqrt N\varepsilon_N^2\to0.
\]

Moreover, note that by the assumptions on the kernel function and the density, we have
\begin{align}
\sup_{d\in\mathcal{D}} |E[K_h(D-d)] - f_D^0(d)| = O(h^2).
\end{align}
Therefore,
\begin{align}
\sup_{d\in\mathcal{D}}\big|E[\partial_{f}\psi_h(Z,\theta_{0h}(d), f_D^0(d),\eta_0(d))]b_h(d)\bar\xi_N\big|=O_p(h^2)=o_p((Nh)^{-1/2})
\end{align}
where we have used the fact that $E[\partial_{f}\psi_h(Z,\theta_{0h}(d), f_D^0(d),\eta_0(d))] = -\theta_{0h}(d)/f_D^0(d)$, the boundedness assumption, $\bar\xi_N=1+O_p(N^{-1/2})$, and that the kernel bandwidth is assumed to be undersmoothing. This establishes $\sup_{d\in\mathcal{D}} |R2(d)| = o_p((Nh)^{-1/2})$.\\

\noindent \textit{Step 3: ``Neyman" Term R1}

To bound (\ref{unif eq:2}), note that
\begin{align}
|E_{n,k}^*[\psi_h(Z,\theta_{0h}(d),f_D^0(d),\hat{\eta}_k(d))] - E_{n,k}^*[\psi_h(Z,\theta_{0h}(d),f_D^0(d),\eta_0(d))]| \leq R_{11} + R_{12}.
\end{align}
Specifically,
\begin{align}
    R_{11} = & \Big|E_{n,k}^*[\psi_h(Z,\theta_{0h}(d),f_D^0(d),\hat{\eta}_k(d))] - E[\psi_h(Z,\theta_{0h}(d),f_D^0(d),\hat{\eta}_k(d))|I_k^c] \notag \\
     &- E_{n,k}^*[\psi_h(Z,\theta_{0h}(d),f_D^0(d),\eta_0(d))] + E[\psi_h(Z,\theta_{0h}(d),f_D^0(d),\eta_0(d))]\Big|\notag \\
    := & |\dot{P}_{n,k}^*(\psi_h(Z,\theta_{0h}(d),f_D^0(d),\hat{\eta}_k(d))) - \dot{P}_{n,k}^*(\psi_h(Z,\theta_{0h}(d),f_D^0(d),\eta_0(d)))|
\end{align}
with $\dot{P}_{n,k}^*(f) = \frac{1}{n}\sum_{i=1}^n \xi_if(Z_i) - E[f(Z_i)]$ denoting the centered sample average with multiplier. With some abuse of notation, we will also use $\dot{P}_{n,k}^*$ to denote the centered process conditional on the auxiliary sample. Note that since $\xi_i\perp Z_i$ and $E[\xi_i]=1$, then $E[f(Z_i)] = E[\xi_i f(Z_i)]$. Moreover,
\begin{align}
R_{12} := |E[\psi_h(Z,\theta_{0h}(d),f_D^0(d),\hat{\eta}_k(d))|I_k^c] - E[\psi_h(Z,\theta_{0h}(d),f_D^0(d),\eta_0(d))]|.
\end{align}

First, we bound $R_{11}$. To simplify notation, denote %$\psi_\eta(d)^i(d) := \psi_h(Z, \theta_{0h}(d), f_D^0(d), \eta(d))$ and 
\begin{align}
\Delta_i(\hat{\eta}_k(d), d) := \psi_h(Z_i, \theta_{0h}(d), f_D^0(d), \hat{\eta}_k(d)) - \psi_h(Z_i, \theta_{0h}(d), f_D^0(d), \eta_0(d))
\end{align}
and we can write 
\begin{align}
    R_{11}(d) = |\dot{P}_{n,k}^*[\Delta_i(\hat{\eta}_k(d), d)]|.
\end{align}
Conditional on the auxiliary sample, we can take $\hat{\eta}_k(d)$ as fixed. Then, we can define the following function class
\begin{align}
    \mathcal{F}_N := \{\Delta(\hat{\eta}_k(d), d): d\in\mathcal{D}\}
\end{align}
and establish that
\begin{align}
    \sup_{d\in\mathcal{D}} R_{11}(d) = \|\dot{P}_{n,k}^*[\Delta(\hat{\eta}_k(d), d)]\|_{\mathcal{F}_N} = o_p( (Nh)^{-1/2} ). \tag{e} \label{ep:3}
\end{align}
We will show (\ref{ep:3}) at the end of the proof.

To bound $R_{12}$, first, note that by definition $E[\psi_h(Z_i, \theta_{0h}(d), f_D^0(d), \eta_0(d))] = 0$ for all $d$, then it suffices to consider only $E[\psi_h(Z_i, \theta_{0h}(d), f_D^0(d), \hat{\eta}_k(d))|I_k^c]$. For notation simplicity, suppressing other arguments in the score and define
\begin{align}
h_k(r,d) := E[\psi_h(\eta_0(d) + r(\hat{\eta}_k(d)-\eta_0(d)), d)|I_k^c]
\end{align}
where by definition $h_k(0,d) = E[\psi_h(\eta_0(d), d)|I_k^c]= 0$ for all $d\in\mathcal{D}$, and $h_k(1,d) = E[\psi_h(\hat{\eta}_k(d), d)|I_k^c]$. Using Taylor's theorem to expand $h_k(1,d)$ around $0$ in its first argument, we have
\begin{align}
h_k(1,d) = h_k(0,d) + h_k'(0,d) + \frac{1}{2}h_k^{''}(\bar{r},d),\quad \bar{r}\in(0,1).  
\end{align}
Note that, by Neyman orthogonality, for all $d\in\mathcal{D}$
\begin{align}
h_k'(0,d) = \partial_\eta(d) E[\psi_h(\eta_0(d),d)][\hat{\eta}_k(d) - \eta_0(d)] = 0
\end{align}
and use that fact that $h_k(0,d)=0$, we have
\begin{align}
\sup_{d\in\mathcal{D}} R_{12}(d) &= \sup_{d\in\mathcal{D}}|h_k(1,d)| = \frac{1}{2}\sup_{d\in\mathcal{D}} |h_k^{''}(\bar{r},d )| \notag \\
& \leq \sup_{r\in(0,1), d\in\mathcal{D}, \eta(d)\in T_N} |\partial_r^2E[\psi_h(\eta_0(d) + r(\eta(d)-\eta_0(d)), d)]| \notag \\
&\lesssim h^{-1/2}\varepsilon_N^2. \tag{f} \label{rate:3}
\end{align}
We will show (\ref{rate:3}) at the end of this proof. Note that since $\varepsilon_N = o(N^{-1/4})$, we have $\sup_{d\in\mathcal{D}} R_{12}(d) = o( (Nh)^{-1/2})$.

Therefore, we conclude that $\sup_{d\in\mathcal{D}} |R1(d)| = o_p((Nh)^{-1/2})$.\\

\noindent \textit{Step 4: Auxiliary Results}

We focus on showing that (\ref{ep:3}) holds. We adapt and credit this part of the proof to \cite{FHLZ22} (see Lemma 8.2 and its proof in the supplementary material). Recall (\ref{ep:3}) states the following
\[
\|\dot{P}_{n,k}^*[\Delta(\hat{\eta}_k(d), d)]\|_{\mathcal{F}_N} = o_p( (Nh)^{-1/2} )
\]
where $\mathcal{F}_N := \{\Delta(\hat{\eta}_k(d), d): d\in\mathcal{D}\}$ and recall that 
\[
\Delta_i(\hat{\eta}_k(d), d) := \psi_h(Z_i, \theta_{0h}(d), f_D^0(d), \hat{\eta}_k(d)) - \psi_h(Z_i, \theta_{0h}(d), f_D^0(d), \eta_0(d)).
\]
Let $\epsilon >0$ be given and let $A_N(\epsilon)$ denote the event that the nuisance parameters $\hat{\eta}_k(d)$ estimated using the auxiliary sample $I_k^c$ satisfy Assumption \ref{cdidas4} (c)--(d) with probability $1-\epsilon$.

Conditional on the auxiliary sample, consider the multiplier class $\mathcal G_N=\{(Z,\xi)\mapsto\xi\Delta(\hat\eta_k(d),d):d\in\mathcal D\}$ with envelope $F(Z,\xi)=C|\xi|h^{-1}$. Since $\xi$ is normally distributed, subexponential tails give
\[
\left\|\max_{i\in I_k}F(Z_i,\xi_i)\right\|_2\lesssim h^{-1}\log N.
\]
On $A_N(\epsilon)$, bounded derivatives yield $|\partial_d\Delta(\hat\eta_k(d),d)|\lesssim h^{-2}$. A grid of mesh proportional to $h\epsilon$ therefore gives
\[
\sup_Q\log N_c(\epsilon\|F\|_{Q,2},\mathcal G_N,L^2(Q))\lesssim\log\frac{C}{h\epsilon}.
\]

Moreover, on $A_N(\epsilon)$,
\begin{align}
&\sup_{d\in\mathcal{D}} E[\Delta_i^2(\hat{\eta}_k(d), d)|I_k^c] \notag \\
\leq& \sup_{d\in\mathcal{D}, \eta(d)\in T_N} E[\Delta_i^2(\eta(d), d)|I_k^c] \notag \\
\leq& \sup_{d\in\mathcal{D}, \eta(d)\in T_N} E[\Delta_i^2(\eta(d), d)] \notag \\
=& \sup_{d\in\mathcal{D}, \eta(d)\in T_N} E[(\psi_h(Z_i, \theta_{0h}(d), f_D^0(d), \eta(d)) - \psi_h(Z_i, \theta_{0h}(d), f_D^0(d), \eta_0(d)))^2] \notag \\
\lesssim&  h^{-1}\varepsilon_N^2.
\end{align}

Using $E[\xi_i^2]=2$ and, without loss of generality, $\varepsilon_N\ge N^{-1/2}$, by \cite{CCK2014a} Corollary 5.1, we have
\begin{align}
    E\big[ \| \dot{P}_{n,k}^*[\Delta(\hat{\eta}_k(d),d)]\|_{\mathcal{F_N}}|I_k^c \big]\mathbf{1}\{A_N(\epsilon)\} \lesssim \sqrt{\frac{h^{-1}\varepsilon_N^2\log(N)}{N}} + \frac{\log^2(N)h^{-1}}{N}.
\end{align}

Then, let $\delta >0$ be given, as $n\to\infty$,
\begin{align}
    &P\Big( \| \dot{P}_{n,k}^*[\Delta(\hat{\eta}_k(d),d)]\|_{\mathcal{F_N}} \geq \delta (Nh)^{-1/2} \Big)\\
    \leq& \epsilon +  P\Big( \| \dot{P}_{n,k}^*[\Delta(\hat{\eta}_k(d),d)]\|_{\mathcal{F_N}}\mathbf{1}\{A_N(\epsilon)\} \geq \delta (Nh)^{-1/2} \Big)\\
    \leq& \epsilon +  E\Big[P\Big( \| \dot{P}_{n,k}^*[\Delta(\hat{\eta}_k(d),d)]\|_{\mathcal{F_N}} \geq \delta (Nh)^{-1/2} |I_k^c \Big)\mathbf{1}\{A_N(\epsilon)\}\Big]\\
    \leq& \epsilon +  E\Bigg[\frac{E[ \| \dot{P}_{n,k}^*[\Delta(\hat{\eta}_k(d),d)]\|_{\mathcal{F_N}} |I_k^c]}{\delta (Nh)^{-1/2}} \mathbf{1}\{A_N(\epsilon)\}\Bigg]\\
    \lesssim & \epsilon + \frac{\sqrt{\frac{h^{-1}\varepsilon_N^2\log(N)}{N}} + \frac{\log^2(N)h^{-1}}{N}}{\delta (Nh)^{-1/2}}\\
    \lesssim & 2\epsilon.
\end{align}
Since $\varepsilon_N\sqrt{\log N}\to0$ and $\log^2(N)/\sqrt{Nh}\to0$, (\ref{ep:3}) follows. The same maximal inequality gives (\ref{ep:2}) and the weighted kernel bound used above. The bound (\ref{rate:3}) follows from the argument for auxiliary result (d) in the proof of Theorem~\ref{thm:asymp}, applied uniformly over $d\in\mathcal D$ using Assumption~\ref{cdidas4}.

Therefore, we conclude that
\begin{align}
&\hat{\theta}_h^*(d) - \theta_{0h}(d)\notag \\
&= \frac{1}{K}\sum_{k=1}^K E_{n,k}^*\Big[\psi_h(Z,\theta_{0h}(d),f_D^0(d),\eta_0(d)) - \frac{\theta_{0h}(d)}{f_D^0(d)}\Big(K_h(D - d) - E[K_h(D-d)]\Big)\Big]\notag\\
&+ R^*(d)
\end{align}
where $\sup_{d\in\mathcal{D}}|R^*(d)| = o_p((Nh)^{-1/2})$.

Note that we can establish a similar result for $\hat{\theta}_h(d) - \theta_{0h}(d)$ by replacing multiplier $\xi_i$ with constant $1$:
\begin{align}
&\hat{\theta}_h(d) - \theta_{0h}(d)\notag \\
&= \frac{1}{K}\sum_{k=1}^K E_{n,k}\Big[\psi_h(Z,\theta_{0h}(d),f_D^0(d),\eta_0(d)) - \frac{\theta_{0h}(d)}{f_D^0(d)}\Big(K_h(D - d) - E[K_h(D-d)]\Big)\Big]\notag \\ 
&+ R'(d)
\end{align}
where $\sup_{d\in\mathcal{D}}|R'(d)| = o_p((Nh)^{-1/2})$.

Then, taking the difference
\begin{align}
    &\hat{\theta}_h^*(d) - \hat{\theta}_h(d)\\
    = & \hat{\theta}_h^*(d) - \theta_{0h}(d) - [\hat{\theta}_h(d) - \theta_{0h}(d)] \\
    = & \frac{1}{N}\sum_{i=1}^N(\xi_i - 1)\Bigg[\psi_h(Z_i,\theta_{0h}(d),f_D^0(d),\eta_0(d)) - \frac{\theta_{0h}(d)}{f_D^0(d)}\big(K_h(D_i-d)-E[K_h(D-d)]\big)\Bigg]\\
    +& R^*(d) - R'(d).
\end{align}
By the remainder bounds, we conclude that
\begin{align}
\sup_{d\in\mathcal{D}} |R^*(d)-R'(d)|=o_p((Nh)^{-1/2}).
\end{align}
\hfill$\square$

\end{document}